\documentclass[11pt, a4paper]{article}
\usepackage{jheppub}
\usepackage{epstopdf}
\input{epsf.tex}
\usepackage{amsfonts}
\usepackage{amsmath}
\usepackage{colortbl}
\usepackage{booktabs}
\usepackage{multirow}
\usepackage{anysize}
\usepackage{array}
\usepackage{latexsym, amscd, amsthm}
\usepackage{pdfsync}
\usepackage{epsf}
\usepackage[all]{xy}
\usepackage[usenames,dvipsnames]{xcolor}
\usepackage{enumerate}
\usepackage{float}
\restylefloat{table}
\usepackage{upgreek}
\usepackage{subcaption}
\usepackage{array}
\usepackage{multirow}

\DeclareSymbolFont{extraup}{U}{zavm}{m}{n}
\DeclareMathSymbol{\varheart}{\mathalpha}{extraup}{86}
\DeclareMathSymbol{\vardiamond}{\mathalpha}{extraup}{87}

\usepackage{fancyhdr}

\newtheorem{Theorem}{Theorem}[section]
\newtheorem{Lemma}[Theorem]{Lemma}
\newtheorem{Proposition}[Theorem]{Proposition}
\theoremstyle{definition}
\newtheorem{Definition}[Theorem]{Definition}
\theoremstyle{remark}

\newcommand{\Hom}{{\rm Hom}}

\newcommand{\Sym}{\mathrm{Sym}}

\newcommand{\im}{\mathrm{im}}

\newcommand{\bcE}{\boldsymbol{\check{E}}}

\newcommand{\bp}{\begin{Proposition}}
\newcommand{\ep}{\end{Proposition}}
\newcommand{\bl}{\begin{Lemma}}
\newcommand{\el}{\end{Lemma}}
\newcommand{\bt}{\begin{Theorem}}
\newcommand{\et}{\end{Theorem}}
\newcommand{\bd}{\begin{Definition}}
\newcommand{\ed}{\end{Definition}}

\newcommand{\ev}{\mathrm{ev}}
\newcommand{\eqdef}{\stackrel{{\rm def.}}{=}}

\newcommand{\cinf}{{{\cal \cC}^\infty(M,\R)}}

\DeclareFontFamily{U}{rsf}{}
\DeclareFontShape{U}{rsf}{m}{n}{<5> <6> rsfs5 <7> <8> <9> rsfs7 <10-> rsfs10}{}
\DeclareMathAlphabet\Scr{U}{rsf}{m}{n}

\newcommand{\KA}{K\"{a}hler-Atiyah~}
\def\cU{\mathcal{U}}
\def\cW{\mathcal{W}}
\def\hV{{\hat V}}

\def\N{\mathbb{N}}
\def\Z{\mathbb{Z}}

\def\R{\mathbb{R}}

\def\H{\mathbb{H}}

\def\rk{{\rm rk}}
\def\corank{{\rm corank}}
\def\deg{{\rm deg}}

\def\dd{\mathrm{d}}

\def\AdS{\mathrm{AdS}}

\def\supp{\mathrm{supp}}

\def\Int{\mathrm{Int}}
\def\Fr{\mathrm{Fr}}
\def \fr{\mathrm{fr}}
\def\sp{\mathrm{sp}}
\def\nsp{\mathrm{nsp}}
\def\Sing{\mathrm{Sing}}

\def\cy{\mathrm{cf}}

\def\c{\mathrm{c}}
\def\nc{\mathrm{nc}}
\def\gen{\mathrm{gen}}
\def\ind{\mathrm{ind}}
\def\bcF{\bar{\cF}}

\newcommand{\be}{\begin{equation*}}
\newcommand{\ee}{\end{equation*}}
\newcommand{\ben}{\begin{equation}}
\newcommand{\een}{\end{equation}}
\newcommand{\beqa}{\begin{eqnarray*}}
\newcommand{\eeqa}{\end{eqnarray*}}
\newcommand{\beqan}{\begin{eqnarray}}
\newcommand{\eeqan}{\end{eqnarray}}

\newcommand{\nn}{\nonumber}

\newcommand{\id}{\mathrm{id}}

\newcommand{\tr}{\mathrm{tr}}

\def\Hess{\mathrm{Hess}}
\def\Diff{\mathrm{Diff}}
\def\per{\mathrm{per}}

\def\cC{{\mathcal C}}
\def\cB{\Scr B}

\def\cK{\mathrm{\cal K}}
\def\cM{\mathrm{\cal M}}

\def\Spin{\mathrm{Spin}}

\def\Pin{\mathrm{Pin}}

\def\SO{\mathrm{SO}}

\def\cD{\mathcal{D}}

\def\cE{\mathcal{E}}

\def\cN{\mathcal{N}}
\def\cG{\mathcal{G}}

\def\cF{\mathcal{F}}
\def\cC{\mathcal{C}}
\def\cH{\mathcal{H}}
\def\SU{\mathrm{SU}}

\def\G_2{\mathrm{G_2}}

\def\cL{\mathcal{L}}
\def\cS{\mathcal{S}}

\def\cV{\mathcal{V}}
\def\btu{\bigtriangleup}

\def\mb{\mathbf{b}}

\def\mf{\mathbf{f}}
\def\mF{\mathbf{F}}
\def\momega{{\boldsymbol{\omega}}}

\def\f{\mathfrak{f}}
\def\m{\mathrm{min}}
\def\M{\mathrm{max}}
\def\H{\mathfrak{H}}
\def\dotC{\overset{\bullet}{C}}
\def\nc{\mathrm{nc}}
\def\nsp{\mathrm{nsp}}
\def\sp{\mathrm{sp}}

\title{Singular foliations for M-theory compactification}

\author{Elena Mirela Babalic$^{1,2}$, Calin Iuliu Lazaroiu$^3$}

\affiliation{$^1$ Department of Theoretical Physics, National
  Institute of Physics and Nuclear Engineering, Str. Reactorului
  no.30, P.O.BOX MG-6, Postcode 077125, Bucharest-Magurele, Romania  \\ 
$^2$ Department of Physics, 
University of Craiova, 13 Al. I. Cuza Str., Craiova 200585, Romania\\
$^3$ Center for Geometry and Physics, Institute for Basic Science (IBS), Pohang 790-784, Republic of Korea}

\emailAdd{mbabalic@theory.nipne.ro, calin@ibs.re.kr} 

\abstract{We use the theory of singular foliations to study $\cN=1$
  compactifications of eleven-dimensional supergravity on
  eight-manifolds $M$ down to $\AdS_3$ spaces, allowing for the
  possibility that the internal part $\xi$ of the supersymmetry
  generator is chiral on some locus $\cW$ which does not coincide with
  $M$. We show that the complement $M\setminus \cW$ must be a {\em dense}
  open subset of $M$ and that $M$ admits a singular foliation ${\bar
    \cF}$ endowed with a longitudinal $G_2$ structure and defined by a
  closed one-form $\momega$, whose geometry is determined by the
  supersymmetry conditions. The singular leaves are those leaves which
  meet $\cW$.  When $\momega$ is a Morse form, the chiral locus is a
  finite set of points, consisting of isolated zero-dimensional leaves
  and of conical singularities of seven-dimensional leaves. In that
  case, we describe the topology of $\bcF$ using results from
  Novikov theory.  We also show how this description fits in with
  previous formulas which were extracted by exploiting the
  $\Spin(7)_\pm$ structures which exist on the complement of $\cW$.}

\begin{document}

\maketitle 

\pagebreak

\vskip .6in

\section*{Introduction}

$\cN=1$ flux compactifications of eleven-dimensional supergravity on
eight-manifolds $M$ down to $\AdS_3$ spaces \cite{MartelliSparks,
  Tsimpis} provide a vast extension of the better studied class of
compactifications down to 3-dimensional Minkowski space \cite{Becker1,
  Becker2, Constantin}, having the advantage that they are already
consistent at the classical level \cite{MartelliSparks}.  They form a
useful testing ground for various proposals aimed at providing unified
descriptions of flux backgrounds \cite{Grana} and may be relevant to
recent attempts to gain a better understanding of F-theory
\cite{Bonetti}.  When the internal part $\xi$ of the supersymmetry
generator is everywhere non-chiral, such backgrounds can be studied
\cite{g2} using foliations endowed with longitudinal $G_2$ structures,
an approach which permits a geometric description of the supersymmetry
conditions while providing powerful tools for studying the topology of
such backgrounds.
 
In this paper, we extend the results of \cite{g2} to the general case
when the internal part $\xi$ of the supersymmetry generator is allowed
to become chiral on some locus $\cW\subset M$. Assuming that $\cW\neq
M$, i.e. that $\xi$ is not everywhere chiral, we show that, at the
classical level, the Einstein equations imply that the chiral locus
$\cW$ must be a set with empty interior, which therefore is negligible
with respect to the Lebesgue measure of the internal space. As a
consequence, the behavior of geometric data along this locus can be
obtained from the non-chiral locus $\cU\eqdef M\setminus \cW$ through
a limiting process. The geometric information along the non-chiral
locus $\cU$ is encoded \cite{g2} by a regular foliation $\cF$ which
carries a longitudinal $G_2$ structure and whose geometry is
determined by the supersymmetry conditions in terms of the
supergravity four-form field strength.  When $\emptyset \neq
\cW\subsetneq M$, we show that $\cF$ extends to a singular foliation
$\bcF$ of the whole manifold $M$ by adding leaves which are
allowed to have singularities at points belonging to $\cW$. This
singular foliation ``integrates'' a cosmooth\footnote{Note that
  $\cD$ is {\em not} a singular distribution in the sense of
  Stefan-Sussmann \cite{Stefan, Sussmann} (it is cosmooth rather than
  smooth). See Appendix \ref{app:gendist}.}  \cite{BulloLewis, Drager, Michor, Ratiu}
singular distribution $\cD$ (a.k.a. generalized sub-bundle of $TM$),
defined as the kernel distribution of a closed one-form $\momega$
which belongs to a cohomology class $\f\in H^1(M,\R)$ determined by
the supergravity four-form field strength. The set of zeroes of
$\momega$ coincides with the chiral locus $\cW$. In the most general
case, $\bcF$ can be viewed as a Haefliger structure
\cite{Haefliger} on $M$. The singular foliation $\bcF$ carries a
longitudinal $G_2$ structure, which is allowed to degenerate at the
singular points of singular leaves. On the non-chiral locus $\cU$,
the problem can be studied using the approach of \cite{g2} or the
approach advocated in \cite{Tsimpis}, which makes use of two
$\Spin(7)_\pm$ structures.  We show explicitly how one can translate
between these two approaches and prove that the results of \cite{g2}
agree with those of \cite{Tsimpis} along this locus.

While the topology of singular foliations defined by a closed one-form
can be extremely complicated in general, the situation is better
understood in the case when $\momega$ is a Morse one-form. The Morse
case is generic in the sense that such 1-forms constitute an open and
dense subset of the set of all closed one-forms belonging to the
cohomology class $\f$.  In the Morse case, the singular foliation
$\bcF$ can be described using the {\em foliation graph}
\cite{MelnikovaThesis, MelnikovaGraph, FKL} associated to the
corresponding decomposition of $M$ (see \cite{Melnikova2, Melnikova3,
  FKL} and \cite{Gelbukh1}--\cite{Gelbukh9}), which provides a
combinatorial way to encode some important aspects of the foliation's
topology --- up to neglecting the information contained in the
so-called {\em minimal components} of the decomposition, components
which should possess an as yet unexplored non-commutative geometric
description. This provides a far-reaching extension of the picture
found in \cite{g2} for the everywhere non-chiral case $\cU=M$, a case
which corresponds to the situation when the foliation graph is reduced
to either a circle (when $\cF$ has compact leaves, being a fibration
over $S^1$) or to a single so-called exceptional vertex (when $\cF$
has non-compact dense leaves, being a minimal foliation). In the
minimal case of the backgrounds considered \cite{g2}, the exceptional
vertex corresponds to a noncommutative torus which encodes the
noncommutative geometry \cite{ConnesFol, ConnesNG} of the leaf space.

The paper is organized as follows. Section 1 gives a brief review of
the class of compactifications we consider, in order to fix notations
and conventions. Section 2 discusses a geometric characterization of
Majorana spinors $\xi$ on $M$ which is inspired by the rigorous
approach developed in \cite{ga1,ga2, gf} for the method of bilinears
\cite{Tod}, in the case when the spinor $\xi$ is allowed to be chiral
at some loci. It also gives the \KA parameterizations of this spinor
which correspond to the approach of \cite{g2} and to that of
\cite{Tsimpis} and describes the relevant $G$-structures using both
spinors and idempotents in the \KA algebra of $M$. In the same
section, we give the general description of the singular foliation
$\bcF$ as the Haefliger structure defined by the closed one-form
$\momega$.  Section 3 describes the relation between the $G_2$ and
$\Spin(7)_\pm$ parameterizations of the fluxes as well as the relation
between the torsion classes of the leafwise $G_2$ structure and the
Lee form and characteristic torsion of the $\Spin(7)_\pm$ structures
defined on the non-chiral locus. The same section gives the comparison
of the approach of \cite{g2} with that of \cite{Tsimpis} along that
locus. Section 4 discusses the topology of the singular foliation
$\bcF$ in the Morse case while Section 5 concludes. The
appendices contain various technical details.

\paragraph{Notations and conventions. } 
Throughout this paper, $M$ denotes an oriented, connected and compact
smooth manifold (which will mostly be of dimension eight), whose
unital commutative $\R$-algebra of smooth real-valued functions we
denote by $\Omega^0(M)=\cinf$. Given a subset $A$ of $M$, we let
${\bar A}$ denote the closure of $A$ in $M$ (taken with respect to the
manifold topology of $M$). The {\em large topological frontier} (also
called {\em topological boundary}) of $A$ is defined as $\Fr(A)\eqdef
{\bar A}\setminus \Int(A)$, where $\Int(A)$ denotes the interior of
$A$. The {\em small topological frontier} is $\fr(A)\eqdef {\bar
  A}\setminus A$. Notice that $\fr(A)\subseteq\Fr(A)$ and that
$\fr(A)=\Fr(A)$ when $A$ is open, in which case we speak simply of the
{\em frontier} of $A$.  All fiber bundles we consider are
smooth\footnote{The ``generalized bundles''\cite{Drager, BulloLewis}
  considered occasionally in this paper are {\em not} fiber
  bundles.}. We use freely the results and notations of
\cite{ga1,ga2,gf,g2}, with the same conventions as there. To simplify
notation, we write the geometric product $\diamond$ of
\cite{ga1,ga2,gf} simply as juxtaposition while indicating the wedge
product of differential forms through $\wedge$.  If $\cD\subset TM$ is
a singular (a.k.a. generalized) distribution on $M$ and $\cU$ is an
open subset of $M$ such that $\cD|_\cU$ is a regular Frobenius
distribution (see Appendix \ref{app:gendist}), we let
$\Omega_\cU(\cD)=\Gamma(\cU,\wedge (\cD|_\cU)^\ast)$ denote the
$\cC^\infty(\cU,\R)$-module of $\cD|_\cU$-longitudinal differential
forms defined on $\cU$.  When $\dim M=8$, then for any 4-form
$\omega\in \Omega^4(M)$ we let $\omega^\pm\eqdef \frac{1}{2}(\omega\pm
\ast \omega)$ denote the selfdual and anti-selfdual parts of $\omega$
(namely, $\ast \omega^\pm=\pm \omega^\pm$). When $M$ is
eight-dimensional, we let $\Omega^{4\pm}(M)$ denote the spaces of
selfdual and anti-selfdual four-forms, respectively.  We use the
``{\rm Det}'' convention for the wedge product and the corresponding
``{\rm Perm}'' convention for the symmetric product. Hence given a
local coframe $e^a$ of $M$, we have:
\ben
\label{DetPerm}
\begin{split}
&e^{a_1}\wedge \ldots \wedge e^{a_k}\eqdef \sum_{\sigma\in S_k}\epsilon(\sigma) e^{a_{\sigma(1)}}\otimes \ldots \otimes e^{a_{\sigma(k)}}~~,\\
&e^{a_1}\odot \ldots \odot e^{a_k}\eqdef \sum_{\sigma\in S_k}e^{a_{\sigma(1)}}\otimes \ldots \otimes e^{a_{\sigma(k)}}~~,
\end{split}
\een
{\em without} prefactors of $\frac{1}{k!}$ in the right hand side,
where $S_k$ is the symmetric group on $k$ letters and
$\epsilon(\sigma)$ denotes the signature of a permutation $\sigma$.
This is the convention used, for example, in \cite{Spivak}. We use
$\Sym^2_0(T^\ast M)$ to denote the space of traceless symmetric covariant
2-tensors on $M$ and $\Sym^2_{\cU,0}(\cD^\ast)$ to denote the space of
traceless symmetric covariant 2-tensors defined on $\cU$ and which are
longitudinal to the Frobenius distribution $\cD|_\cU$, when $\cD$ is
as above. By definition, a $\Spin(7)_+$ structure on $M$ is a
$\Spin(7)$ structure with respect to the orientation chosen for $M$
while a $\Spin(7)_-$ structure is a $\Spin(7)$ structure with respect
to the opposite orientation.

\section{Basics}
\label{sec:basics}
We start with a brief review of the set-up, in order to fix notation. 
As in \cite{MartelliSparks, Tsimpis}, we consider 11-dimensional
supergravity \cite{sugra11} on an eleven-dimensional connected and paracompact 
spin manifold $\mathbf{M}$ with Lorentzian metric $\mathbf{g}$ (of `mostly plus'
signature).  Besides the metric, the classical action of the theory contains the
three-form potential with four-form field strength
$\mathbf{G}\in\Omega^4(\mathbf{M})$ and the gravitino
$\mathbf{\Psi}$, which is a Majorana spinor of spin $3/2$. The bosonic part of the action takes the form: 
\be
S_{\rm bos}[\mathbf{g}, \mathbf{C}]=
\frac{1}{2\mathbf{\kappa}_{11}^2}\int_{\mathbf M}R\boldsymbol{\nu}-
\frac{1}{4\mathbf{\kappa}_{11}^2}\int_{\mathbf M}\big(\mathbf{G}\wedge \star \mathbf{G}+\frac{1}{3}\mathbf{C}\wedge \mathbf{G}\wedge \mathbf{G}\big)~~,
\ee
where $\mathbf{\kappa}_{11}$ is the gravitational coupling constant in eleven dimensions, $\boldsymbol{\nu}$ and $R$ are the volume form and the scalar curvature of $\mathbf{g}$ and 
$\mathbf{G}=\dd \mathbf{C}$.
For supersymmetric bosonic classical backgrounds, both the gravitino and
its supersymmetry variation must vanish, which requires that there
exist at least one solution $\boldsymbol{\eta}$ to the equation:
\ben
\label{susy}
\delta_{\boldsymbol{\eta}} \mathbf{\Psi} \eqdef  \mathfrak{D} \boldsymbol{\eta} = 0~~,
\een
where $\mathfrak{D}$ denotes the supercovariant connection. The
eleven-dimensional supersymmetry generator $\boldsymbol{\eta}$ is a
Majorana spinor (real pinor) of spin $1/2$ on $\mathbf{M}$.

As in \cite{MartelliSparks, Tsimpis}, consider compactification down
to an $\AdS_3$ space of cosmological constant $\Lambda=-8\kappa^2$,
where $\kappa$ is a positive real parameter --- this includes the
Minkowski case as the limit $\kappa\rightarrow 0$.  Thus $\mathbf{
  M}=N\times M$, where $N$ is an oriented 3-manifold diffeomorphic
to $\R^3$ and carrying the $\AdS_3$ metric $g_3$ while $M$ is an oriented, 
compact and connected Riemannian eight-manifold whose metric we denote by $g$. The metric on
$\mathbf{M}$ is a warped product:
\beqan
\label{warpedprod}
\dd \mathbf{s}^2  & = & e^{2\Delta} \dd s^2~~~{\rm where}~~~\dd s^2=\dd s^2_3+ g_{mn} \dd x^m \dd x^n~~.
\eeqan
The warp factor $\Delta$ is a smooth real-valued function defined on $M$ while $\dd s_3^2$ is the
squared length element of the $\AdS_3$ metric $g_3$. For the field strength $\mathbf{G}$, we use the ansatz:
\ben
\label{Gansatz}
\mathbf{ G} = \nu_3\wedge \mathbf{f}+\mF~~,~~~~\mathrm{with}~~ 
\mF\eqdef e^{3\Delta}F~~,~~\mathbf{f}\eqdef e^{3\Delta} f~~,
\een
where $f\in \Omega^1(M)$, $F\in \Omega^4(M)$ and $\nu_3$ is the
volume form of $(N,g_3)$. For $\boldsymbol{\eta}$, we use the ansatz:
\be
\boldsymbol{\eta}=e^{\frac{\Delta}{2}}(\zeta\otimes \xi)~~,
\ee
where $\xi$ is a Majorana spinor of spin $1/2$ on the internal space
$(M,g)$ (a section of the rank 16 real vector bundle $S$ of indefinite chirality real 
pinors) and $\zeta$ is a Majorana spinor on $(N,g_3)$. 

Assuming that $\zeta$ is a Killing spinor on the $\AdS_3$ space $(N,g_3)$, the
supersymmetry condition \eqref{susy} is equivalent with the following
system for $\xi$:
\ben
\label{par_eq}
\boxed{\mathbb{D}\xi = 0~~,~~Q\xi = 0}~~,
\een 
where 
\be
\mathbb{D}_X=\nabla_X^S+\frac{1}{4}\gamma(X\lrcorner F)+\frac{1}{4}\gamma((X_\sharp\wedge f) \nu) +\kappa \gamma(X\lrcorner \nu)~~,~~X\in \Gamma(M,TM)
\ee
is a linear connection on $S$ (here $\nabla^S$ is the connection induced on $S$ by the Levi-Civita
connection of $(M,g)$, while $\nu$ is the volume form of $(M,g)$) and 
\be
Q=\frac{1}{2}\gamma(\dd \Delta)-\frac{1}{6}\gamma(\iota_f\nu)-\frac{1}{12}\gamma(F)-\kappa\gamma(\nu)
\ee 
is a globally-defined endomorphism of $S$. As in \cite{MartelliSparks,
  Tsimpis}, {\em we do not require that $\xi$ has definite chirality}.

The set of solutions of \eqref{par_eq} is a finite-dimensional
$\R$-linear subspace $\cK(\mathbb{D},{Q})$ of the infinite-dimensional
vector space $\Gamma(M,S)$ of smooth sections of $S$. Up to rescalings by
smooth nowhere-vanishing real-valued functions defined on $M$, the
vector bundle $S$ has two admissible pairings $\cB_\pm$ (see \cite{gf,
  AC1, AC2}), both of which are symmetric but which are distinguished
by their types $\epsilon_{\cB_\pm}=\pm 1$. Without loss of generality,
we choose to work with $\cB\eqdef \cB_+$.  We can in fact take $\cB$
to be a scalar product on $S$ and denote the corresponding norm by
$||~||$ (see \cite{ga1,ga2} for details). Requiring that the
background preserves exactly $\cN=1$ supersymmetry amounts to asking
that $\dim \cK(\mathbb{D},Q)=1$. It is not hard to check \cite{ga1} that
$\cB$ is $\mathbb{D}$-flat:
\ben
\label{flatness}
\dd \cB(\xi',\xi'')=\cB(\mathbb{D}\xi', \xi'')+\cB(\xi',\mathbb{D}\xi'')~~,
~~\forall \xi',\xi''\in \Gamma(M,S)~~.
\een
Hence any solution of \eqref{par_eq} which has unit $\cB$-norm at a
point will have unit $\cB$-norm at every point of $M$ and we can take
the internal part $\xi$ of the supersymmetry generator to be
everywhere of norm one.

\section{Parameterizing a Majorana spinor on $M$}
\label{sec:fierz}

\subsection{Globally valid parameterization}

Fixing a Majorana spinor $\xi\in \Gamma(M,S)$ which is everywhere of
$\cB$-norm one, consider the inhomogeneus differential form:
\ben
\label{checkE}
\check{E}_{\xi,\xi}=\frac{1}{16} \sum_{k=0}^8 \bcE^{(k)}_{\xi,\xi}\in \Omega(M)~~,
\een
whose rescaled rank components have the following expansions in any
local orthonormal coframe $(e^a)_{a=1\ldots 8}$ of $M$ defined on some
open subset $U$:
\be
\bcE^{(k)}_{\xi,\xi}=_U\frac{1}{k!}\cB(\xi,\gamma_{a_1...a_k}\xi)e^{a_1...a_k} \in \Omega^k(M)~~.
\ee
The conditions:
\ben
\label{Esquare}
\check{E}^2=\check{E}~~,~~~\cS(\check{E})=1~~,~~\tau(\check{E})=\check{E}~~
\een
encode the fact that an inhomogeneous form $\check{E}\eqdef\check{E}_{\xi,\xi} $ is of the type \eqref{checkE} for some Majorana spinor $\xi$ which is everywhere of norm one. 
As a result of the last condition in \eqref{Esquare}, the non-zero components of $\check{E}$ have ranks $k=0,1,4,5$ and we have
$\cS(\check{E}_{\xi,\xi})=\bcE^{(0)}_{\xi,\xi}=||\xi||^2=1$, where
$\cS$ is the canonical trace of the \KA algebra. Hence:
\ben
\label{Eexp}
\boxed{\check{E}=\frac{1}{16}(1+V+Y+Z+b\nu)}~~,
\een
where we introduced the notations: 
\ben
\label{forms8}
V\eqdef \bcE^{(1)}  ~~,~~ Y\eqdef\bcE^{(4)} ~,~~ Z\eqdef \bcE^{(5)}~~,~~ b\nu\eqdef \bcE^{(8)}~~.
\een
Here, $b$ is a smooth real valued function defined on $M$ and $\nu$ is
the volume form of $(M,g)$, which satisfies $||\nu||=1$; notice the
relation $\cS(\nu\check{E}_{\xi,\xi})=b$. On a small enough open subset
$U\subset M$ supporting a local coframe $(e^a)$ of $M$, one has the
expansions:
\beqan
\label{forms8alt}
&&V=_U\cB(\xi,\gamma_a\xi)e^a ~~,~~ Y=_U\frac{1}{4!}\cB(\xi,\gamma_{a_1\ldots a_4}\xi) e^{a_1\ldots a_4}~,\nn\\
&& Z=_U\frac{1}{5!} \cB(\xi,\gamma_{a_1\ldots a_5}\xi) e^{a_1\ldots a_5}~~,
~~ b=_U\cB(\xi, \gamma(\nu)\xi)~~.
\eeqan
One finds \cite{ga1} that \eqref{Esquare} is equivalent with the following relations which hold globally on $M$:
\ben
\label{SolMS}
\boxed{
\begin{split}
& ||V||^2=1-b^2\geq0~~,~~||Y^\pm||^2=\frac{7}{2}(1\pm b)^2~~,\\
& \iota_V(\ast Z)=0~~,~~\iota_V Z=Y-b\ast Y~~,\\
&(\iota_\alpha (\ast Z)) \wedge (\iota_\beta (\ast Z)) \wedge  (\ast Z) 
= - 6 \langle \alpha\wedge V, \beta\wedge V\rangle \iota_V \nu~~,~~\forall \alpha,\beta\in \Omega^1(M)~~.
\end{split}}
\een
Notice that the first relation in the second row is equivalent with $V\wedge Z=0$, 
which means that $V$ and $Z$ commute in the \KA algebra of $(M,g)$. 
\paragraph{Remark.} 
Let (R) denote the second relation (namely $\iota_V Z=Y-b\ast Y$) on
the second row of \eqref{SolMS}.  Separating the selfdual and
anti-selfdual parts shows that (R) is {\em equivalent} with the
following two conditions:
\ben
\label{iVZpm}
(\iota_V Z)^\pm=(1\mp b)Y^\pm~~.
\een

\paragraph{Proposition.} Relations \eqref{SolMS} imply that the following normalization
conditions hold globally on $M$:
\ben
\label{Ynorms}
||Y||^2=7(1+b^2)~~,~~||Z||^2=7 (1-b^2)~~.  
\een
\noindent{\bf Proof.}  The first equation in \eqref{Ynorms} follows
from the last relations on the first row of \eqref{SolMS} by noticing
that $||Y||^2=||Y^+||^2+||Y^-||^2$ (since $\langle Y^+,Y^-\rangle=0$).
We have:
\ben
\label{intmd}
||\iota_VZ||^2=||\ast \iota_V Z||^2=||V\wedge (\ast
Z)||^2=||V||^2||\ast Z||^2=||V||^2||Z||^2~~, 
\een 
where in the middle equality we used the first equation on the second
row of \eqref{SolMS}, which tells us that $\ast Z$ is orthogonal on
$V$.  The second equation in \eqref{Ynorms} now follows from
\eqref{intmd} and from the identity: 
\be
||\iota_VZ||^2=(1-b)^2||Y^+||^2+(1+b)^2||Y^-||^2=7(1-b^2)=7||V||^2~~,
\ee 
where we used \eqref{iVZpm} and both relations in the first row of
\eqref{SolMS}. $\blacksquare$

\paragraph{The twisted selfdual and twisted anti-selfdual parts of $\check{E}$.}
The identity $\nu^2=1$ implies that the elements: 
\be
R^\pm\eqdef \frac{1}{2}(1\pm \nu)~~ 
\ee
are complementary idempotents in the \KA algebra:
\ben
\label{pirels}
(R^\pm)^2=R^\pm~~,~~R^\pm R^\mp=0~~,~~R^++R^-=\id_{\Omega(M)}~~.
\een
The (anti)selfdual part of a four-form $\omega\in \Omega^4(M)$ can be expressed as: 
\be
\omega_\pm=R^\pm\omega~~.
\ee
Notice that this relation also gives the twisted (anti)selfdual parts \cite{ga1} of an inhomogeneous form $\omega\in \Omega(M)$.
The identities: 
\be
YR^\pm=R^\pm Y=Y^\pm~~,~~(1+b\nu)R^\pm=(1\pm b)R^\pm~~
\ee
allow us to compute the twisted selfdual part $\check{E}^+$ and twisted anti-selfdual part $\check{E}^-$ of $\check{E}$:
\ben
\label{Epm}
\check{E}^\pm=\check{E}R^\pm=\frac{1}{16}\left[(1\pm b+V+Z)R_\pm + Y^\pm\right]\in \Omega(M)~~.
\een
The following decomposition holds globally on $M$:
\be
\check{E}=\check{E}^++\check{E}^-~~.
\ee

\subsection{The chirality decomposition of $M$} 
Let $S^\pm\subset S$ be the rank eight subbundles of $S$ consisting of
positive and negative chirality spinors (the eigen-subbundles of
$\gamma(\nu)$ corresponding to the eigenvalues $+1$ and $-1$).  Since
$\gamma(\nu)$ is $\cB$-symmetric, $S^+$ and $S^-$ give a
$\cB$-orthogonal decomposition $S=S^+\oplus S^-$.  Decomposing a
normalized spinor as $\xi=\xi^++\xi^-$ with $\xi^\pm\eqdef
\frac{1}{2}(\id_S\pm \gamma(\nu))\xi\in \Gamma(M,S^\pm)$, we have:
\be
||\xi||^2=||\xi^+||^2+||\xi^-||^2=1
\ee
and: 
\be
b=\cB(\xi,\gamma(\nu)\xi)=||\xi^+||^2-||\xi^-||^2~~.
\ee
These two relations give: 
\ben
\label{xipmnorms}
\boxed{||\xi^\pm||^2=\frac{1}{2}(1\pm b)}~~.
\een
Notice that $b$ equals $\pm 1$ at a point $p\in M$ iff $\xi_p\in
S^\pm_p$.  Since $||V||^2=1-b^2$, the one-form $V$ vanishes at $p$
iff $\xi_p$ is chiral i.e. iff $\xi_p\in S_p^+\cup S_p^-$. Consider
the {\em non-chiral locus} (an open subset of $M$):
\be
\cU \eqdef \{p\in M|\xi\not \in S_p^+\cup S_p^-\}=\{p\in M|\xi^+_p\neq 0~\mathrm{and}~~\xi_p^-\neq 0\}=\{p\in M|V_p\neq  0\}=\{p\in M||b(p)| < 1\}~~,
\ee
and its closed complement, the {\em chiral locus}:
\be
\cW\eqdef M\setminus \cU=\{p\in M|\xi_p\in S_p^+\cup S^-_p\}=\{p\in M|\xi^+_p=0~\mathrm{or}~\xi^-_p=0\}=\{p\in M|V_p=0\}=\{p\in M| |b(p)|=1\}~~.
\ee
The chiral locus $\cW$ decomposes further as a disjoint union of two closed
subsets, the {\em positive and negative chirality loci}:
\be
\cW=\cW^+\sqcup \cW^-~~,
\ee
where:
\beqa
\cW^\pm\eqdef \{p\in M|\xi_p\in S^\pm_p\}=\{p\in M| b(p)=\pm 1\}=\{p\in M|\xi_p^\mp=0\}~~.
\eeqa
The extreme cases $\cW^+=M$ or $\cW^-=M$, as well as $\cW^+=\cW^-=\emptyset$
are allowed. However, the case $\cU=\emptyset$ with both $\cW^+$ and
$\cW^-$ nonempty (then $M=\cW^+\sqcup \cW^-$) is forbidden (recall that $b$ is
smooth and hence continuous while $M$ is connected). Since $\xi$ does not 
vanish on $M$, we have: 
\be
\cU^\pm\eqdef \cU\cup \cW^\pm=\{p\in M|\xi^\pm_p\neq 0\}~~.
\ee

\paragraph{Remark.} 
Since $|b|\leq 1$ on $M$, the sets $\cW^\pm$ (when non-empty) consist
of critical points of $b$, namely the absolute maxima and minima of
$b$ on $M$. Hence the differential of $b$ vanishes at every point of
$\cW$. In general $\cW^\pm$ can be quite `wild' (they can
be very far from being immersed submanifolds of $M$).

\subsection{A topological no-go theorem}
\label{subsec:nogo}

Recall that $M$ is compact. The following result clarifies the kind of
topologies of the chiral loci which are of physical interest.

\paragraph{Theorem}  
Assume that the supersymmetry conditions, the Bianchi identity and
equations of motion for $G$ as well as the Einstein equations are
satisfied. There exist only the following four possibilities:
\begin{enumerate}
\item The set $\cW^+$ coincides with $M$ and hence $\cW^-$ and $\cU$
  are empty. In this case, $\xi$ is a chiral spinor of positive
  chirality which is covariantly constant on $M$ and we have
  $\kappa=f=F=0$ while $\Delta$ is constant on $M$.
\item The set $\cW^-$ coincides with $M$ and hence $\cW^+$ and $\cU$
  are empty. In this case, $\xi$ is a chiral spinor of negative
  chirality which is covariantly constant on $M$ and we have
  $\kappa=f=F=0$ while $\Delta$ is constant on $M$.
\item The set $\cU$ coincides with $M$ and hence $\cW^+$ and $\cW^-$
  are empty.
\item At least one of the sets $\cW^+$ or $\cW^-$ is non-empty but
  both of these sets have empty interior. In this case, $\cU$ is dense
  in $M$ and the union $\cW=\cW^+\cup\cW^-$ coincides with the
  topological frontier $\Fr(\cU)=\fr(\cU)={\bar \cU}\setminus \cU$ of $\cU$.
\end{enumerate}

\noindent The proof of the theorem is given in Appendix \ref{app:nogo}.

\paragraph{Remarks.} 
\begin{itemize}
\itemsep 0.0em
\item The theorem is a strengthening of an observation originally made in \cite{MartelliSparks} in the case when $\xi$ is nowhere-chiral.
\item The theorem holds in classical supergravity only. 
One may be able to avoid its conclusions by considering quantum corrections. 
\item Cases 1 and 2 correspond to the classical limit of the compactifications studied in \cite{Becker1, Becker2, Constantin}. Case 3 was studied in 
\cite{MartelliSparks,g2}. 
\end{itemize}

\begin{figure}
\centering
\!\!\!\!\!\begin{subfigure}{.5\textwidth}
\centering
\includegraphics[width=0.9\linewidth]{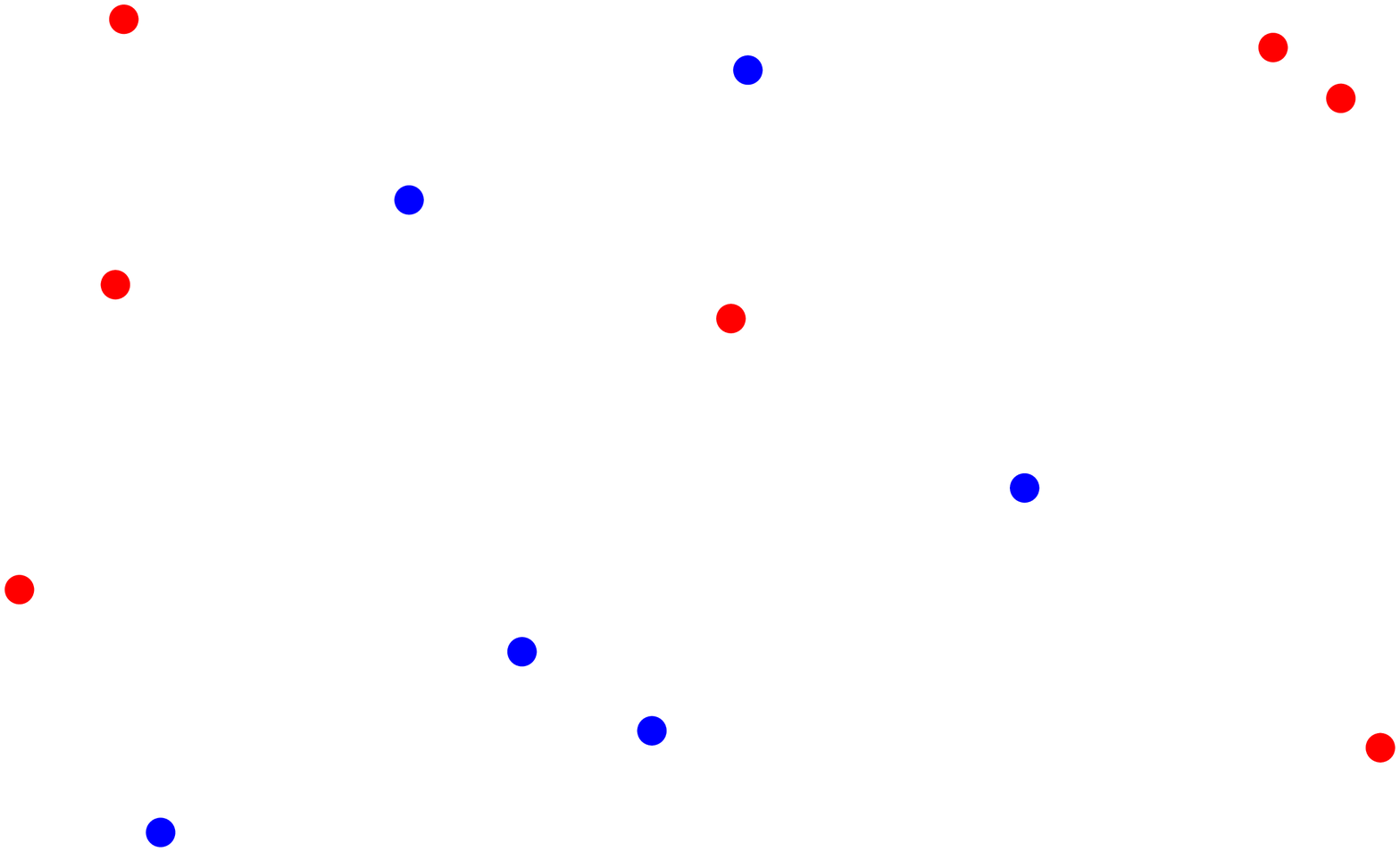}

\

\caption{Sketch of the chiral loci in the Morse sub-case of Case 4 of
  the Theorem. In this case, each of $\cW^+$ and $\cW^-$ is a finite
  set of points, with the points of $\cW^+$ indicated in red and those
  of $\cW^-$ indicated in blue.}
\end{subfigure}~~~~~~~~~
\begin{subfigure}{.5\textwidth}
\centering
\includegraphics[width=0.9\linewidth]{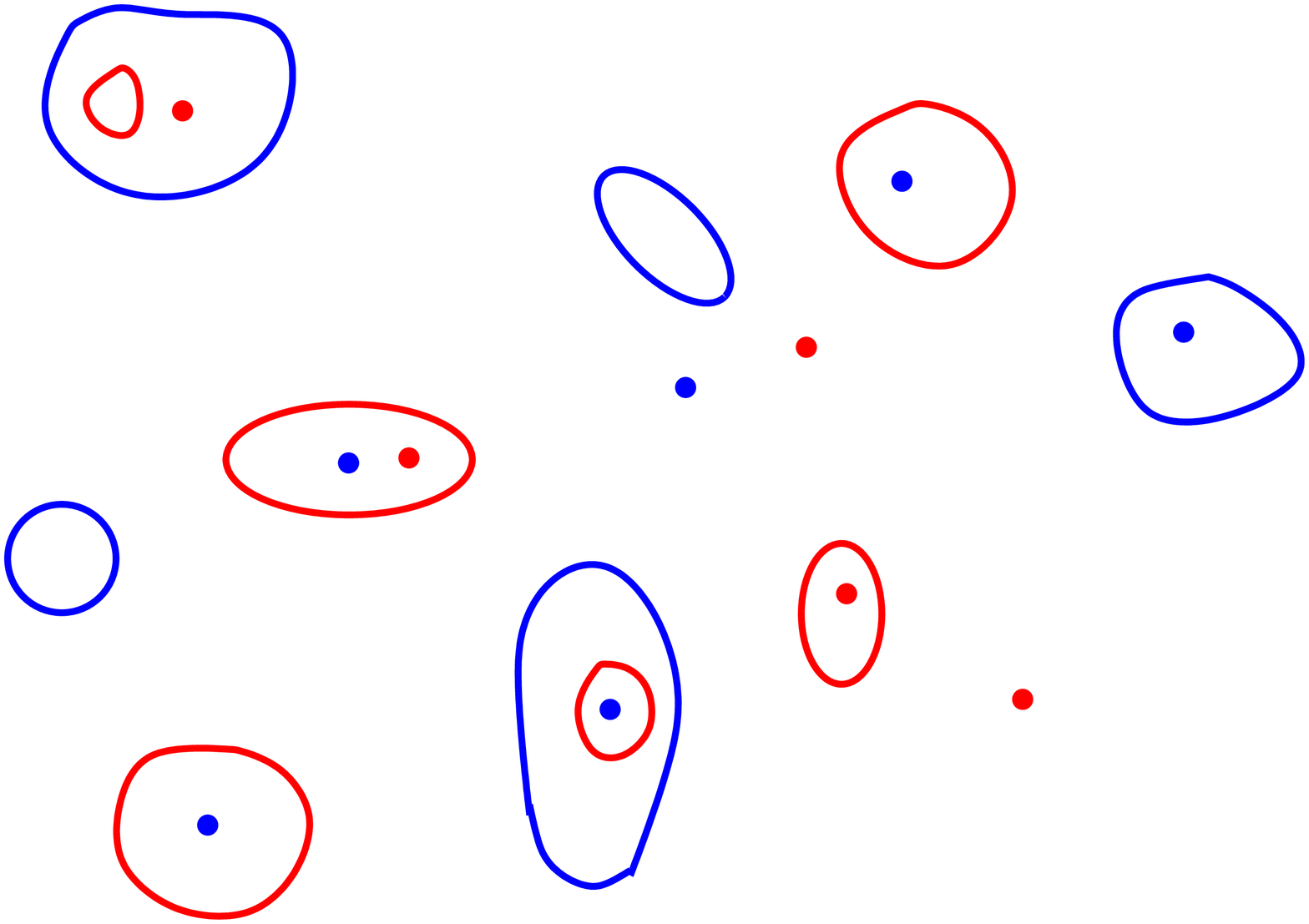}
\caption{Sketch of $\cW^\pm$ in the Bott-Morse sub-case of Case 4 of
  the Theorem. The connected components of $\cW$ are submanifolds of
  various dimensions, shown respectively in red and blue for
  $\cW^+$ and $\cW^-$.}
\end{subfigure}
\caption{Sketch of chiral loci in two sub-cases of Case 4 of the
  Theorem, for the case of a two-dimensional manifold $M$. The
  non-chiral locus $\cU$ is the complement of $\cW$ in $M$ and is
  indicated by white space, after performing appropriate cuts which
  allow one to map $M$ to some region of the plane which is not
  indicated explicitly. The figures should be interpreted with care in
  our case $\dim M=8$.}
\label{fig:loci}
\end{figure}

The study of Case 4 is the focus of the present paper. Due to the
theorem, we shall from now on assume that we are in this case,
i.e. that $\cW$ is non-empty and that it coincides with the frontier
of $\cU$; in particular, we can assume that the closure of $\cU$
coincides with $M$:
\be
M={\bar \cU}=\cU\sqcup \cW~~,~~\cW=\Fr \cU~~.
\ee
In Figure \ref{fig:loci}, we sketch the chirality decomposition in two
sub-cases of Case 4, which correspond to the assumptions that the
one-form $\momega\eqdef 4\kappa e^{3\Delta}V$ is of Morse and
Bott-Morse type, respectively.

\!\!\!\!\!

\subsection{The singular distribution $\cD$}

The one-form $V$ determines a singular (a.k.a. generalized) distribution $\cD$ (generalized sub-bundle of $TM$) which is defined through:
\be
\cD_p\eqdef{\ker V_p}~~,~~\forall p\in M~~.
\ee
This singular distribution is {\em cosmooth} (rather than smooth) in
the sense of \cite{Drager} (see Appendix \ref{app:gendist}). Notice
that $\cD$ is smooth iff $\xi$ is everywhere non-chiral --- i.e. iff
$\cW=\emptyset$, which is the case studied in \cite{g2}; in that case,
$\cD$ is a regular Frobenius distribution. Since in this paper we
assume $\cW\neq \emptyset$, it follows that $\cD$ is {\em not} a
singular distribution in the sense of Stefan-Sussmann
\cite{Stefan,Sussmann}. The set of regular points of $\cD$ equals the
non-chiral locus $\cU$ and we have:
\beqa
\rk\cD_p&=&7~~\mathrm{when}~~p\in\cU~~,\\
\rk\cD_p&=&8~~\mathrm{when}~~p\in \cW~~.
\eeqa
In particular, the restriction $\cD|_\cU$ is a regular Frobenius
distribution on the non-chiral locus $\cU$.  As in \cite{g2}, we endow
$\cD|_\cU$ with the orientation induced by that of $M$ using the unit
norm vector field $n\eqdef {\hat V}^\sharp=\frac{1}{||V||}V^\sharp$,
which corresponds to the $\cD|_\cU$-longitudinal volume form:
\be
\nu_\top\eqdef \iota_{\hat V}\nu|_\cU=n\lrcorner \nu|_\cU\in \Omega^7_{\cU}(\cD)~~.
\ee
Let $\ast_\perp:\Omega_\cU(\cD)\rightarrow \Omega_\cU(\cD)$ denote the
corresponding Hodge operator along the Frobenius distribution
$\cD|_\cU$:
\ben
\label{AstV}
\ast_\perp \omega=\ast(\hV\wedge \omega)=-\iota_\hV(\ast \omega)=\tau(\omega)\nu_\top~~,~~\forall \omega\in \Omega_\cU(\cD)~~.
\een
\subsection{Spinor parameterization and $G_2$ structure on the non-chiral locus}

\paragraph{Proposition \cite{g2}.} 

Relations \eqref{Esquare} are equivalent on $\cU$ with the following conditions:
\ben
\label{fsol}
V^2|_\cU=1-b^2~~,~~Y|_\cU=(1+b\nu)|_\cU \psi~~,~~Z|_\cU=V|_\cU \psi~~,
\een
where $\psi\in \Omega^4_\cU(\cD)$ is the canonically normalized
coassociative form of a $G_2$ structure on the Frobenius distribution $\cD|_\cU$
which is compatible with the metric $g|_\cD$ induced by $g$ and with the
orientation of $\cD|_\cU$.

\

\noindent Let $\varphi\eqdef \ast_\perp \psi\in \Omega^3_\cU(\cD)$ be
the associative form of the $G_2$ structure on $\cD|_\cU$ mentioned in the
proposition. We have \cite{g2}:
\beqan
\psi&=&\frac{1}{1-b^2} VZ=\frac{1}{1-b^2}(1-b\nu)Y\in \Omega^4_\cU(\cD)~~, \label{psidef}\\
\varphi&=&\frac{1}{||V||}\ast Z=\frac{1}{\sqrt{1-b^2}} Z\nu \in \Omega^3_\cU(\cD)~~.\label{phidef}
\eeqan
On the non-chiral locus, one can parameterize $\check{E}$ as \cite{g2}: 
\ben
\label{Enr}
\check{E}|_\cU=\frac{1}{16}(1+V+b\nu)(1+\psi)=P|_\cU\Pi~~,
\een
where: 
\be
P\eqdef\frac{1}{2}(1+V+b\nu)\in \Omega(M)~~,~~\Pi\eqdef
\frac{1}{8}(1+\psi)\in \Omega_\cU(\cD)
\ee
and where $P|_\cU$ and $\Pi$ are commuting idempotents in the \KA
algebra of $\cU$.  Notice the relations:
\ben
\varphi=\ast_\perp\psi=\ast ({\hat V}\wedge \psi)~~,~~
\ast\varphi=-{\hat V}\wedge \psi~~,~~\ast\psi=\hV\wedge \varphi~~
\een
and:
\ben
\label{phipsirels}
V\varphi=-\varphi V =V\wedge \varphi~~,~~V\psi=\psi V=V\wedge \psi~~.
\een

\paragraph{The selfdual and anti-selfdual parts of $\psi$.} 
We have: 
\ben
\label{psipm}
\psi^\pm=\frac{1}{2}(\psi\pm \ast \psi)=\frac{1}{2}(\psi\pm {\hat V}\wedge \varphi)\in \Omega(\cU)~~.
\een

\paragraph{Lemma.} The four-forms $\psi^\pm\in \Omega(\cU)$ satisfy the relations: 
\beqan
&&{\hat V}\psi^\pm{\hat V}=\psi^\mp~~,\label{psipmrel}~\\
&&\psi^+\psi^-=\psi^-\psi^+=0~~,\label{psipmprod}~\\
&& \psi^\pm=\frac{Y^\pm}{1\pm b}|_\cU~~,\label{psiY}~\\
&& ||\psi^+||^2=||\psi^-||^2=\frac{7}{2}~~.\label{psipmnorms}~
\eeqan

\paragraph{Proof.}
Using $\psi^\pm=R^\pm \psi$, relation \eqref{psipmprod} follows
immediately from the fact that $\nu$ commutes with $\psi$.  The last
relation in \eqref{phipsirels} gives:
\ben
\label{psirel}
{\hat V}\psi{\hat V}=\psi~~\mathrm{on}~~\cU~~.
\een
Using the fact that ${\hat V}$ and $\nu$ anti-commute in the \KA
algebra while $\psi$ and $\nu$ commute (because $\nu$ is twisted
central), relation \eqref{psirel} implies \eqref{psipmrel}.
Separating $Y$ into its selfdual and anti-selfdual parts and using
the fact that $\nu Y=Y\nu=\ast Y$, the last equality in \eqref{psidef}
implies \eqref{psiY}, which implies \eqref{psipmnorms} when combined
with the first relation in \eqref{Ynorms}. $\blacksquare$

\paragraph{Proposition.} 
The inhomogeneous differential forms: 
\be
\Pi^\pm\eqdef R^\pm|_\cU\Pi=\Pi R^\pm|_\cU=\frac{1}{8}(R^\pm|_\cU+\psi^\pm)=\frac{1}{16}(1\pm \nu|_\cU+2\psi^\pm)\in \Omega(\cU)
\ee 
satisfy $\Pi=\Pi^++\Pi^-$ and $\hV\Pi^\pm\hV=\Pi^\mp$ and are orthogonal
idempotents in the \KA algebra of $\cU$:
\be
(\Pi^\pm)^2=\Pi^\pm~~,~~\Pi^\pm\Pi^\mp=0~~.
\ee
Furthermore, we have: 
\ben
\label{EpmDec}
\check{E}^\pm|_{\cU}=P|_\cU\Pi^\pm~~.
\een
Notice that $\Pi^\pm$ are twisted (anti-)selfdual: 
\be
\Pi^\pm \nu=\pm \Pi^\pm~~.
\ee
\noindent{\bf Proof.} Notice that $\psi$ and $R^\pm$ commute since
$\psi$ and $\nu$ commute. The conclusion now follows immediately using
the properties of $\Pi$ and $R^\pm$.
$\blacksquare$

\subsection{Spinor parameterization and $\Spin(7)_\pm$ structures on the loci $\cU^\pm$}
\label{subsec:L}

\paragraph{Extending $\psi^\pm$ to $\cU^\pm$.}

Notice that $P\in \Omega(M)$ is globally defined on $M$ while $\Pi\in
\Omega(\cU)$ is only defined on the non-chiral locus. 

\paragraph{Proposition.} 
The four-form $\psi^\pm$ has a continuous extension to the locus
$\cU^\pm$, which we denote through
$\bar{\psi}^\pm\in\Omega^4(\cU^\pm)$. Namely:
\be
\bar{\psi}^\pm\eqdef \frac{1}{1\pm b}(Y^\pm|_{\cU^\pm})\in \Omega^4(\cU^\pm)~~.
\ee 
Furthermore, the idempotents $\Pi^\pm\in \Omega(\cU)$ have continuous
extensions to idempotents $\bar{\Pi}^\pm\in \Omega(\cU^\pm)$, which
are given by:
\ben
\label{barPi}
{\bar \Pi}^\pm\eqdef \frac{1}{8}(R^\pm|_{\cU^\pm}+\bar{\psi}^\pm)=\frac{1}{16}(1+2\psi^\pm\pm \nu)\in \Omega(\cU^\pm)~~
\een
and which are twisted (anti-)selfdual:
\be
{\bar \Pi}^\pm R^\pm|_{\cU^\pm}={\bar \Pi}^\pm~~,~~{\bar \Pi}^\pm R^\mp|_{\cU^\pm}=0~~.
\ee 

\paragraph{Remarks.}
\begin{enumerate}
\itemsep 0.0em
\item Notice that \eqref{psiY} does not provide any information about
  the limit of $\psi^\mp$ along $\cW^\pm$, so $\psi^\mp$ (and hence
  also $\Pi^\mp$) will not generally have an extension to
  $\cU^\pm$. However, \eqref{psipmnorms} tells us that $\psi^\mp$ is
  bounded on $M$. In particular, we have:
\ben
\label{limpsi}
 \lim_{b\rightarrow \pm 1}(V\psi^\mp)= \lim_{b\rightarrow \pm 1}(\psi^\mp V)=0~~.
\een
\item On the locus $\cW^\pm$ we have:
\ben
\label{WpmRes}
b|_{\cW^\pm}=\pm 1~~,~~V|_{\cW^\pm}=Z|_{\cW^\pm}=Y^\mp|_{\cW^\pm}=0~~,
\een
where the last relations follow from the last equation in
\eqref{SolMS} and from \eqref{psiY}. The remaining conditions in
\eqref{SolMS} are automatically satisfied.
\item Notice the relation:
\be
Y^\pm|_{\cW^\pm}=2{\bar \psi}^\pm|_{\cW^\pm}~~,
\ee
which follows from the fact that $b|_{\cW^\pm}=\pm 1$. 
\end{enumerate}

\noindent{\bf Proof.}  Since $Y^\pm\in \Omega(M)$ is well-defined on
$M$, the conclusion follows immediately from relation \eqref{psiY} and
from the fact that $1\pm b$ does not vanish on $\cU^\pm$. The
relations satisfied by $\bar{\Pi}^\pm$ on $\cU^\pm$ follow by
continuity from the similar relations satisfied by $\Pi^\pm$ on $\cU$.
$\blacksquare$

\

\noindent While $\Pi^\mp$ does not generally have an extension to
$\cW^\pm$, the product $P\Pi^\mp$ has zero limit on $\cW^\pm$:

\paragraph{Proposition.}
We have $P|_{\cW^\pm}=R^\pm$ as well as: 
\ben
\label{EW}
\exists \lim_{b\rightarrow \pm 1} P\Pi^\mp=\check{E}^\mp|_{\cW^\pm}=0~~,~~\check{E}^\pm|_{\cW^\pm}={\bar \Pi}^\pm|_{\cW^\pm}=\frac{1}{8}(R^\pm+\bar{\psi}^\pm)|_{\cW^\pm}=\frac{1}{16}(1\pm \nu +2\bar{\psi}^\pm)|_{\cW^\pm}~~.
\een
\noindent{\bf Proof.} 
The relation $P|_{\cW^\pm}=R^\pm$ is obvious. The other statements
follow from \eqref{Epm} and \eqref{EpmDec} using
\eqref{WpmRes}. $\blacksquare$

\paragraph{The $\Spin(7)_\pm$ structures on $\cU^\pm$.}

\paragraph{Lemma.} 
Let $(e^a)_{a=1\ldots 8}$ be a local coframe defined over an open
subset $U\subset M$ and let $\eta\in \Gamma(U,S)$. Then:
\be
\cB(\gamma^a\eta,\gamma^b\eta)=g^{ab}||\eta||^2~~,
\ee
where $\gamma^a=\gamma(e^a)$ and $g^{ab}=\langle e^a,e^b\rangle$. 

\

\noindent{\bf Proof.} 
Using the property $(\gamma^a)^t=\gamma^a$ and the fact that
$(\gamma^a\gamma^b)^t=\gamma^b\gamma^a$, compute:
\be
\cB(\gamma^a\eta,\gamma^b\eta)=\cB(\eta,\gamma^a\gamma^b\eta)=
\cB(\eta,\gamma^b\gamma^a\eta)=\frac{1}{2}\cB(\eta,\{\gamma^a,\gamma^b\}\eta)=g^{ab}\cB(\eta,\eta)=g^{ab}||\eta||^2~~.
\ee
$\blacksquare$

\

\noindent When $\eta$ is non-vanishing everywhere on $U$, the
proposition implies that the spinors $\gamma^a\eta$ form a
linearly-independent set of sections of $S$ above $U$. Taking $\eta$
to have chirality $\pm 1$ and recalling that $\gamma^a$ map $S^\pm$
into $S^\mp$ and that $\rk S^+=\rk S^-=8$, this gives:

\paragraph{Corollary.} Let $(e^a)_{a=1\ldots 8}$ be a local orthonormal coframe defined over an open subset 
$U\subset M$ and $\eta\in \Gamma(U,S^\pm)$ be a spinor of chirality
$\pm 1$ which is nowhere vanishing on $U$. Then
$(\gamma^a\eta)_{a=1\ldots 8}$ is a $\cB$-orthogonal local frame of
$S^\mp$ above $U$. Every local section $\xi\in \Gamma(U,S^\mp)$
expands in this frame as:
\be
\xi=\frac{1}{||\eta||^2}\sum_{a=1}^8 \cB(\xi,\gamma_a\eta) \gamma^a\eta~~.
\ee

\paragraph{Proposition.} Let $U$ be an open subset of $M$ which supports 
an orthonormal coframe $e^a$ of $(M,g)$. Then:

\

\noindent 1. If $\xi^+$ is everywhere non-vanishing on $U$, then
$\xi^-$ expands above $U$ as $\xi^-= \sum_{a=1}^8
L^+_a\gamma^a\xi^+=\gamma(L^+)\xi^+$, where $L^+_a$ are the
coefficients of the one-form $L^+=L_a^+\dd x^a=\frac{1}{1+b}V$.

\

\noindent 2. If $\xi^-$ is everywhere non-vanishing on $U$, then
$\xi^+$ expands above $U$ as $\xi^+=\sum_{a=1}^8
L^-_a\gamma^a\xi^-=\gamma(L^-)\xi^-$, where $L^-_a$ are the
coefficients of the one-form $L^-=L_a^-\dd x^a=\frac{1}{1-b}V$.

\

\noindent {\bf Proof.} Assume that $\xi^+$ (respectively $\xi^-$)
vanishes nowhere on $U$.  The corollary shows that $\xi^\mp$ expands
as $\xi^\mp=\sum_{a=1}^8 L^\pm_a\gamma^a\xi^\pm$ where:
\ben
\label{Ld}
L^\pm_a=\frac{1}{||\xi^\pm||^2}\cB(\xi^\mp,\gamma_a\xi^\pm)~~.
\een
Recalling that $S^+$ and $S^-$ are $\cB$-orthonormal while $\gamma^a$
are $\cB$-symmetric, we find:
\be
\cB(\xi^+,\gamma_a\xi^-)=\cB(\xi^-,\gamma_a\xi^+)=\frac{1}{2}\cB(\xi,\gamma_a\xi)=\frac{1}{2}V_a~~.
\ee 
Using this and \eqref{xipmnorms}, equation \eqref{Ld} becomes
$L^\pm_a=\frac{1}{1\pm b} V_a$.~~$\blacksquare$

\paragraph{Remarks.}
\begin{enumerate}
\itemsep 0.0em
\item The ``+'' case of \eqref{Ld} was used in \cite{Tsimpis}, where
  no explicit expression for $L^+$ (which is denoted by $L$ in
  loc. cit.)  was given\footnote{Notice that $L^+$ is not a quadratic
    function of $\xi$, since it involves the denominator $1+b$ and
    thus it is not homogeneous under rescalings $\xi\rightarrow
    \lambda\xi$ with $\lambda\neq 0$.}.
\item Notice that $L^+$ and $L^-$ are not independent (they are
  proportional to each other) and that each of them contains the same
  information as $V$ and $b$.
\end{enumerate}

\noindent Recalling \eqref{xipmnorms}, consider the unit norm spinors (of chirality $\pm 1$):
\ben
\label{etadef}
\boxed{\eta^\pm=\sqrt{1+||L^\pm||^2}\xi^\pm=\sqrt{\frac{2}{1\pm b}}\xi^\pm\in \Gamma(\cU^\pm,S^\pm)}~.
\een
Using the fact that $||\eta^\pm||=1$ while $\cB(\eta^\pm,\gamma_{a_1\ldots a_k}\eta^\pm)$ vanishes unless $k\equiv_4 0$, we find: 
\ben
\label{Eeta}
\check{E}_{\eta^\pm,\eta^\pm}=\frac{1}{16}(1+\Phi^\pm\pm \nu)\in \Omega(\cU^\pm)~~,
\een
where:
\ben
\label{PhiDef}
\Phi^\pm\eqdef \frac{1}{4!}\cB(\eta^\pm,\gamma_{a_1\ldots a_4}\eta^\pm)e^{a_1\ldots a_4}
=\bcE^{(4)}_{\eta^\pm,\eta^\pm}=\frac{2}{1\pm b}\bcE^{(4)}_{\xi^\pm,\xi^\pm}\in \Omega^4(\cU^\pm)~~
\een
and where we noticed that $\cB(\eta^\pm,\gamma(\nu)\eta^\pm)=\pm 1$. 

\paragraph{Proposition.}  
The four-form $\Phi^+$ is selfdual while the four-form $\Phi^-$ is
anti-selfdual. They satisfy the following relations on the locus $\cU^\pm$:
\ben
\boxed{\Phi^\pm= 2\bar{\psi}^\pm}~.~\label{PhiT}\\
\een
In particular, the inhomogeneous form \eqref{Eeta} coincides with the
extension \eqref{barPi} of $\Pi^\pm$ to this locus:
\be
\check{E}_{\eta^\pm,\eta^\pm}=\bar{\Pi}^\pm
\ee
and we have: 
\ben
||\Phi^\pm||^2 = 14~~\label{PhiNorms}~~.
\een
Moreover, the restriction of $\Phi^+$ is the canonically-normalized
calibration defining a $\Spin(7)$ structure on the open submanifold
$\cU$ of $M$ while the restriction of $\Phi^-$ is the
canonically-normalized calibration defining a $\Spin(7)$ structure on
the orientation reversal of $\cU$.

\paragraph{Proof.}
Recalling that $\xi^\pm=\frac{1}{2}(1\pm \gamma(\nu))\xi$, the
identities $\check{E}_{\xi,\gamma(\nu)\xi}=\check{E}_{\xi,\xi}\nu$ and
$\check{E}_{\gamma(\nu)\xi,\xi}=\nu \check{E}_{\xi,\xi}$ of \cite{ga1} 
and the fact that $\nu$ is involutive and twisted central give:
\be
\check{E}_{\xi^\pm,\xi^\pm}=\frac{1}{4}(\check{E}_{\xi,\xi}\pm \nu \check{E}_{\xi,\xi}\pm \check{E}_{\xi,\xi}\nu+\nu\check{E}_{\xi,\xi}\nu)=
\frac{1}{4}(\check{E}_{\xi,\xi}+\pi(\check{E}_{\xi,\xi}))(1\pm \nu)=\frac{1}{2}\check{E}_{\xi,\xi}^\ev(1\pm\nu)=
\frac{1}{2}(\check{E}^\ev_{\xi,\xi}\pm \ast \tau(\check{E}_{\xi,\xi}^\ev))~~
\ee
Since the Hodge operator preserves $\Omega^4(M)$ and since the
reversion $\tau$ of the \KA algebra restricts to the identity on the
space of four-forms, this implies:
\be
\bcE^{(4)}_{\xi^\pm,\xi^\pm}=\frac{1}{2}(\bcE^{(4)}_{\xi,\xi}\pm \ast \bcE^{(4)}_{\xi,\xi})=\frac{1}{2}(Y\pm \ast Y)=Y^\pm~~,
\ee
where the superscript $\pm$ indicates the selfdual/anti-selfdual part. Substituting this
into \eqref{PhiDef} gives relation \eqref{PhiT}. The statements of the
proposition regarding the restrictions of $\Phi^\pm$ to the open
submanifold $\cU$ follow from the fact that $\eta_\pm$ is a
Majorana-Weyl spinor of norm one and of chirality $\pm 1$; it is
well-known \cite{Joyce} that giving such a spinor on an eight-manifold
$\cU$ induces $\Spin(7)$ structures on the underlying manifold or on
its orientation reversal, whose normalized calibrations are given by
\eqref{PhiDef}. In particular, \eqref{PhiNorms} holds on $\cU$ since
there it amounts to the condition that $\Phi^\pm$ are canonically
normalized. By continuity, this implies that \eqref{PhiNorms} also
holds on $\cW^\pm$. $\blacksquare$

\paragraph{Remarks.} 
\begin{enumerate}
\itemsep 0.0em
\item The proposition implies that the following relation holds on the non-chiral locus:
\be
\check{E}_{\xi,\xi}|_\cU=P|_{\cU}(\check{E}_{\eta^+,\eta^+}+\check{E}_{\eta^-,\eta^-})~~.
\ee
This shows how the idempotent $\check{E}_{\xi,\xi}|_{\cU}$ which
characterizes the normalized Majorana spinor $\xi$ on the locus $\cU$
relates to the two idempotents
$\check{E}_{\eta^\pm,\eta^\pm}|_\cU=\Pi^\pm$ which characterize the
Majorana-Weyl spinors $\eta^\pm$ and which encode the
$\Spin(7)_\pm$ structures through the \KA algebra. While $\check{E}_{\eta^+,\eta^+}$
depends only on the positive chirality spinor $\eta^+$ and
$\check{E}_{\eta^-,\eta^-}$ depends only on the negative chirality
spinor $\eta^-$, the idempotent $P$ contains the quantities $b$ and
$V$, each of which involves both chirality components of the spinor
$\xi$:
\be
b=||\xi^+||^2-||\xi^-||^2~~,~~V=2 \cB(\xi^+,\gamma_m\xi^-)e^m=(1-b^2)\cB(\eta^+,\gamma_m\eta^-)e^m~~.
\ee
The object $P$ encodes in the \KA algebra the $\SO(7)$ structure which
corresponds to the distribution $\cD$ on $\cU$. Finally, notice that the idempotent $\Pi$ encodes 
the $G_2$ structure along the distribution $\cD$. Notice that $P$ and $\Pi$ commute, while $P$ and 
$\Pi_\pm$ do not commute. 
\item Equation \eqref{PhiT} implies that $\Phi^\pm$ coincides with
  $\pm Y^\pm$ on the locus $\cW^\pm$ since $b=\pm 1$ there. Notice
  that \eqref{PhiNorms} agrees via \eqref{PhiT} with the last
  equations in \eqref{SolMS}. 
\end{enumerate}

\paragraph{Spinor parameterization on the loci $\cU^\pm$.}
On the locus $\cU$, relations \eqref{fsol} and \eqref{PhiT} give:
\ben
\label{YZT}
\boxed{
\begin{split}
&Z|_\cU=\frac{1}{2}V(\Phi^++\Phi^-)~~,\\
&Y|_\cU=\frac{1}{2}[(1+b)\Phi^++(1-b)\Phi^-]~~.
\end{split}}
\een
In these relations, $\Phi^+$
and $\Phi^-$ are not independent but related through:
\be
\Phi^\mp={\hat V}\Phi^\pm{\hat V}
\ee
as a consequence of \eqref{psipmrel}. Hence on the non-chiral locus we
can eliminate $\Phi^\mp$ in terms of $\Phi^\pm$ to obtain the
following non-redundant parameterizations:
\be
Z|_\cU=\frac{1}{2}\sqrt{1-b^2}(\hV\Phi^\pm+\Phi^\pm \hV)~~,~~
Y|_\cU=\frac{1}{2}\left[(1\pm b)\Phi^\pm +(1\mp b)\hV\Phi^\pm\hV\right]~~,
\ee
which give:
\be
16 \check{E}|_\cU=P|_{\cU}(\Pi_\pm+{\hat V}\Pi_\pm {\hat V})=1+V+\frac{1}{2}\left[(1\pm b)\Phi^\pm+(1\mp b)\hV\Phi^\pm \hV\right]+
\frac{1}{2} \sqrt{1-b^2}(\hV \Phi^\pm+\Phi^\pm \hV)+b\nu~~.
\ee
This imply the following parameterizations on the loci $\cU^\pm$:
\be
\boxed{16 \check{E}|_{\cU^\pm}=1+V+\frac{1}{2}\left[(1\pm b)\Phi^\pm+\frac{1}{1\pm b}V\Phi^\pm V\right]+
\frac{1}{2} (V \Phi^\pm+\Phi^\pm V)+b\nu }~~,
\ee
where it is understood that (see \eqref{limpsi}): 
\be
\lim_{b\rightarrow \pm 1}{V\Phi^\mp}=\lim_{b\rightarrow \pm 1}{\Phi^\mp V}=0~~
\ee
and hence (see \eqref{EW}):
\be
16 \check{E}|_{\cW^\pm}={\bar \Pi}^\pm|_{\cW^\pm}=\frac{1}{16}(1+\Phi^\pm \pm \nu)|_{\cW^\pm}~~.
\ee
Up to expressing $V$ and $b$ through $L^\pm$, this is the parameterization which corresponds to the
approach of \cite{Tsimpis}. 

\subsection{Comparing spinors and G structures on the non-chiral locus}

Equation \eqref{psipm} gives:
\be
\Phi^\pm|_\cU=2\psi^\pm=\psi\pm {\hat V}\wedge \varphi~~,
\ee
i.e.:
\ben
\label{PhiDec}
\boxed{(\Phi^\pm|_\cU)_\top=\pm \varphi~~,~~(\Phi^\pm|_\cU)_\perp=\psi}~~.
\een
The relation $\xi^\mp=\gamma(L^\pm)\xi^\pm$ gives
$\eta^\mp=\gamma({\hat V})\eta^\pm$, which shows that the everywhere
normalized spinor:
\ben
\label{eta0}
\eta_0\eqdef \frac{1}{\sqrt{2}}(\eta^++\eta^-)\in \Gamma(\cU,S)
\een
is a Majorana spinor along $\cD$ in the seven-dimensional sense,
i.e. we have $D(\eta_0)=\eta_0$ where $D\eqdef \gamma({\hat V})$ is
the real structure of $S$, when the latter is viewed as a complex
spinor bundle over $\cD$ (see \cite{g2}). The identity
$\bcE^{(4)}_{\eta^\pm,\eta^\mp}=0$ implies the following
spinorial expression for $\psi$:
\ben
\label{psieta}
\psi=\bcE^{(4)}_{\eta_0,\eta_0}=\frac{1}{4!}\cB(\eta_0,\gamma_{a_1\ldots a_4}\eta_0)e^{a_1\ldots a_4}~~.
\een
The relation $\xi^\mp=\gamma(L^\pm)\xi^\pm$ gives $\eta^\mp=\gamma({\hat V})\eta^\pm$, which implies:
\be
\eta_0=\frac{1}{\sqrt{2}}(\id_S+\gamma({\hat V}))\eta^+= \frac{1}{\sqrt{2}}(\id_S+\gamma({\hat V}))\eta^-~~.
\ee
Notice that $\frac{1}{2}(\id_S+\gamma(\hV))$ is an idempotent
endomorphism of $S$. As explained in \cite{g2}, the spinor $\eta_0$
induces the $G_2$ structure of the distribution $\cD$. The situation
is summarized in Table \ref{table:G}.

\!\!\!\!\!\!\!\begin{table}[tt]
\centering
\!\!\!\!\!\!\!\begin{tabular}{|c|c|c|c|c|}
\hline
G structure & $\Spin(7)_+$ & $\Spin(7)_{-}$ & $G_2$ (on $\cD|_\cU$) & $\SO(7)$ ($\cD|_\cU$)\\
\hline\hline
spinor    & $\eta^+$ & $\eta^-$ & $\eta_0=\frac{1}{\sqrt{2}}(\eta^+\!\!+\eta^-)$ & ---\\
\hline
idempotent & \!\!$\Pi^+\!\!=\!\!\frac{1}{16}(1+\Phi^+\!\!+\nu)$ & \!\!$\Pi^-\!\!=\!\!\frac{1}{16}(1+\Phi^-\!\!-\nu)$ & 
\!\!$\Pi=\Pi^+\!\!+\Pi^-\!\!=\frac{1}{8}(1+\psi)$& \!\!$P=\frac{1}{2}(1+V+b\nu)$ \\
\hline
forms & $\Phi^+=2\psi^+$ & $\Phi^-=2\psi^-$ & $\varphi$ and $\psi=\ast_\perp\varphi$ & $b$ and $V$ \\
\hline
extends to & $\cU^+$ & $\cU^-$ & $\cU$ & $\cU$ \\
\hline
\end{tabular}
\
\caption{Summary of various $G$ structures and of their reflections in the \KA algebra.}
\label{table:G}
\end{table}~~~~~~~~~~~~~~

\paragraph{Remarks.}
\begin{enumerate}
\itemsep 0.0em
\item None of the $G$ structures in Table \ref{table:G} extends to $M$. In
fact, the structure group $\SO(8)$ of the frame bundle of $M$ {\em does not
globally reduce, in general, to any proper subgroup}. As pointed out in
\cite{Tsimpis}, this is due to the fact that the action of $\Spin(8)$
on the fibers $S_p\simeq \R^{16}$ of $S$ (which is the action of
$\Spin(8)$ on the direct sum ${\bf 8}_s\oplus {\bf 8}_c$ of the
positive and negative chirality spin $1/2$ representations) is not
transitive when restricted to the unit sphere $S^{15}\subset
\R^{16}$. As shown in loc. cit, one can in some sense ``cure'' this problem by
considering the manifold ${\hat M}\eqdef M\times S^1$, using the fact that $\Spin(9)$
acts transitively on $S^{15}$. However, such an approach does not
immediately provide useful information on the geometry of $M$, in
particular the geometry of the singular foliation $\bcF$
discussed in the next subsection is not immediately visible in that
approach. It was also shown in loc. cit. that one can repackage the
information contained in the $\Spin(7)_\pm$ structures into a
generalized $\Spin(7)$ structure on ${\hat M}$ in the sense of
\cite{Witt}. In particular, it is easy to check that relations (4.8)
of \cite{Tsimpis} are equivalent with some of the exterior
differential constraints which can be obtained by expanding equation
(3.5) of \cite{g2} into its rank components --- exterior differential
constraints which were discussed at length in \cite{ga1} and in the
appendix of \cite{g2}.  As shown in detail in \cite{g2}, those exterior
differential constraints do not suffice to encode the full supersymmetry
conditions for such backgrounds.
\item The fact that the structure group of $TM$ does not globally
  reduce beyond $\SO(8)$ in this class of examples illustrates some
  limits of the philosophy that flux compactifications can be
  described using reductions of structure group.  That philosophy is
  based on the observation that a collection of (s)pinors defines a
  {\em local} reduction of structure group over any open subset of the
  compactification manifold $M$ along which the stabilizer of the
  pointwise values of those spinors is fixed up to conjugacy in the
  corresponding $\Spin$ or $\Pin$ group.  However, such a reduction
  does {\em not} generally hold globally on $M$, since the local
  reductions thus obtained can ``jump'' --- in our class of examples,
  the jump occurs at the points of the chiral locus $\cW$. The
  appropriate notion is instead that of {\em generalized} reduction of
  structure group, of which the class of compactifications considered
  here is an example.  In this respect, we mention that the cosmooth
  generalized distribution $\cD$ can be viewed as providing a
  generalized reduction of structure group of $M$, which is an
  ordinary reduction from $\SO(8)$ to $\SO(7)$ only when restricted to
  its regular subset $\cU$, on which $\cD|_\cU$ provides \cite{g2} an
  almost product structure. We also mention that the conditions
  imposed by supersymmetry can be formulated globally by using an
  extension of the language of Haefliger structures (see Section
  \ref{subsec:Haefliger}), an approach which can in fact be used to
  give a fully general approach to flux compactifications. It is such
  concepts, rather than the classical concept of $G$ structures
  \cite{Chern}, which provide the language appropriate for giving
       {\em globally valid} descriptions of the most general flux
       compactifications.
\end{enumerate}

\subsection{The singular foliation of $M$ defined by $\cD$}
\label{subsec:Haefliger}
As in \cite{g2}, one can show that the one-form: 
\be
\momega\eqdef 4\kappa e^{3\Delta} V~~
\ee
satisfies the following relations which hold globally on $M$ as a
consequence of the supersymmetry conditions \eqref{par_eq}:
\beqan
\label{meq}
\dd\momega&=&0~~,~~\\
\momega~&=&\mathbf{f}-\dd\mathbf{b}~~,~~\mathrm{where}~~\mathbf{b}\eqdef e^{3\Delta}b~~.\nn
\eeqan
As a result of the first equation, the generalized distribution
$\cD=\ker V=\ker\momega$ determines a singular foliation $\bcF$
of $M$, which degenerates along the chiral locus $\cW$, since that locus
coincides with the set of zeroes of $\momega$.  The second
equation implies that $\momega$ belongs to the cohomology class $\f\in
H^1(M,\R)$ of $\mathbf{f}$.

Since $\cD$ is cosmooth rather than smooth, the notion of singular
foliation which is appropriate in our case\footnote{Notice that this
  is not the notion of singular foliation considered in
  \cite{Androulidakis1,Androulidakis2}, which is instead based on
  Stefan-Sussmann (i.e. smooth, rather than cosmooth) distributions.}
is that of Haefliger structure \cite{Haefliger}. More precisely,
$\bcF$ can be described as the Haefliger structure defined as
follows. Consider an open cover $(U_\alpha)_{\alpha\in I}$ of $M$ such
that each $U_\alpha$ is simply-connected and let $\momega_\alpha\eqdef
\momega|_{U_\alpha}\in \Omega^1(U_\alpha)$. We have
$\momega_\alpha=\dd {\mathbf h}_\alpha$ for some ${\mathbf h}_\alpha\in
\Omega^0(U_\alpha)$, where ${\mathbf h}_\alpha$ are determined up to shifts:
\ben
\label{hshifts}
{\mathbf h}_\alpha\rightarrow \mathbf{h}'_\alpha+{\mathbf c}_\alpha~~,~~{\mathbf c}_\alpha\in \R~~.
\een
For any $\alpha,\beta\in I$ and any $p\in U_\alpha\cap U_\beta$,
consider the orientation-preserving diffeomorphism
$\boldsymbol{\phi}_{\alpha\beta}(p)\in \Diff_+(\R)$ of the real line
given by the translation:
\be
\boldsymbol{\phi}_{\alpha\beta}(p)(x)\eqdef x+{\mathbf h}_\beta(p)-{\mathbf h}_\alpha(p)~~\forall x\in \R~~.
\ee
Then $\boldsymbol{\phi}_{\alpha\beta}(p)({\mathbf
  h}_\alpha(p))={\mathbf h}_\beta(p)$. The germ
$\hat{\boldsymbol{\phi}}_{\alpha\beta}(p)$ of
$\boldsymbol{\phi}_{\alpha\beta}(p)$ at ${\mathbf h}_\alpha(p)$ is an
element of the Haefliger groupoid $\Gamma_1^\infty$ and
it is easy to check that
$\hat{\boldsymbol{\phi}}_{\alpha\beta}:U_\alpha\cap U_\beta\rightarrow
\Gamma_1^\infty$ is a Haefliger cocycle on $M$:
\be
\hat{\boldsymbol{\phi}}_{\beta\gamma}(p)\circ \hat{\boldsymbol{\phi}}_{\alpha\beta}(p)=\hat{\boldsymbol{\phi}}_{\alpha\gamma}(p)~~~~\forall \alpha,\beta,\gamma\in I~~,~~\forall p\in U_\alpha\cap U_\beta\cap U_\gamma~~.
\ee
Moreover, the shifts \eqref{hshifts} correspond to transformations:
\be
\hat{\boldsymbol{\phi}}_{\alpha\beta}\rightarrow \hat{\boldsymbol{\phi}}'_{\alpha\beta}=\hat{{\mathbf q}}_\beta\circ \hat{\boldsymbol{\phi}}_{\alpha\beta}\circ \hat{{\mathbf q}}_\alpha^{-1}~~,
\ee
where $\hat{\mathbf{q}}_\alpha:U_\alpha\rightarrow \Gamma_1^\infty$ are defined by
declaring that $\hat{\mathbf{q}}_{\alpha}(p)$ is the germ
at $p\in U_\alpha$ of the orientation-preserving diffeomorphism
$\mathbf{t}_\alpha\in \Diff_+(\R)$ given by the following translation
of the real line:
\be
\mathbf{t}_\alpha(x)=x+{\mathbf c}_\alpha~~\forall x\in \R~~.
\ee
It follows that the closed one-form $\momega$ determines a
well-defined element of the non-Abelian cohomology $\in
H^1(M,\Gamma_1^\infty)$, which is the Haefliger structure defined by
$\momega$. The singular foliation $\bcF$ which ``integrates''
$\cD$ can be identified with this element.

The approach through Haefliger structures allows one to define
rigorously the singular foliation $\bcF$ in the most general
case, i.e. without making any supplementary assumptions on the closed
one-form $\momega$. In general, such singular foliations can be
extremely complicated and little is known about their topology and
geometry. However, the description of $\bcF$ simplifies when
$\momega$ is a closed one-form of Morse or Bott-Morse type.  In
Section \ref{sec:Morse}, we discuss the Morse case, recalling some
results which apply to $\bcF$ in that situation.

\section{Relating the $G_2$ and $\Spin(7)$ approaches on the non-chiral locus}
\label{sec:G2Spin7}
On the non-chiral locus $\cU$, we have the regular foliation $\cF$
which is endowed with a longitudinal $G_2$ structure having
associative and coassociative forms $\varphi$ and $\psi$. We also have
a $\Spin(7)_+$ and a $\Spin(7)_-$ structure, which are determined
respectively by the calibrations $\Phi^\pm=2\psi^\pm=\psi\pm {\hat
  V}\wedge \varphi$. Given this data, one can relate various
quantities determined by $(\cD,\varphi)$ to quantities determined by
$\Phi^\pm$ as we explain below.  We stress that the results of this
subsection are independent of the supersymmetry conditions
\eqref{par_eq} and hence they hold in the general situation described
above.  We mention that the relation between the type of $G_2$
structure induced on an oriented submanifold of a $\Spin(7)$ structure
manifold and the intrinsic geometry of such submanifolds was studied
in \cite{Gray,Cabrera}.

\subsection{The $\G_2$ and $\Spin(7)_\pm$ decompositions of $\Omega^4(\cU)$}

The group $G_2$ has a natural fiberwise rank-preserving action on the
graded vector bundle $\wedge (\cD|_\cU)^\ast$, which is given at every $p\in
\cU$ by the local embedding of $G_2$ as the stabilizer $G_{2,p}$ in
$\SO(\cD_p)$ of the 3-form $\varphi_p\in \wedge^3(\cD_p^\ast)$. Since
$\SO(\cD_p)$ embeds into $\SO(T_pM)$ as the stabilizer of the 1-form
$V_p\in T_p^\ast M$, this induces a rank-preserving action of
$G_{2,p}$ on $\wedge T_p^\ast\cU$ which can be described as
follows. Decomposing any form $\omega\in \wedge T_p^\ast \cU$ as
$\omega=\omega_\perp+{\hat V}\wedge \omega_\top$, the action of an
element of $g$ of $G_2$ on $\omega$ is given by the simultaneous
action of $g$ on the components $\omega_\perp$ and $\omega_\top$, both
of which belong to $\wedge \cD_p^\ast$. The corresponding
representation of $G_2$ at $p$ is equivalent with the direct sum of
the representations in which the components $\omega_\top$ and
$\omega_\perp$ transform at $p$. In particular, $F_{\perp,p}$ and
$F_{\top,p}$ transform in a $G_2$ representation which is equivalent
with the direct sum $\wedge^3 \cD^\ast_p\oplus \wedge^4 \cD^\ast_p$.
The group $\Spin(7)$ is embedded inside $\SO(T_pM)$
in two ways, namely as the stabilizers $\Spin(7)_{\pm,p}$ of the
selfdual 4-forms $\Phi^{\pm}_p$.  Then \eqref{PhiT} shows that
$G_{2,p}$ is the stabilizer of $V_p$ in $\Spin(7)_{\pm,p}$. The action
of $G_{2,p}$ on $\wedge T^\ast_p M$ is obtained from that of
$\Spin(7)_{\pm,p}$ by restriction. Hence the irreducible components of
the action of $\Spin(7)_{\pm,p}$ on $\wedge^k(T^\ast_p M)$ decompose
as direct sums of the irreducible components of the action of
$G_{2,p}$ on the same space. We have the following decompositions into
irreps. (see, for example, \cite{Fernandez, Kthesis}):
\beqan
\label{irreps}
\wedge^4 T^\ast_p M &=& \wedge^4_{{\bf 1},\pm}T^\ast_p M\oplus  \wedge^4_{{\bf 7},\pm}T^\ast_p M\oplus  \wedge^4_{{\bf 27},\pm}T^\ast_p M\oplus  \wedge^4_{{\bf 35},\pm}T^\ast_p M~~~\mathrm{for}~~\Spin(7)_{\pm,p}~~,\nn\\
\wedge^4 T^\ast_p M &=& \wedge^4_{1}T^\ast_p M\oplus  \wedge^4_{7}T^\ast_p M\oplus  \wedge^4_{27}T^\ast_p M~~~\mathrm{for}~~G_{2,p}~~,\\
\wedge^3 T^\ast_p M &=& \wedge^3_{1}T^\ast_p M\oplus  \wedge^3_{7}T^\ast_p M\oplus  \wedge^3_{27}T^\ast_p M~~~\mathrm{for}~~G_{2,p}~~,\nn
\eeqan 
where the numbers used as lower indices indicate the dimension of the
corresponding irrep. The last two of these decompositions imply
similar decompositions into irreps. of $G_{2,p}$ for the spaces of
selfdual and anti-selfdual three- and four-forms:
\ben
(\wedge^4 T^\ast_p M)^\pm = \wedge^4_{1}T^\ast_p M\oplus  \wedge^4_{7}T^\ast_p M\oplus  \wedge^4_{27}T^\ast_p M~~~\mathrm{for}~~G_{2,p}~~.
\een
Furthermore, we have: 
\beqan
\label{Spin7ASD}
\begin{split}
& (\wedge^4 T^\ast_p M)^\pm=\wedge^4_{{\bf 1},\pm}T^\ast_p M\oplus  \wedge^4_{{\bf 7},\pm}T^\ast_p M\oplus  \wedge^4_{{\bf 27},\pm}T^\ast_p M~~~\mathrm{for}~~\Spin(7)_{\pm,p}~~,\\
& (\wedge^4 T^\ast_p M)^\mp = \wedge^4_{{\bf 35},\pm}T^\ast_p M~~~\mathrm{for}~~\Spin(7)_{\pm,p}~~,
\end{split}
\eeqan
where the $\pm$ superscripts indicate the subspaces of selfdual and
anti-selfdual forms while the $\pm$ subscripts indicate which of the
$\Spin(7)_p$ subgroups of $\SO(T_pM)$ we consider.  Comparing these
two decompositions, one sees immediately that the irreps of
$\Spin(7)_{\pm,p}$ appearing in \eqref{Spin7ASD} decompose as follows
under the $G_2$ action on $\wedge^4 T^\ast_p M$ which was discussed
above:
\beqan
\label{G2Spin7Irreps}
\boxed{
\begin{split}
&\wedge^4_{{\bf k},\pm}T^\ast_p M = \wedge^4_{k}T^\ast_p M~~,~~\mathrm{for}~~k=1,7,27~~\\
&\wedge^4_{{\bf 35},\pm}T^\ast_p M = \wedge^4_{1}T^\ast_p M\oplus  \wedge^4_{7}T^\ast_p M\oplus  \wedge^4_{27}T^\ast_p M
\end{split}}~~.
\eeqan
Let $\omega^{(k)}\in \Omega_k(\cU)$ and $\omega^{{\bf [k]}}_\pm\in
\Omega_{{\bf k}}(\cU)$ denote the (pointwise) projections of a form
$\omega$ on the irreps of $G_2$ and $\Spin(7)_\pm$ respectively. 

\subsection{The $G_2$ and $\Spin(7)_\pm$ parameterizations of $F$}
\label{subsec:Spin7Param}

\paragraph{$G_2$ parameterization.}

Recall from \cite{g2} that $F|_\cU=F_\perp+\hV\wedge F_\top$ and
$f|_\cU=f_\perp+f_\top\hV $, where $f_\top\in \Omega^0(\cU)$,
$f_\perp\in \Omega^1_\cU(\cD)$, $F_\top\in \Omega^3_\cU(\cD)$ and $F_\perp\in
\Omega^4_\cU(\cD)$, with:
\beqan
\label{Fdecomp}
\boxed{
\begin{split}
&F_\perp=F_\perp^{(7)}+F_\perp^{(S)}~~\mathrm{where}~
~F_\perp^{(7)}=\alpha_1\wedge\varphi\in \Omega^4_7(\cD)~~,
~~F_\perp^{(S)}=-\hat{h}_{kl}e^k\wedge\iota_{e^l}\psi\in \Omega^4_{\cU,S}(\cD)~~~\\
&F_\top=F_\top^{(7)}+F_\top^{(S)}~~\mathrm{where}~
~F_\top^{(7)}=-\iota_{\alpha_2}\psi\in \Omega^3_{\cU,7}(\cD)~~,
~~F_\top^{(S)}=\chi_{kl}e^k\wedge\iota_{e^l}\varphi\in \Omega^3_{\cU,S}(\cD)~
\end{split}
}~~.~~~~~~~
\eeqan
Here $\alpha_1,\alpha_2\in\Omega^1_\cU(\cD)$, while ${\hat h}=\frac{1}{2}{\hat
  h}_{ij}e^i\odot e^j$ and $\chi=\frac{1}{2}\chi_{ij}e^i\odot e^j$ are
sections of the bundle $\Sym^2_\cU(\cD^\ast)$. We have
$F_\top^{(S)}=F_\top^{(1)}+F_\top^{(27)}$ with $F_\top^{(1)}\in
\Omega^3_1(\cD)~,~F_\top^{(27)}\in \Omega^3_{\cU,27}(\cD)$ and a similar
decomposition for $F_\perp^{(S)}$.  The last relations correspond to
the decompositions of $\chi$ and ${\hat h}$ into their homothety parts
$\tr(\chi)g|_\cD$, $\tr({\hat h})g|_\cD$ and traceless parts:
\be
\chi^{(0)}\eqdef \chi-\frac{1}{7}\tr(\chi)g|_\cD~~,~~h^{(0)}={\hat h}-\frac{1}{7}\tr({\hat h})g|_\cD~~. 
\ee
Let $h,{\hat \chi}\in \Sym^2_\cU(\cD^\ast)$ denote the symmetric
tensors defined through:
\be
h_{ij}\eqdef {\hat h}_{ij}-\frac{1}{3}\tr_g({\hat h})g_{ij}~~,~~{\hat \chi}_{ij}\eqdef \chi_{ij}-\frac{1}{4}\tr_g(\chi)g_{ij}~~,
\ee
where:
\be
\tr_g(\chi)=-\frac{4}{3}\tr_g({\hat \chi})~~,~~\tr_g({\hat h})=-\frac{3}{4}\tr_g( h)~~.
\ee
The situation is summarized in Table 2.
\begin{table}[tt]
\centering
\begin{tabular}{|c|c|c|c|}
\hline
$G_2$ representation & $1$ & $7$ & $27$  \\
\hline\hline
$F_\perp\in \Omega^4_\cU(\cD)$ & $\tr_g({\hat h})$ & $\alpha_1 \in \Omega^1_\cU(\cD)$ & $h^{(0)}\in \Sym^2_{\cU,0}(\cD^\ast)$  \\
\hline
$F_\top\in \Omega^3_\cU(\cD)$ & $\tr_g({\hat \chi})$ & $\alpha_2 \in \Omega^1_\cU(\cD)$ & $\chi^{(0)} \in \Sym^2_{\cU,0}(\cD^\ast)$ \\
\hline
\end{tabular}
\
\caption{The $G_2$ parameterization of $F$ on the non-chiral locus.}
\label{table:G2param}
\end{table}
\paragraph{$\Spin(7)_\pm$ parameterization.}
The discussion of the previous subsection gives the following decompositions of the selfdual and anti-selfdual parts of $F$:
\be
F^\pm=F^{[\bf{1}]}_\pm+F^{[\bf{7}]}_\pm+F^{[\bf{27}]}_\pm\in \Omega^{4\pm}(\cU)~~,~~F^\mp=F^{[\bf{35}]}_\pm\in \Omega^{4\mp}(\cU)~~.
\ee
Since the Hodge operator intertwines $\Spin(7)_\pm$ representations, we have:
\beqa
&&(F^{[\bf{k}]}_\pm)_\perp=\pm \ast_\perp (F^{[\bf{k}]}_\pm)_\top~~~\mathrm{for}~~k=1,7,27~~,\\
&&(F^{[\bf{35}]}_\pm)_\perp=\mp \ast_\perp (F^{[\bf{35}]}_\pm)_\top~~.
\eeqa
One can parameterize $F^{[{\bf k}]}_\pm$ through a zero-form
$\cF^{[\bf{1}]}_\pm\in \Omega^0(\cU)$, a 2-form $\cF^{[{\bf 7}]}_\pm\in
\Omega^2(\cU)$, a $\cD$-longitudinal traceless symmetric covariant tensor $\cF^{[{\bf
      27}]}_\pm\in \Sym^2_{\cU,0}(\cD^\ast)$ and a traceless symmetric
covariant tensor $\cF^{[{\bf 35}]}_\pm\in \Sym^2_0(T^\ast \cU)$, which are
defined by:
\beqan
\label{TF1F7}
&&F^{[\bf{1}]}_\pm=\frac{1}{42}\cF^{[\bf{1}]}_\pm \Phi^\pm~~,\nn\\ 
&&F^{[\bf{7}]}_\pm=\frac{1}{96}\Phi\btu_1\cF^{[\bf{7}]}_\pm~~,\nn\\ 
&&F^{[\bf{27}]}_\pm=\frac{1}{24}(\cF^{[{\bf 27}]}_\pm)_{ij}e^i\wedge \iota_{e^j}\Phi^\mp~~,\\ 
&&F^{[\bf{35}]}_\pm=\frac{1}{24}(\cF^{[\bf{35}]}_{\pm})_{ab}e^a\wedge \iota_{e^b}\Phi^\pm\nn~~.  
\eeqan
The quantities $\cF^{[{\bf k}]}$ with $k=1,7,35$ can be recovered
from $F$ through the relation:
\ben
\label{cFF}
6(\iota_{e^a}F)\btu_3(\iota_{e^b}\Phi^\pm)=g_{ab}\cF^{[{\bf 1}]}_\pm+(\cF^{[{\bf 7}]}_\pm)_{ab}+(\cF^{[{\bf 35}]}_\pm)_{ab}~~.
\een
Define: 
\ben
\label{Tparam}
\boxed{
\begin{split}
& \beta_{1\pm}\eqdef (\cF^{[\bf{7}]}_\pm)_\top\in \Omega^1_\cU(\cD)~~,\\
& \beta_{2\pm}\eqdef n~\lrcorner~{\cal F}_\pm^{[{\bf 35}]}=(\cF^{[{\bf 35}]}_\pm)_{1j}e^j\in \Omega^1_\cU(\cD)~~,\\
& \sigma_\pm\eqdef\frac{1}{2}(\cF^{[\bf{35}]}_\pm)_{ij}e^i\odot e^j\in \Sym^2_{\cU}(\cD^\ast)~~,
\end{split}
}~~
\een
where $e_a$ is a local orthonormal frame such that $e_1=n\eqdef {\hat
  V}^\sharp$ and $j=2,\ldots, 8$. The fact that $F^{[{\bf 7}]}_\pm$ is
(anti-)selfdual implies:
\ben
\label{cF7perp}
(\cF^{[\bf{7}]}_\pm)_\perp=\mp\iota_{\beta_{1\pm}}\varphi~~.
\een
\begin{table}[tt]
\centering
\begin{tabular}{|c|c|c|c|c|}
\hline
$\Spin(7)_\pm$ representation & $\mathbf{1}$ & $\mathbf{7}$ & $\mathbf{27}$  & $\mathbf{35}$\\
\hline\hline
component   & $F^{[{\bf 1}]}_\pm\in \Omega^{4\mp}(\cU)$ & $F^{[{\bf 7}]}_\pm\in \Omega^{4\mp} (\cU)$ & $F^{[{\bf 27}]}_\pm\in \Omega^{4\mp}(\cU)$ & $F^{[{\bf 35}]}_\pm\in \Omega^{4\pm}(\cU)$\\
\hline
$\cU$-tensors & $\cF^{[{\bf 1}]}_\pm\in \Omega^0(\cU)$ & $\cF^{[{\bf 7}]}_\pm\in \Omega^2(\cU)$ & $\cF^{[{\bf 27}]}_\pm\in \Sym^2_{\cU,0}(\cD^\ast)$ & $\cF^{[{\bf 35}]}_\pm\in \Sym^2_0(T^\ast \cU)$ \\
\hline
$\cD$-tensors & $\cF^{[{\bf 1}]}_\pm\in \Omega^0(\cU)$ & $\beta_{1\pm}\in \Omega^1_\cU(\cD)$ & $\cF^{[{\bf 27}]}_\pm\in \Sym^2_{\cU,0}(\cD^\ast)$ & $\begin{array}{c}\beta_{2\pm}\in \Omega^1_\cU(\cD)~~ \\ \sigma\in \Sym^2_\cU(\cD^\ast)\end{array}$ \\
\hline
\end{tabular}
\
\caption{The $\Spin(7)_\pm$ parameterization of $F$ on the non-chiral locus and its $\cD$-refined version.}
\label{table:Spin7param}
\end{table}

\noindent Choosing an orthonormal frame with $e_1=n={\hat V}^\sharp$
and recalling \eqref{PhiDec}, relations \eqref{TF1F7} and
\eqref{Tparam} give the following parameterization of $F$, which
refines the parameterization used in \cite{Tsimpis} by taking into
account the decomposition into directions parallel and perpendicular
to ${\hat V}$:
\ben
\label{FSpin7TParam}
\!\!\!\!\!\!\!\!\!\boxed{
\begin{split}
&(F^{[\bf{1}]}_\pm)_\top=\pm \frac{1}{42}\cF^{[\bf{1}]}_\pm \varphi~~,~~~~~~~~~~~~~~~~~~~(F^{[\bf{1}]}_\pm)_\perp=\frac{1}{42}\cF^{[\bf{1}]}_\pm\psi~~,\\
&(F^{[\bf{7}]}_\pm)_\top=\frac{1}{24}\iota_{\beta_{1\pm}}\psi~~,~~~~~~~~~~~~~~~~~~
~~~~(F^{[\bf{7}]}_\pm)_\perp=\mp \frac{1}{24}\beta_{1\pm}\wedge \varphi~~,\\
& (F^{[\bf{27}]}_\pm)_\top=\mp\frac{1}{24}(\cF^{[{\bf 27}]}_\pm)_{ij}e^i\wedge \iota_{e^j}\varphi~~,~~~(F^{[\bf{27}]}_\pm)_\perp=\frac{1}{24}(\cF^{[{\bf 27}]}_\pm)_{ij}e^i\wedge \iota_{e^j}\psi~~,\\
&(F^{[\bf{35}]}_\pm)_\top\!=\pm\frac{1}{24}\big[\iota_{\beta_{2\pm}} \psi-\frac{4}{7}(\tr\sigma_\pm)\varphi+(\sigma_\pm^{(0)})_{ij}e^i\wedge \iota_{e^j}\varphi\big]~~,\\
& (F^{[\bf{35}]}_\pm)_\perp\!=\frac{1}{24}\big[{\beta_{2\pm}}\wedge \varphi+\frac{4}{7}(\tr\sigma_\pm)\psi+(\sigma_\pm^{(0)})_{ij}e^i\wedge \iota_{e^j}\psi\big]~~.
\end{split}}
\een
To arrive at the above, we used the relations:
\be
\varphi\btu_1 (\cF^{[\bf{7}]}_\pm)_\perp=\mp 3\iota_{\beta_{1\pm}}\psi~~,~~\psi\btu_1(\cF^{[\bf{7}]}_\pm)_\perp =\mp 3\beta_{1\pm}\wedge \varphi~~,
\ee
which follow from \eqref{cF7perp} and the identities given in the
Appendix of \cite{Kflows}. 

\paragraph{Relating the $G_2$ and $\Spin(7)_\pm$ parameterizations of $F$.}
Relation \eqref{G2Spin7Irreps} implies:
\ben
\label{FSpin7TopPerp}
\!\!\!\!\!\!\begin{split}
&(F_\pm^{\bf [k]})_\top=\frac{1}{2}(F_\top^{(k)}\pm \ast_\perp F_\perp^{(k)})~~~~,~~~(F_\pm^{\bf [k]})_\perp=\frac{1}{2}(F_\perp^{(k)}\pm \ast_\perp F_\top^{(k)}) ~~~\mathrm{for}~~k=1,7,27~~,\\
&(F_\pm^{\bf [35]})_\top=\frac{1}{2}(F_\top\mp \ast_\perp F_\perp)~~~~~,~~~(F_\pm^{\bf [35]})_\perp=\frac{1}{2}(F_\perp\mp \ast_\perp F_\top)~~.
\end{split}
\een
Comparing \eqref{FSpin7TParam} with \eqref{FSpin7TopPerp} and using
the $G_2$ parameterization of $F_\top$ and $F_\perp$ given in
\eqref{Fdecomp}, one can express the quantities in the last row of
Table \ref{table:Spin7param} in terms of $\alpha_1,\alpha_2$ and
${\hat h},{\hat \chi}$:
\ben
\label{Spin7G2Param}
\boxed{
\begin{split}
&\cF^{[{\bf 1}]}_\pm=-12\tr_g({\hat h}\pm{\hat \chi})~~,\\
&\sigma_\pm~~~=-12({\hat h}\mp{\hat \chi})~~,\\
&\cF^{[{\bf 27}]}_\pm=-12({\hat h}^{(0)}\pm{\hat \chi}^{(0)})~~,\\
&\beta_{1\pm}~~=-12(\alpha_2\pm\alpha_1)~~,\\
&\beta_{2\pm}~~=+12(\alpha_1\mp \alpha_2)~~.
\end{split}}
\een
These simple relations provide the connection between the $G_2$
parameterization \eqref{Fdecomp} and the refined $\Spin(7)_\pm$
parameterizations \eqref{FSpin7TParam}, thus allowing one to relate
the $G_2$ and $\Spin(7)_\pm$ decompositions of $F$.

\subsection{Relating the $G_2$ torsion classes to the Lee form and characteristic torsion of the $\Spin(7)_\pm$ structures}

Recall that the {\em Lee form} of the $\Spin(7)_\pm$ structure
determined by $\Phi^\pm$ on $\cU$ is the one-form defined through:
\ben
\label{Lee}
\boxed{\theta_\pm\eqdef \pm \frac{1}{7}\ast (\Phi^\pm\wedge \updelta\Phi^\pm) =-\frac{1}{7}\ast[\Phi^\pm\wedge (\ast \dd\Phi^\pm)]\in \Omega^1(\cU)~\Longrightarrow ~\Phi^\pm\wedge \updelta\Phi^\pm=\mp 7\ast \theta_\pm}~~,
\een
where we use the conventions of \cite{Ivanov} and the fact that $\ast
\Phi^\pm=\pm\Phi^\pm$. Also recall from loc. cit. that there exists a
unique $g$-compatible connection $\nabla^c$ with skew-symmetric
torsion such that $\nabla^c\Phi^\pm=0$. This connection is called the
{\em characteristic connection} of the $\Spin(7)_\pm$ structure.  Its
torsion form (obtained by lowering the upper index of the torsion
tensor of $\nabla^c$) is given by:
\ben
\label{Spin7Torsion}
\boxed{T_\pm=-\updelta\Phi^\pm\mp \frac{7}{6}\ast(\theta_\pm\wedge\Phi^\pm)=-\updelta\Phi^\pm
-\frac{7}{6}\iota_{\theta_\pm}\Phi^\pm=\pm \ast (\dd\Phi^\pm - \frac{7}{6}\theta_\pm\wedge\Phi^\pm)\in \Omega^3(\cU)}
\een
and is called the {\em characteristic torsion} of the $\Spin(7)_\pm$
structure. The normalization relation $||\Phi^\pm||^2=14$,
i.e. $\Phi^\pm\wedge \Phi^\pm=\pm 14\nu$ implies $\Phi^\pm\wedge
\iota_{\theta_\pm}\Phi^\pm=\pm 7\ast\theta_\pm$.  Thus $\Phi^\pm\wedge
T_\pm=\mp\frac{7}{6}\ast \theta_\pm$, where we used \eqref{Lee} and
\eqref{Spin7Torsion}.  It follows that the Lee form is determined by
the characteristic torsion through the equation:
\ben
\label{thetaT}
\boxed{\theta_\pm=\pm \frac{6}{7}\ast(\Phi^\pm\wedge T_\pm)}~~.
\een
Relation \eqref{Spin7Torsion} shows that the exterior derivative of $\Phi^\pm$ takes the form:
\ben
\label{Spin7TorsionEq}
\dd \Phi^\pm= \frac{7}{6}\theta_\pm\wedge \Phi^\pm\mp \ast T_\pm=\pm[\ast(\Phi^\pm\wedge T_\pm)]\wedge \Phi^\pm \mp \ast T_\pm~~.
\een
Recall the relation (see \cite{g2}): 
\be
D_n\psi=-3\vartheta\wedge \varphi~~,
\ee
where $\vartheta\in \Omega^1(\cD)$. Together with \eqref{PhiDec} and with the formula for the exterior
derivative of longitudinal forms (see Appendix C. of \cite{g2}), this gives:
\beqa
(\dd \Phi^\pm)_\top &=&\pm (H_\sharp \mp3\vartheta  - 3\tau_1) \wedge \varphi -(\frac{4}{7}\tr A\pm\tau_0 )\psi 
- A^{(0)}_{jk}e^j\wedge \iota_{e^k}\psi \mp \ast_\perp\tau_3~~, \\
(\dd \Phi^\pm)_\perp &=& 4\tau_1\wedge \psi+\ast_\perp\tau_2~~,
\eeqa
which implies:
\beqan
(\ast \dd \Phi^\pm)_\top&=&-\tau_2- 4\iota_{\tau_1}\varphi~~,\nn\\
(\ast \dd \Phi^\pm)_\perp&=&\mp ~\iota_{(H_\sharp \mp3\vartheta-3\tau_1)}\psi - (\frac{4}{7}\tr A\pm\tau_0 )\varphi + A^{(0)}_{jk}e^j\wedge \iota_{e^k}\varphi \mp\tau_3~~.
\eeqan
Using this relation and \eqref{PhiDec}, we can compute $\theta_\pm$ from
\eqref{Lee} and then determine $T_\pm$ from equation
\eqref{Spin7Torsion}. We find:
\ben
\label{Ttheta}
\!\!\!\!\boxed{
\begin{split}
&(\theta_\pm)_\top\!\!=-\frac{4}{7}\tr A \mp\tau_0~~~~~~~,
~~~~~~~(\theta_\pm)_\perp\!\!=- \frac{4}{7}(H_\sharp \mp 3\vartheta-6\tau_1)~~,\\
&(T_\pm)_\top\!\!=- \frac{2}{3}\iota_{(\pm H_\sharp-3\vartheta)}\varphi \mp \tau_2~~,
~~(T_\pm)_\perp\!\!=- \frac{1}{6}(\frac{4}{7}\tr A \pm \tau_0)\varphi
- \frac{1}{3}\iota_{(H_\sharp \mp 3\vartheta+ 3\tau_1)}\psi \pm A_{jk}^{(0)}e^j\wedge \iota_{e^k}\varphi - \tau_3~.
\end{split}}
\een
To arrive at the last two relations, we used the identities: 
\be
\iota_{\tau_2}\varphi=\iota_{\tau_3}\psi=\langle \tau_3,\varphi\rangle=0~~,~~
\ee
which follow from relations (B.13) and (B.14) given in Appendix B
of \cite{g2} upon using the fact that $\tau_3\in \Omega^3_{\cU,27}(\cD)$.

\subsection{Relation to previous work}
The problem of determining the fluxes $f,F$ in terms of the geometry
along the locus $\cU^+$ was considered in reference \cite{Tsimpis},
where the quantities denoted here by $L^+,\Phi^+$ were denoted simply
by $L,\Phi$.  Using the results of the previous subsections, one can
show that the relations given in Theorem 3 of \cite{g2} are equivalent,
on the non-chiral locus $\cU$, with equations (3.16), (3.17) and
(3.18) of \cite{Tsimpis}. This solves the problem of comparing the
approach of loc. cit. with that of \cite{MartelliSparks, g2}. The
major steps of the comparison with loc. cit. are given in Appendix
\ref{app:T}.

\section{Description of the singular foliation in the Morse case}
\label{sec:Morse}

In this section, we consider the case when the closed one-form
$\momega\in \Omega^1(M)$ is Morse. This case is generic in the sense
that Morse one-forms form a dense open subset of the set of all closed
one-forms belonging to the fixed cohomology class $\f$ --- hence a form
which satisfies equations \eqref{meq} can be replaced by a Morse form
by infinitesimally perturbing $b$.  Singular foliations defined by
Morse one-forms were studied in \cite{Gelbukh1}--\cite{Gelbukh9} and \cite{Levitt1}--\cite{Honda}.
Let $\Pi_f=\im (\per_\f)\subset \R$ be the period group of the
cohomology class $\f$ and $\rho(\f)=\rk \Pi_\f$ be its irrationality
rank.  The general results summarized in the following subsection hold
for any smooth, compact and connected manifold of dimension $d$ which
is strictly bigger than two, under the assumption that the set of
zeroes of $\momega$ (which in Novikov theory \cite{Farber} is called
the set of {\em singular points}):
\be
\Sing(\momega)\eqdef \{p\in M|\omega_p=0\}
\ee
is non-empty. Notice that $\Sing(\momega)$ is a finite set since $M$
is compact and since the zeroes of a Morse 1-form are isolated.
The complement:
\be
M^\ast\eqdef M\setminus \Sing(\momega)
\ee
is a non-compact open submanifold of $M$. Below, we shall use the notations 
$\cF_\momega$ for the regular foliation induced by $\momega$ on $M^\ast$ 
and $\bcF_\momega$ for the singular foliation induced on $M$.
In our application we have $n=8$ and: 
\be
\Sing(\momega)=\cW~~,~~M^\ast=\cU~~,~~\cF_\momega=\cF~~,~~\bcF_\momega=\bcF~~.
\ee

\subsection{Types of singular points} 
\label{subsec:singtypes}
Let $\ind_p(\momega)$ denote the Morse index of a point $p\in \Sing
(\momega)$, i.e. the Morse index at $p$ of a Morse function $h_p\in
\cC^\infty(U_p,\R)$ such that $\dd h_p$ equals $\momega|_{U_p}$, where
$U_p$ is some vicinity of $p$. This index does not depend on the
choice of $U_p$ and $h_p$. Let:
\beqa
&&\Sing_k(\momega)\eqdef \{p\in \Sing(\momega)|\ind_p(\momega)=k\}~~,~~k=1,\ldots, d\\
&&\Sigma_k(\momega)\eqdef \{p\in \Sing(\momega)|\ind_p(\momega)=k~\mathrm{or}~\ind_p(\momega)=d-k\}~~,~~k=1,\ldots,\left[\frac{d}{2}\right]~~. 
\eeqa
Thus $\Sigma_k(\momega)=\Sing_k(\momega)\cup\Sing_{n-k}(\momega)$ for
$k<\frac{d}{2}$ and $\Sigma_{d_0}(\momega)=\Sing_{d_0}(\momega)$ when $d=2d_0$
is even. In a small enough vicinity of $p\in
\Sing_k(\momega)$ (which we can assume to equal $U_p$ by shrinking the
latter if necessary), the Morse lemma applied to $h_p$ implies that
there exists a local coordinate system $(x_1,\ldots,x_d)$ such that:
\be
h_p=-\sum_{j=1}^k{x_j^2}+\sum_{j=k+1}^{d} x_j^2~~.
\ee
\paragraph{Definition.} The elements of $\Sigma_0(\momega)$ are called {\em centers}
while all other singularities of $\momega$ are called {\em saddle
  points}. The elements of $\Sigma_1(\momega)$ are called {\em strong
  saddle points}, while all other saddle points are called {\em weak}.

\paragraph{Remark.} Strong saddle points are sometimes called ``conical 
points''.  That terminology can lead to confusion, since all singular
points which are not centers are conical singularities of the singular
leaf to which they belong (see below), in the sense that the singular
leaf can be modeled by a cone (with one or two sheets) in a vicinity
of such a singular point. In other references, a ``conical point'' means 
any singularity which is not a center, i.e. what we call a saddle point.

\subsection{The regular and singular foliations defined by a Morse 1-form}

\paragraph{The regular foliation $\cF_\momega$.}

The Morse 1-form $\momega$ defines a regular foliation $\cF_\momega$
of the open submanifold $M^\ast$, namely the foliation which, by the
Frobenius theorem, integrates the regular Frobenius distribution
$\ker(\momega)|_{M^\ast}$. Following \cite{Gelbukh7}, we say that a
singular point $p\in \Sing(\momega)$ {\em adjoins} a leaf
$L$\footnote{This should not be confused with the quantity $L^\pm$
  discussed in Subsection 3.4 (or with the quantity denoted by $L$ in
  \cite{Tsimpis}).} of $\cF_\momega$ if the union $\{p\}\cup L$ is
connected; notice that a center cannot adjoin any leaf of
$\cF_\momega$. Let:
\be
s(L)\eqdef \{p\in \Sing(\momega)|p~\mathrm{adjoins}~L\}\subset \Sing(\momega)~~.
\ee
The set $s(L)$ is contained in the intersection of $\Sing(\momega)$ with 
the small topological frontier $\fr(L)$ of $L$:
\ben
\label{inclusion}
s(L)\subseteq \fr(L)\cap \Sing(\momega)~~.
\een
Notice that this inclusion can be strict; a beautifully drawn example
illustrating this in the two-dimensional case can be found in
\cite{Gelbukh9} (see Figure 2(c) of loc. cit.). We have $s(L)\cap
\Sigma_0(\momega)=\emptyset$ and hence
$s(L)=\sqcup_{k=1}^{\left[\frac{d}{2}\right]}s_k(L)$, where:
\be
s_k(L)\eqdef s(L)\cap \Sigma_k(L)~~.
\ee

\paragraph{Classification of the leaves of $\cF_\momega$.}
\begin{itemize}
\itemsep 0.0em
\item{\bf Compactifiable and non-compactifiable leaves.} We say that a
  leaf $L$ of $\cF_\momega$ is {\em compactifiable} if the set $L\cup
  \Sing(\momega)$ is compact, which amounts to the condition that the
  small topological frontier $\fr(L)\eqdef {\bar L}\setminus L$ of $L$
  in $M$ is a (possibly void) subset of $\Sing(\momega)$ and hence a
  finite set. With this definition, compact leaves of $\cF_\momega$
  are compactifiable, but not all compactifiable leaves are compact. A
  {\em non-compactifiable leaf} of $\cF_\momega$ is a leaf which is
  not compactifiable; obviously such a leaf is also non-compact. The
  closure of a non-compactifiable leaf is a set with non-empty
  interior \cite{ImanishiSing, Levitt1}, so the small frontier of such
  a leaf is an infinite set.
\item{\bf Ordinary and special leaves.} The leaf $L$ is called {\em
  ordinary} if $s(L)$ is empty and {\em special} if $s(L)$ is
  non-empty.  An ordinary leaf is either compact or
  non-compactifiable.  Any non-compact but compactifiable leaf is a
  special leaf, but there also exist non-compactifiable special leaves
  (see Table \ref{table:leaves}). 
\end{itemize}

\noindent 
The foliation $\cF_\momega$ has only a finite number of special leaves,
because the local form of leaves near the points of $\Sing(\momega)$
(see below) shows that at most two special leaves can contain each
such point in their closures (recall that we assume $d\geq 3$) . We
shall see later that each non-compactifiable leaf (whether special or
not) covers densely some open and connected subset of $M^\ast$. Notice
that every singular point which is not a center adjoins some special leaf. 
Hence:
\ben
\label{SigmakDec}
\Sigma_k(\momega)=\cup_{L=\mathrm{special~leaf~of~}\cF_\momega}s_k(L)~~,~~\forall k=1\ldots \left[\frac{d}{2}\right]~~.
\een

\begin{table}[tt]
\centering
\begin{tabular}{|c|c|c|c|}
\hline
\multirow{2}{*}{type of $L$} &  \multicolumn{2}{c}{compactifiable} \vline & \multirow{2}{*}{non-compactifiable}\\
\cline{2-3}
& compact & non-compact  &  \\
\hline\hline
ordinary & Y  & ---  & Y \\
\hline
special & --- & Y & Y\\
\hline
$\mathrm{Card}(\fr L)$ &\multicolumn{2}{c}{finite} \vline & infinite\\
\hline
\end{tabular}
\caption{Classification of the leaves of $\cF_\momega$, where the
  allowed combinations are indicated by the letter ``Y''. A
  compactifiable leaf is ordinary iff it is compact and it is special
  iff it is non-compact. A non-compactifiable leaf may be either
  ordinary or special. Non-compactifiable leaves coincide
  \cite{ImanishiSing,Levitt1} with those leaves whose small frontier is an
  infinite set, while compactifiable leaves are those leaves whose
  small frontier is finite.}
\label{table:leaves}
\end{table}

\paragraph{The singular foliation $\bcF_\momega$.}

One can describe \cite{FKL, Farber} the singular
  foliation $\bcF_\momega$ of $M$ defined by $\momega$ as
the partition of $M$ induced by the equivalence relation $\sim$
defined as follows. We put $p\sim q$ if there exists a smooth curve
$\gamma:[0,1]\rightarrow M$ such that:
\be
\gamma(0)=p~~,~~\gamma(1)=q~~\mathrm{and}~~\momega(\dot{\gamma}(t))=0~~\forall t\in [0,1]~~.
\ee
The {\em leaves of $\bcF_\momega$} are the equivalence classes of this
relation; they are connected subsets of $M$ (which need not be
topological manifolds when endowed with the induced topology). Any such
leaf is either of the form $\{p\}$ where $p\in \Sigma_0(\momega)$ is a center
or is a topological subspace of $M$ of Lebesgue covering dimension
equal to $n-1$.  

\paragraph{Remark.} 
We stress that $\bcF_\momega$ is not generally a foliation of $M$ in
the ordinary sense of foliation theory but (as explained in the
previous section) it should be viewed as a Haefliger structure. It is
not even a $\cC^0$-foliation, i.e. a foliation in the category of
topological manifolds (locally Euclidean Hausdorff topological
spaces), because singular leaves of $\bcF_\momega$ which pass through
strong saddle points can be locally disconnected by removing those
points and hence are not topological manifolds.

\paragraph{Regular and singular leaves of $\bcF_\momega$.}

A leaf $\cL$ of $\bcF_\momega$ is called {\em singular} if it
intersects $\Sing(\momega)$ and {\em regular} otherwise. The regular
leaves of $\bcF_\momega$ coincide with the ordinary leaves of $\cF_\momega$; notice 
that every center singularity is a singular leaf. 
On the other hand, each singular leaf which is not a center is a
disjoint union of a finite number of special leaves of $\cF_\momega$
and of some subset $s(\cL)\eqdef \cL\cap \Sing(\momega)$ of
$\Sing(\momega)$, which we shall call the {\em set of singular points
  of $\cL$}. We have:
\be
\cL\setminus s(\cL)=L_1\sqcup \ldots \sqcup L_r\sqcup L'_1\sqcup \ldots \sqcup L'_t~~,
\ee
where $L_1,\ldots, L_r$ are compactifiable special leaves while
$L'_1,\ldots, L'_t$ are non-compactifiable special leaves of
$\cF_\momega$. We also have $s(\cL)=s^\c(\cL)\cup s^\nc(\cL)$ (generally
a non-disjoint union), with:
\be
s^c(\cL)\eqdef \cup_{i=1}^r s(L_i)~~,~~s^\nc(\cL)\eqdef \cup_{j=1}^t(L'_j)~~.
\ee
The singular leaf $\cL$ decomposes as: 
\ben
\label{cLdec}
\cL=\cL^{\rm c}\sqcup \cL^\nc~~,
\een
where the {\em compact part} and {\em non-compact part} of $\cL$ are defined through: 
\beqa
\cL^c~&\eqdef& \bar{L}_1\cup \ldots \cup \bar{L}_r=L_1\sqcup \ldots \sqcup L_r\sqcup s^c(\cL)\nn\\
\cL^\nc&\eqdef& \cL\setminus \cL^c=(L'_1\sqcup \ldots \sqcup L'_t)\sqcup (s(\cL)\setminus s^c(\cL))~~.
\eeqa
The set $s^c(\cL)$ consists of those singular points of $\cL$ which
lie on the compact part $\cL^c$. Notice that both the compact and
non-compact parts of $\cL$ can be void and that a non-compactifiable
special leaf component $L'_j$ of $\cL$ can adjoin points from
$s^c(\cL)$ as well as from $s(\cL)\setminus s^c(\cL)$ simultaneously;
furthermore, $\cL^\nc$ may meet itself at certain points of
$s(\cL)\setminus s^c(\cL)$\footnote{We thank I. Gelbukh for drawing
  our attention to these points.}.  When $\cL$ is a center leaf
$\{p\}$, we define $\cL^{\rm c}\eqdef s(\cL)=\{p\}$ and $\cL^\nc
\eqdef \emptyset$. Notice that any non-empty subset $A$ of $\cL$
determines $\cL$ as the saturation of $A$ with respect to the
equivalence relation $\sim$.  If $S_\momega$ denotes the union of
$\Sing(\momega)$ with all special leaves of $\cF_\momega$, then the
singular leaves of $\bcF_\momega$ (including the centers) coincide
with the connected components of $S_\momega$. Notice that
$\fr(L_i)=s(L_i)={\bar L}_i\cap \Sing(\momega)$ for each
compactifiable leaf component $L_i$, $i=1\ldots r$. The compact sets
${\bar L}_i$ meet themselves or each other only in strong saddle
points.  In particular, we have:
\be
{\bar L}_{i_1}\cap {\bar L}_{i_2}=s(L_{i_1})\cap s(L_{i_2})=s_1(L_{i_1})\cap s_1(L_{i_2})\subset \Sigma_1(\momega)~~\mathrm{for}~~1\leq i_1<i_2\leq r~~.
\ee

\noindent The following definition generalizes the notion of generic
Morse function:

\paragraph{Definition.} 
The Morse form $\momega$ is called {\em generic} if every singular
leaf of $\bcF_\momega$ contains exactly one singular point $p\in
\Sing(\momega)$. 

\subsection{Behavior of the singular leaves near singular points}

In a small enough vicinity of $p\in \Sing_k(\momega)$, the singular
leaf $\cL_p$ passing through $p$ is modeled by the locus $Q_k\subset
\R^n$ given by the equation $h_p=0$, where $p$ corresponds to the
origin of $\R^n$. One distinguishes the cases (see Tables
\ref{table:singtypes} and \ref{table:strongsaddles}):

\begin{itemize}
\itemsep 0.0em
\item $k\in \{0, n\}$, i.e. $p$ is a center. Then $\cL_p=\{p\}$ and
  the nearby leaves of $\cF_p$ are diffeomorphic to $S^{n-1}$.
\item $2\leq k\leq n-2$, i.e. $p$ is a weak saddle point. Then $Q_k$
  is diffeomorphic to a cone over $S^{k-1}\times S^{n-k-1}$ and
  $\R^n\setminus Q_k$ has two connected components while
  $Q_k\setminus\{p\}$ is connected. Removing $p$ does not {\em
    locally} disconnect $\cL_p$.
\item $k\in \{1,n-1\}$, i.e. $p$ is a strong saddle point. Then $Q_k$
  is diffeomorphic to a cone over $\{-1,1\}\times S^{n-2}$ and
  $\R^n\setminus Q_k$ has three connected components while
  $Q_k\setminus\{0\}$ has two components.  Removing $p$ {\em locally}
  disconnects $\cL_p$. A strong saddle point $p\in \Sigma_1(\momega)$
  is called {\em splitting} \cite{Gelbukh7} (or {\em blocking}
  \cite{Levitt3}) if it adjoins two different (special) leaves of the regular
  foliation $\cF_\momega$ and it is called {\em non-splitting} (or a
  {\em transformation point} \cite{Gelbukh7}) if it adjoins a single (special)
  leaf of $\cF_\momega$ (see Table \ref{table:strongsaddles}). If a
  singular leaf $\cL$ contains only one splitting point, then removing
  it disconnects $\cL$. If a singular leaf $\cL$ contains more than
  one splitting point, then removing it may not disconnect $\cL$ (an
  example of such behavior is given in \cite[Figure 7(b)]{Gelbukh7}).
\end{itemize}

\begin{table}[h!]
\centering
\begin{tabular}{ | c | c| c| c| }
\hline
Name & Morse index & Local form of $\cL_p$ & Local form of regular leaves\\ 
\hline
Center & $0$ or $n$ & $\bullet=\{p\}$ &
\begin{minipage}{.3\textwidth}
\centering
\vspace{0.2em}
\includegraphics[scale=0.25]{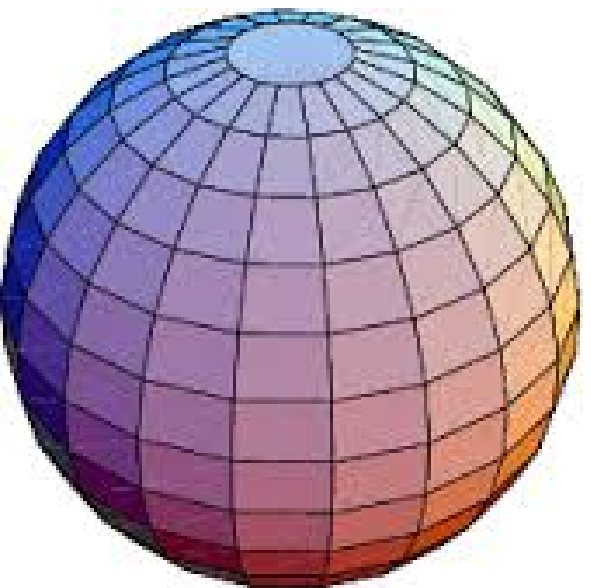}
\vspace{0.2em}
\end{minipage}\\ 
\hline
Weak saddle &  between $2$ and $n-2$ & 
\begin{minipage}{.23\textwidth}
\centering
\vspace{0.2em}
\includegraphics[scale=0.3]{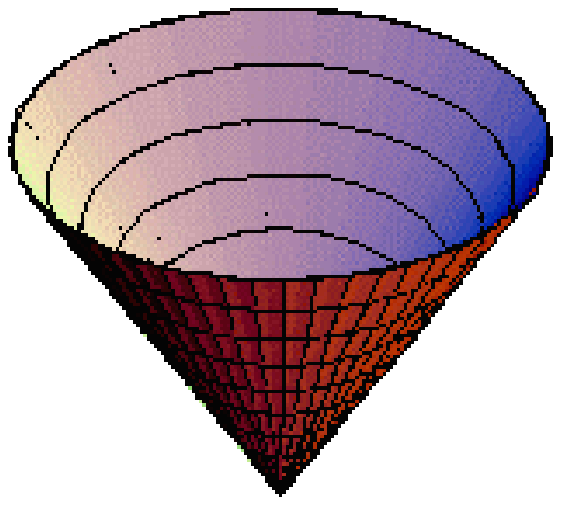}
\vspace{0.2em}
\end{minipage} 
& \begin{minipage}{.23\textwidth}
\centering
\vspace{0.2em}
\includegraphics[scale=0.3]{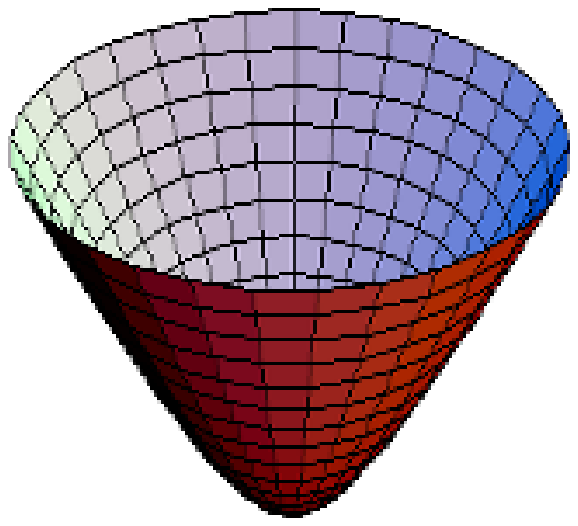}
\vspace{0.2em}
\end{minipage}\\ 
\hline
Strong saddle &  $1$ or $n-1$ &
\begin{minipage}{.23\textwidth}
\centering
\vspace{0.2em}
\includegraphics[scale=0.3]{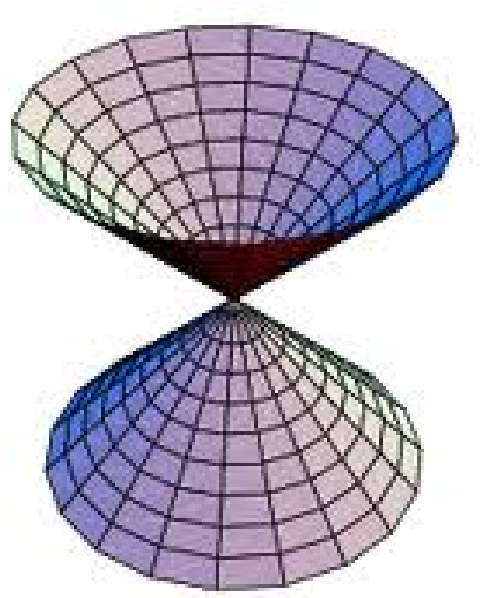}
\vspace{0.2em}
\end{minipage} 
& 
\begin{minipage}{.3\textwidth}
\centering
\vspace{0.2em}
\includegraphics[scale=0.3]{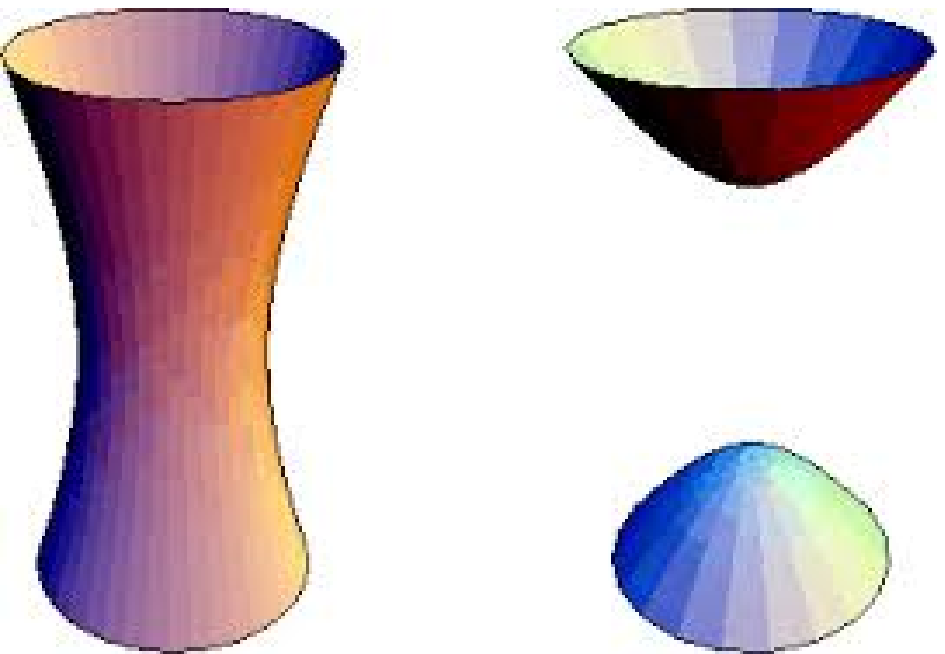}
\vspace{0.2em}
\end{minipage}\\ 
\hline
\end{tabular}
\caption{Types of singular points $p$. The first and third figure on
  the right depict the case $d=3$ for centers and strong saddles,
  while the second figure attempts to depict the case $d>3$ for a weak
  saddle (notice that weak saddles do not exist unless $d>3$).  In
  that case, the topology of the leaves does not change locally when
  they ``pass through" the weak saddle point. $\cL_p$ denotes the
  singular leaf of $\bcF_\momega$ which passes through $p$.}
\label{table:singtypes}
\end{table}
\noindent We have a decomposition
$\Sigma_1(\momega)=\Sigma_1^\sp(\momega)\sqcup\Sigma_1^\nsp (\momega)$
of the set of strong saddle points, where:
\beqa
&&\Sigma_1^\sp(\momega)\eqdef\{p\in \Sigma_1(\momega)|p~~\mathrm{is~a~splitting~singularity}\}\\
&&\Sigma_1^\nsp(\momega)\eqdef\{p\in \Sigma_1(\momega)|p~~\mathrm{is~a~non-splitting~singularity}\}~~.
\eeqa
Taking into account the local behavior of leaves near the various
types of singular points, we find that the decomposition
\eqref{SigmakDec} is disjoint for $k\neq 1$:
\be
\Sigma_k(\momega)=\sqcup_{L=\mathrm{special~leaf~of~}\cF_\momega}s_k(L)~~,~~\forall k=2\ldots \left[\frac{d}{2}\right]~~.
\ee
while the decomposition for $k=1$ may fail to be disjoint: 
\ben
\label{Sigma1Dec}
\Sigma_1(\momega)=\cup_{L=\mathrm{special~leaf~of~}\cF_\momega}s_1(L)~~.
\een
More precisely: 
\beqa
\Sigma^\nsp_1(\momega)&=&\sqcup_{L=\mathrm{special~leaf~of~}\cF_\momega}s^\nsp_1(L)~~\\
\Sigma_1^\sp(\momega)&=&\cup_{L=\mathrm{special~leaf~of~}\cF_\momega}s_1^\sp(L)~~,
\eeqa
where we defined: 
\be
s^\nsp_1(L)\eqdef s_1(L)\cap \Sigma_1^\nsp(\momega)~~,~~s^\sp_1(L)\eqdef s_1(L)\cap \Sigma_1^\sp(\momega)
\ee
and where the second union may be non-disjoint. This is because two
distinct special leaves of $\cF_\momega$ can meet each other only at a
strong saddle point which is a splitting singularity.
\begin{table}[h!]
\centering
\begin{tabular}{ | c | c | }
\hline
Singularity type & Example of global shape for $\cL_p$ \\ 
\hline
Splitting & 
\begin{minipage}{.3\textwidth}
\centering
\vspace{0.6em}
\includegraphics[scale=0.08, angle=90]{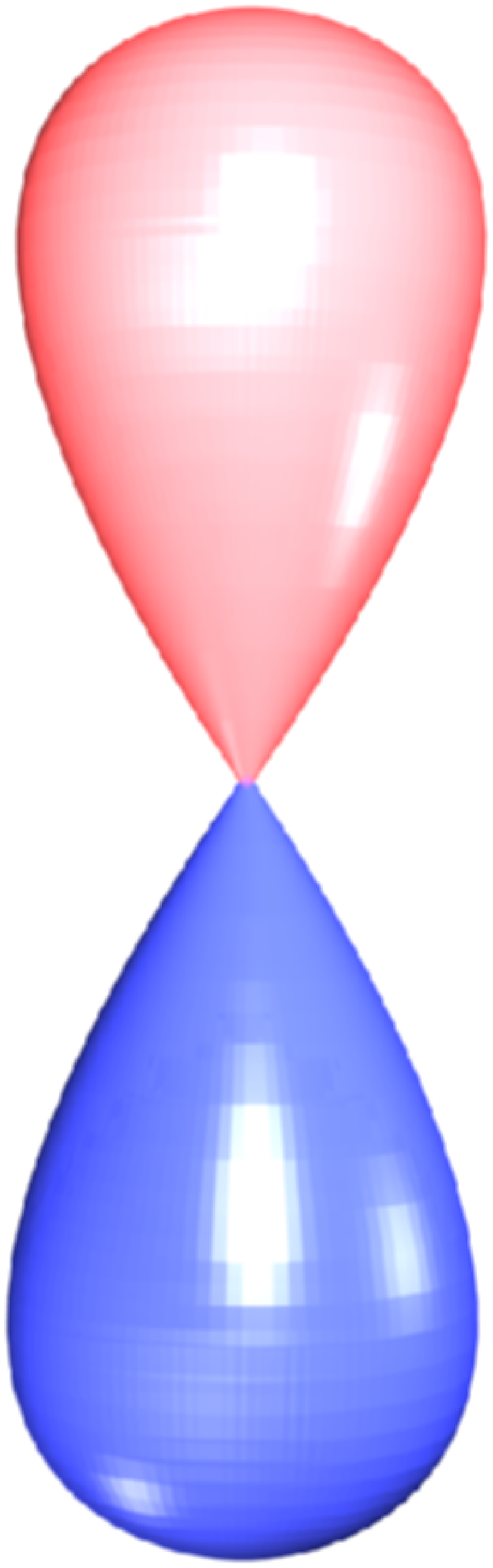}
\vspace{0.6em}
\end{minipage}\\ 
\hline
Non-splitting &  
\begin{minipage}{.3\textwidth}
\centering
\vspace{0.2em}
\includegraphics[scale=0.4]{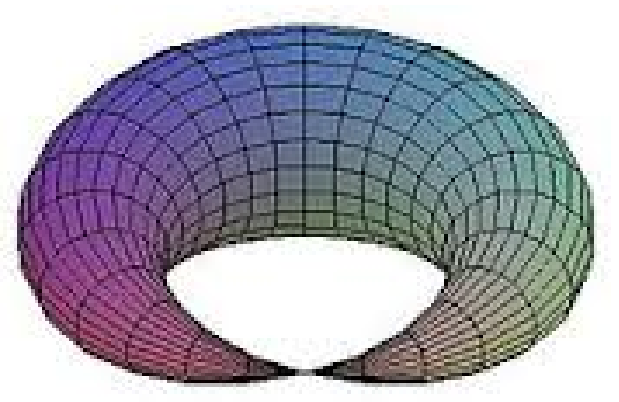}
\vspace{0.2em}
\end{minipage}\\ 
\hline
\end{tabular}
\caption{Types of strong saddle points. The figures illustrate the two
  types through two simple examples in the case $d=3$.  The figure in
  the first row uses different colors to indicate two different
  special compactifiable leaves of $\cF_\omega$ which are subsets of
  the same singular leaf of $\bcF_\momega$, each of them adjoining the same splitting
  singular point. The figure in the second row shows a single special
  compactifiable leaf of $\cF_\momega$ which adjoins a single
  non-splitting singular point.}
\label{table:strongsaddles}
\end{table}
\subsection{Combinatorics of singular leaves} 

\

\paragraph{Definition.} 
A singular leaf of $\bcF_\momega$ which is not a center is called a
{\em strong singular leaf} if it contains at least one strong saddle
point and a {\em weak singular leaf} otherwise.

\

\noindent A weak singular leaf is obtained by adjoining weak saddle
points to a single special leaf of $\cF_\momega$. Such singular leaves
are mutually disjoint and their singular points determine a partition
of the set $\Sigma_{>1}(\momega)\eqdef
\cup_{k=2}^{\left[\frac{d}{2}\right]}\Sigma_k(\momega)$.  The
situation is more complicated for strong singular leaves, as we now
describe.

\begin{figure}[h!]
\begin{center}
\scalebox{0.5}{\input{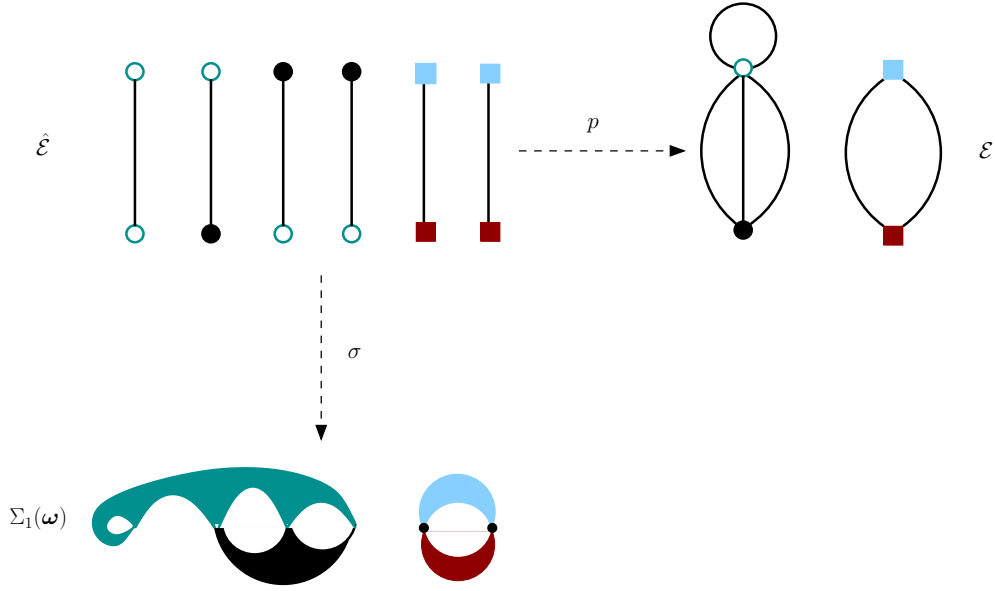}}
\caption{Example of the graphs ${\hat \cE}$ and $\cE$ for a Morse form
  foliation $\bcF_\momega$ with two compact strong singular leaves.  
  The regular foliation $\cF_\momega$ of
  $M^\ast$ has four special leaves, each of which is compactifiable;
  they are depicted using four different colors. At the bottom of the
  picture, we depict $\Sigma_1(\momega)$ as well as the schematic
  shape of the special leaves in the case $d=3$. The strong singular leaves
  of $\bcF_\momega$ correspond to the left and right parts of the
  figure at the bottom; each of them is a union of two special leaves
  of $\cF_\momega$ and of singular points. Each special leaf
  corresponds to a vertex of $\cE$.}
\label{fig:dimer}
\end{center}
\end{figure}

At each $p\in \Sigma_1(\momega)$, consider the strong singular leaf
$\cL$ passing through $p$. The intersection of $\cL\setminus\{p\}$
with a sufficiently small neighborhood of $p$ is a disconnected manifold
diffeomorphic to a union of two cones without apex, whose rays near $p$ determine a
connected cone $C_p\subset T_pM$ inside the tangent space to $M$ at
$p$ (see the last row of Table \ref{table:singtypes}). The set
$\dotC_p\eqdef C_p\setminus\{0_p\}$ (where $0_p$ is the zero vector of
$T_p M$) has two connected components, thus $\pi_0(\dotC_p)$ is a
two-element set. Hence the finite set:
\be
{\hat \Sigma_1}(\momega)\eqdef \sqcup_{p\in \Sigma_1(M)} \pi_0(\dotC_p)
\ee is a
double cover of $\Sigma_1(\momega)$ through the projection $\sigma$
that takes $\pi_0(\dotC_p)$ to $\{p\}$. Consider the complete
unoriented graph having as vertices the elements of ${\hat
  \Sigma_1}(\momega)$.  This graph has a dimer covering given by the
collection of edges:
\be
{\hat \cE}=\{\pi_0(\dotC_p)|p\in \Sigma_1(\momega)\}~~,
\ee
which connect vertically the vertices lying above the same point of
$\Sigma_1(\momega)$ (see Figure \ref{fig:dimer}).  If $L$ is a special
leaf of $\cF_\momega$ and $p\in \Sigma_1(\momega)$ adjoins $L$, then
the connected components of the intersection of $L$ with a
sufficiently small vicinity of $p$ are locally approximated at $p$ by
one or two of the connected components of $\dotC_p$. The second case
occurs iff $p$ is a non-splitting strong saddle point (see Table
\ref{table:strongsaddles}).  Hence $L$ determines a subset ${\hat
  s}_1(L)$ of ${\hat \Sigma}_1(\momega)$ such that $\sigma({\hat
  s}_1(L))=s_1(L)$ and such that the fiber of ${\hat s}_1(L)$ above a
point $p\in s_1(L)$ has one element if $p$ is a splitting singularity
and two elements if $p$ is non-splitting. If $L'$ is a different
special leaf of $\cF_\momega$, then the sets ${\hat s}_1(L')$ and
${\hat s}_1(L)$ are disjoint, even though their projections $s_1(L)$
and $s_1(L')$ through $\sigma$ may intersect in
$\Sigma_1(\momega)$. Hence the special leaves of $\cF_\momega$ define
a partition of ${\hat \Sigma}_1(\momega)$:
\be
{\hat \Sigma}_1(\momega)=\sqcup_{L=\mathrm{special~leaf~of~}\cF_\momega}{\hat s}_1(L)~~,
\ee
which projects through $\sigma$ to the generally non-disjoint
decomposition \eqref{Sigma1Dec}. Viewing ${\hat \cE}$ as a
disconnected graph on the vertex set ${\hat \Sigma}_1(\momega)$, we
let $\cE$ denote the (generally disconnected) graph obtained from
${\hat \cE}$ upon identifying all vertices belonging to ${\hat
  s}_1(L)$ for each special leaf $L$ of $\cF_\momega$, i.e. by
collapsing ${\hat s}_1(L)$ to a point for each special
leaf\footnote{If $s_1(L)$ is empty, this operation does nothing.} $L$.
Let $p:{\hat
  \cE}\rightarrow \cE$ denote the corresponding projection. The graph
$\cE$ has one vertex for each special leaf of $\cF_\momega$ which
adjoins some strong saddle point and an edge for each strong saddle
point. Notice that this edge is a loop when the strong saddle point
is a non-splitting singularity, since a non-splitting singularity 
adjoins a single special leaf.
A strong singular leaf of $\bcF_\momega$ can be written as:
\ben
\label{cLdec2}
\cL=(\sqcup_{\alpha=1}^{r+t}{L''_\alpha})\sqcup s(\cL)~~,
\een
where $L''_\alpha$ are special leaves of $\cF_\momega$ (compactifiable
or not).  Its set of strong saddle singular points $s_1(\cL)=
\cup_{\alpha=1}^{r+t}s_1(L''_\alpha)$ is the projection through
$\sigma$ of the set ${\hat s}_1(\cL)\eqdef
\sqcup_{\alpha=1}^{r+t}{\hat s}_1(L''_\alpha)$.  Let ${\hat \cE}_\cL$
be the (generally disconnected) subgraph of ${\hat \cE}$ consisting of
those edges of ${\hat \cE}$ which meet ${\hat s}_1(\cL)$. Then
$s_1(\cL)$ is obtained from ${\hat \cE}_\cL$ by contracting each edge
to a single point. If all special leaves $L$ of $\cF_\momega$ are
known, then ${\hat \cE}_{\cL}$ uniquely determines the strong singular leaf
$\cL$. Indeed, ${\hat \cE}_\cL$ contains the information about how the
special leaves which form $\cL$ meet themselves and each other at the strong saddle
points.  Since $\cL$ is connected and maximal with this property, the graph $\cE_\cL$
obtained from ${\hat \cE}_\cL$ by identifying to a single point the
vertices of each of the subsets ${\hat s}_1(L''_\alpha)$ is a connected
component of $\cE$. It follows that the strong singular leaves of
$\bcF_\momega$ are in one to one correspondence with the connected
components of the graph $\cE$ --- namely, their subgraphs ${\hat
  \cE}_\cL$ are the preimages through $p$ of those components.

\

\noindent In our application, the set
$\Sing(\momega)=\cW=\cW^+\sqcup\cW^-$ consists of positive and
negative chirality points of $\xi$, which are the points where $b$
attains the values $b=\pm 1$. Relation \eqref{meq} implies that $\mf$
satisfies:
\be
\oint_{\gamma}\mf=0~~
\ee
for any smooth closed curve $\gamma\in \cL\setminus \cW$ and hence
$\f$ restricts to a trivial class in singular cohomology along each
leaf $\cL$ of $\bcF$:
\be
\iota^*(\f)=0\in H^1(\cL,\R)~~,
\ee
where $\iota:\cL\hookrightarrow M$ is the inclusion map while
$H^1(\cL,\R)$ is the first singular cohomology group (which coincides
with the first de Rham cohomology group when $\cL$ is non-singular).
The pull-back of $\mf$ to $\cL\setminus \cW$ is given by:
\be
\mf|_{\cL\setminus\cW}=\mf_\perp=\dd_\perp\mb~~.
\ee
Notice that $f_\perp$ and $b$ have well-defined limits (equal to $f_p$
and $b(p)\in \{-1,1\}$) at each singular point $p\in \cL\cap \cW$ of a
singular leaf $\cL$. If $p_1,p_2\in \cL\cap \cW$ are two singular
points lying on the same singular leaf $\cL$ and
$\gamma:(0,1)\rightarrow \cL\setminus \cW$ is a smooth path which has
limits at $0,1$ given by $p_1$ and $p_2$, then the integral
$\int_\gamma \mf$ is well-defined and given by:
\be
\int_\gamma\mf=e^{3\Delta(p_2)}b(p_2)-e^{3\Delta(p_1)}b(p_1)~~,
\ee
where $b(p_i)\in \{-1,1\}$. 

\subsection{Homology classes of compact leaves} 

Let $H_\momega$ be the (necessarily free) subgroup of $H_{n-1}(M,\Z)$
generated by the compact leaves of $\cF_\momega$ and let
$c(\momega)\eqdef \rk H_\momega$ denote the number of homologically
independent compact leaves.  It was shown in \cite{Gelbukh1} that
$H_\momega$ admits a basis consisting of homology classes $[L_i]$
$(i=1,\ldots, c(\momega))$ of compact leaves\footnote{Such a basis is
  provided by the homology classes of the compact leaves corresponding
  to the edges of any spanning tree of the foliation graph defined
  below.}  and that the homology class of any compact leaf $L$ of
$\cF_\momega$ expands in this basis as:
\be
[L]=\sum_{i=1}^{c(\momega)}n_i[L_i]~~\mathrm{where}~~n_i\in \{-1,1\}~~.
\ee
Furthermore \cite{Gelbukh1,Gelbukh3}, there exists a system of
$\Z$-linearly independent one-cycles $\gamma_i\in H_1(M,\Z)$
$(i=1,\ldots, c(\momega))$ such that $(\gamma_i,[L_j])=\delta_{ij}$
and such that $\gamma_i$ provide a direct sum decomposition:
\be
H_1(M,\Z)=\langle \gamma_1,\ldots, \gamma_{c(\momega)}\rangle \oplus \iota_\ast(H_1(\Delta))~~,
\ee
where $\iota:\Delta \hookrightarrow M$ is the inclusion map. 
Let $\cH_\momega\eqdef H_\momega\cap (\ker\per_\momega)^\perp$. Then
\cite{Gelbukh5} the subgroup $\cH_\momega$ is a direct summand in
$H_\momega$ while $H_\momega$ is a direct summand in
$H_{n-1}(M,\Z)$. Furthermore, only the following
values are allowed for $\rk\cH_\momega$:
\be
\rk\cH_\momega\in \{0,\ldots \rho (\momega)-2\}\cup\{\rho(\momega)\}~~.
\ee

\subsection{The Novikov decomposition of $M$}

What we shall call the ``Novikov decomposition'' is a generalization
of the Morse decomposition \cite{Milnor, Morse1, Morse2}, which was
introduced in \cite{Melnikova2, Melnikova3} (see also \cite{FKL,
  Honda}) and used extensively in \cite{Gelbukh1}--\cite{Gelbukh9};
the name is motivated by analogy with ``Morse decomposition'', due to
the role which this decomposition plays in the modern study of the
topology of closed one-forms \cite{Farber}. Define $C^\M$ to be the
union of all compact leaves and $C^\m$ to be the union of all
non-compactifiable leaves of $\cF_\momega$; it is clear that these two
subsets of $M$ are disjoint.  Then it was shown in
\cite{ImanishiSing,Levitt1} that both $C^\M$ and $C^\m$ are open
subsets of $M$ which have a common topological small frontier
$F$\footnote{This should not be confused with the internal part of the
  flux which is denoted by the same letter.} given by the (disjoint)
union $F_0\cup \Sing(\momega)$, where $F_0$ is the union of all those
leaves of $\cF_\momega$ which are compactifiable but non-compact:
\be
\fr C^\M=\fr C^\m=F\eqdef F_0\sqcup \Sing(\momega)~~.
\ee
Each of the open sets $C^\M$ and $C^\m$ has a finite number of
connected components, which are called the {\em maximal} and {\em
  minimal} components of the set $M\setminus F=C^\M\sqcup C^\m$. We
let:
\begin{itemize}
\itemsep0em 
\item $N_\M(\momega)\eqdef |\pi_0(C^\M)|$ denote the number of maximal components
\item $N_\m(\momega)\eqdef |\pi_0(C^\m)|$ denote the number of minimal components
\end{itemize}
Indexing these by $C^\M_j$ and $C^\m_a$ (where $j=1,\ldots,N_\M(\momega)$
and $a=1,\ldots,N_\m(\momega)$), we have:
\ben
\label{CMm}
C^\M=\sqcup_{j=1}^{N_\M(\momega)}C_j^\M~~,~~C^\m=\sqcup_{a=1}^{N_\m(\momega)}C_a^\m~~
\een
and hence (since \eqref{CMm} are {\em finite} and {\em disjoint}
unions) we also have:
\beqa
&&\overline{C^\M}=\cup_{j=1}^{N_\M(\momega)}\overline{C_j^\M}~~,
~~\overline{C^\m}=\cup_{a=1}^{N_\m(\momega)}\overline{C_a^\m}~~,\\
&&F=\fr C^\M=\cup_{j=1}^{N_\M(\momega)}\fr C_j^\M=\fr C^\m=\cup_{a=1}^{N_\m(\momega)}\fr C_a^\m~~.
\eeqa 
Notice that the unions appearing in these equalities need not be
disjoint anymore, in particular the small frontiers of two distinct maximal
components can intersect each other and similarly for two distinct
minimal components. Let \footnote{$\Delta$ should not be confused with 
the warp factor.}: 
\be
\Delta\eqdef M\setminus C^\M=\overline{C^\m}=C^\m\sqcup F~~
\ee
be the union of all non-compact leaves and singularities. This subset
has a finite number (which we denote by $v(\momega)$) of connected
components $\Delta_s$:
\ben
\label{DeltaDec}
\Delta=\sqcup_{s=1}^{v(\momega)}\Delta_s~~.
\een 
The connected components of $F$ (which are again in finite number) are
finite unions of singular points and of non-compact but compactifiable
leaves of $\cF_\momega$ which coincide with the `compact parts' of
the singular leaves of $\bcF_\momega$ (see \eqref{cLdec}). 

One can show \cite{FKL, Levitt2} that each maximal component $C_j^\M$ is
diffeomorphic to the open unit cylinder over any of the (compact)
leaves $L_j$ of the restricted foliation $\cF_\momega|_{C_j^\M}$,
through a diffeomorphism which maps this restricted foliation to the
foliation of the cylinder given by its sections $L_j\times \{t\}$:
\ben
\label{maxcyl}
C_j^\M\simeq L_j\times (0,1)~~.
\een
In particular, we have:
\be
\rho(\momega|_{C_j^\M})=0~~.
\ee

Being connected, each non-compactifiable leaf $L$ of $\cF_\momega$ is
contained in exactly one minimal component. It was shown in
\cite{ImanishiSing} (see also Appendix of \cite{Levitt1}) that $L$ is
     {\em dense} in that minimal component. Furthermore, one has
     \cite{FKL, Levitt1}:
\be
\rho(\momega|_{C_a^\m})\geq 2~~,~~a=1,\ldots,N_\m(\momega)~~.
\ee
In particular, any minimal component $C_a^\m$ must satisfy $b_1(C_a^\m)\geq 2$. 

\paragraph{Definition.} 
The foliation $\cF_\momega$ is called {\em compactifiable} if each of its 
leaves is compactifiable, i.e. if it has no minimal components.

\subsection{The foliation graph}

Since each maximal component $C_j^\M$ is a cylinder, its frontier
consists of either one or two connected components.  When the frontier
of $C_j^\M$ is connected, there exists exactly one connected component
$\Delta_{s_j}$ of $\Delta$ such that $\fr C_j^\M\subset
\Delta_{s_j}$. When the frontier of $C_j^\M$ has two connected
components, there exist distinct indices $s'_1$ and $s''_j$ such that
these components are subsets of $\Delta_{s'_j}$ and $\Delta_{s''_j}$,
respectively. These observations allow one to define a graph as follows \cite{FKL, Honda}:

\paragraph{Definition.} 
The {\em foliation graph} $\Gamma_\momega$ of $\momega$ is the unoriented
graph whose vertices are the connected components $\Delta_s$ of
$\Delta$ and whose edges are the maximal components $C_j^\M$. An edge
$C_j^\M$ is incident to a vertex $\Delta_s$ iff a connected component
of $\fr C_j^\M$ is contained in $\Delta_s$; it is a loop at $\Delta_s$ iff 
$\fr C_j^\M$ is connected and contained in $\Delta_s$. A vertex $\Delta_s$ of
$\Gamma_\momega$ is called {\em exceptional} (or of {\em type II}) if it contains at least one
minimal component; otherwise, it is called {\em regular} (or of {\em type I}). 

\

\noindent The terminology {\em type I}, {\em type II} for vertices is
used in \cite{Gelbukh7}. Since $M$ is connected, it follows that $\Gamma_\momega$ is a connected
graph. Notice that $\Gamma_\momega$ can have loops and multiple edges as well
as terminal vertices. 
\begin{figure}[h!]
\begin{center}
\includegraphics[scale=0.5]{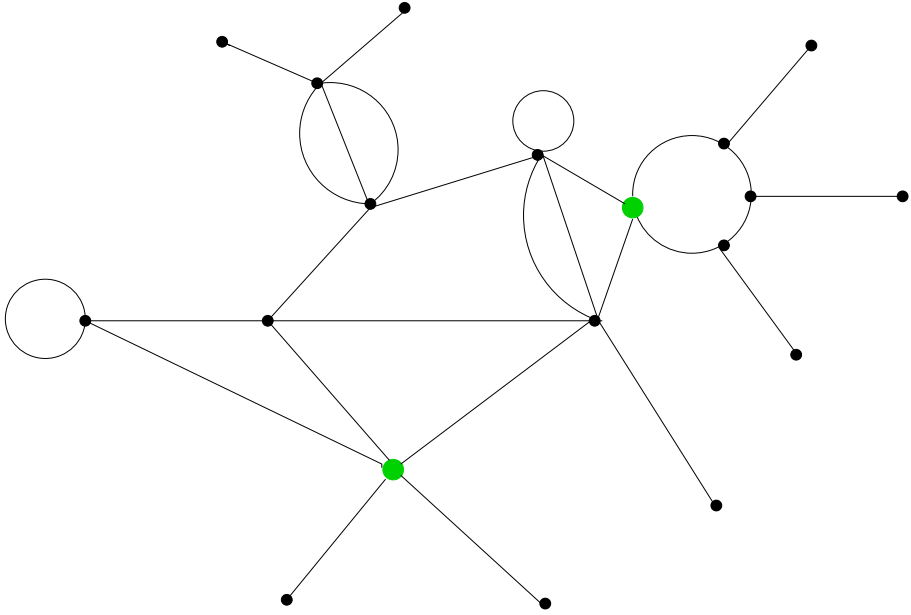}
\caption{An example of foliation graph. Regular (a.k.a type I)
  vertices are represented by black dots, while exceptional
  (a.k.a. type II) vertices are represented by green blobs. All
  terminal vertices are regular vertices and correspond to center
  singularities.  Notice that the graph can have multiple edges as
  well as loops.}
\end{center}
\end{figure}
Let $\deg \Delta_s$ denote the degree (valency) of $\Delta_s$ as a
vertex of the foliation graph. A regular vertex $\Delta_s$ can be of two types: 
\begin{itemize}
\itemsep 0.0em
\item A center singularity $\Delta_s=\{p\}$ (with $p\in \Sigma_0(\momega)$), when $\deg
\Delta_s=1$. In this case, $\Delta_s$ is a terminal vertex of $\Gamma_\momega$.

\item  A compact singular leaf when $\deg \Delta_s \geq 2$. 
\end{itemize}
Every exceptional vertex is a union of minimal components, singular points and compactifiable non-compact leaves of 
$\cF_\momega$. For any vertex $\Delta_s$ of the foliation graph, we have \cite{Gelbukh7}:
\be
|\Delta_s \cap \Sigma_1^\sp(\momega)|\geq \deg \Delta_s+2m_{\Delta_s}-2~~, 
\ee
where $m_{\Delta_s}$ is the number of minimal components contained in
$\Delta_s$. In particular, a regular vertex with $\deg \Delta_s>2$ is a compact singular leaf which
contains at least one splitting strong saddle singularity. The number of edges $e(\Gamma_\momega)$ equals 
$N_\M(\momega)$ while the number of vertices $v(\Gamma_\momega)$ equals
 $v(\momega)$. Furthermore, it was shown in \cite{Gelbukh8} that the
cycle rank $b_1(\Gamma_\momega)$ equals $c(\momega)$.  Thus:
\be
e(\Gamma_\momega)=N_\M(\momega)~~,~~ v(\Gamma_\momega)=v(\momega)\leq |\Sing(\momega)|~~,~~b_1(\Gamma_\momega)=c(\momega)~~.
\ee
The graph Euler identity $e(\Gamma_\momega)=v(\Gamma_\momega)+b_1(\Gamma_\momega)-1$ 
implies: 
\be
N_\M(\momega)=c(\momega)+v(\momega)-1 \leq c(\momega)+|\Sing(\momega)|-1~~,
\ee
where we noticed that $v(\momega)\leq |\Sing(\momega)|$ since each $\Delta_s$ contains at least one 
singular point. An example of foliation graph is depicted in Figure 3.
\paragraph{Constraints on the foliation graph from the irrationality rank of $\momega$.}

\begin{figure}
\centering
\!\!\!\!\!\begin{subfigure}{.5\textwidth}
\centering
\includegraphics[width=0.2\linewidth]{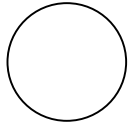}

\

\caption{Foliation graph when $\cW=\emptyset$ and $\rho(\momega)=1$.}
\end{subfigure}~~~~~~
\begin{subfigure}{.5\textwidth}
\centering
\vspace{1.5em}
\includegraphics[width=0.05\linewidth]{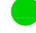}
\vspace{2.6em}
\caption{Foliation graph when $\cW=\emptyset$ and $\rho(\momega)>1$.}
\end{subfigure}
\caption{Degenerate foliation graphs in the everywhere non-chiral case.}
\label{fig:regfolgraph}
\end{figure}

When the chiral locus $\cW$ is empty (i.e. when $\momega$ is
nowhere-vanishing) we have $\Sing(\momega)=\emptyset$ and ${\bar
  \cF}_\momega=\cF_\momega$ is a regular foliation. Even though this
doesn't fit our assumption $\Sing\momega\neq \emptyset$, one can
define a (degenerate) foliation graph also in this situation (which was
considered in \cite{g2}). In this case, knowledge of the irrationality
rank of $\f$ determines the topology of the foliation $\cF_\momega$
for any $\momega\in \f$. Namely, one has only two possibilities (see
Figure \ref{fig:regfolgraph}):
\begin{itemize}
\itemsep 0.0em
\item $\rho(\f)=1$, i.e. $\f$ is projectively rational. Then there
  exists exactly one maximal component (which coincides with $M$) and
  no minimal component. The foliation ``graph'' consists of one loop
  and has no vertices; $M$ is a fibration over $S^1$ as a
  consequence of Tischler's theorem \cite{Tischler}.
\item $\rho(\f)>1$, i.e. $\f$ is projectively irrational. There exists
  exactly one minimal component (which coincides with $M$) and no
  maximal component, i.e. $\cF_\momega$ is a minimal foliation. Then
  the foliation graph consists of a single exceptional vertex and
  every leaf of $\cF_\momega$ is dense in $M$. As explained in \cite{g2}, the 
  noncommutative geometry of the leaf space is described by the $C^\ast$ algebra 
  $C(M/\cF_\momega)$ of the foliation, which is a non-commutative torus of dimensions 
  $\rho(\momega)$. Notice that this refined topological information is not reflected 
  by the foliation graph. 
\end{itemize}

\!\!\!\!\!

\

\noindent The situation is much more complicated when
$\Sing(\momega)$ is non-empty, in that knowledge of
$\rho(\momega)$ does not suffice to specify the topology of the
foliation. In this case, knowledge of $\rho(\f)$ allows one to say
only the following:
\begin{itemize}
\itemsep 0.0em
\item When $\rho(\f)=1$, then the foliation $\cF_\momega$ is
  compactifiable for any $\momega\in \f$ \cite{FKL} and the inequality
  \eqref{crank} below requires $c(\momega)\geq 1$. Hence the foliation
  graph $\Gamma_\momega$ has only regular vertices and must have at
  least one cycle.  Except for this, nothing else can be
  said about $\cF_\momega$ only by knowing that $\rho(\f)=1$.  Indeed,
  it was shown in \cite{Gelbukh8} that any compactifiable Morse form
  foliation $\cF_{\momega'}$ with $c(\momega')\geq 1$ can be realized
  as the foliation defined by a Morse form $\momega$ belonging to a
  projectively rational cohomology class. It was also shown in
  loc. cit. that such a foliation can in fact be realized by a Morse
  form of any irrationality rank lying between $1$ and
  $c(\momega')$, inclusively.
\item When $\rho(\f)>1$, then $\cF_\momega$ may be
  either compactifiable or non-compactifiable, hence the foliation
  graph may or may not have exceptional vertices; when $\cF_\momega$
  is compactifiable, then $\Gamma_\momega$ has no exceptional vertices
  and has a number of cycles at least equal to
  $\rho(\momega)$. Criteria for compactifiability of $\cF_\momega$ can
  be found in \cite{FKL, Gelbukh1,Gelbukh5} and are given below.
\end{itemize}

\paragraph{Theorem \cite{FKL, Gelbukh1,Gelbukh5}.}The following statements are equivalent: 
\begin{enumerate}[(a)]
\itemsep0em 
\item $\cF_\momega$ is compactifiable
\item The period morphism $\per_\f:\pi_1(M)\rightarrow \R$ factorizes through a group morphism $\pi_1(M)\rightarrow K$, where $K$ is a free group
\item $H_\momega^\perp\subset \ker \momega$
\item $\rk \cH_\momega=\rho(\momega)$.
\end{enumerate}
The first criterion above is Proposition 2 in
\cite[Sec. 8.2]{FKL}. Since $\cH_\momega\subset H_\momega$, we have
$\rk \cH_\momega\leq \rk H_\momega=c(\momega)$ and the theorem shows
that compactifiability of $\cF_\momega$ requires:
\ben
\label{crank}
\rho(\momega)\leq c(\momega)~~.
\een

\paragraph{Remark.} 
By its construction, the foliation graph discards topological
information about the restriction of the foliation to the minimal
components of the Novikov decomposition, which are represented in the
graph by exceptional vertices.  As in the case
$\Sing(\momega)=\emptyset$, the $C^\ast$ algebra of the foliation
should provide more refined information about the topology of ${\bar
  \cF}_\momega$ than the foliation graph.  To our knowledge, this
$C^\ast$ algebra has not been computed for foliations given by a Morse
1-form.

\paragraph{The oriented foliation graph.} 
For each maximal component $C_j^\M$, the diffeomorphism \eqref{maxcyl} can be 
chosen\footnote{The sign of $\int_{\gamma_j}\momega$ does not depend on 
the choice of $\gamma$ since $\momega$ vanishes on the leaves of
$\cF_\momega$. If the sign is negative, then it can be made positive
by composing the diffeomorphism \eqref{maxcyl} with $\id_{L_j}\times
R$, where $R\in \Diff_-((0,1))$ is any orientation-reversing
diffeomorphism of the interval $(0,1)$.}  such that the sign of the
integral $\int_{\gamma_j}\momega$ is positive along any smooth curve
$\gamma_j:(0,1)\rightarrow C^\M_j$ which projects to the interval
$(0,1)$.  Identifying the corresponding edge $e_j$ with this interval,
this gives a canonical orientation $\vec{e}_j$ of $e_j$ which
corresponds to ``moving along $e_j$ in the direction of increasing
value if $h_j$'', where $h_j$ is any locally-defined smooth function
on an open subset of $C^\M_j$ whose exterior derivative equals
$\momega$. It follows that the foliation graph $\Gamma_\momega$ admits
a canonical orientation, which makes it into the {\em oriented
  foliation graph} $\vec{\Gamma}_\momega$. 

\paragraph{Weights on the oriented foliation graph.}

Using the canonical orientation, the integrals:
\ben
\label{wdef}
w_j\eqdef \int_{\gamma_j}\momega
\een
(whose value does not depend on the choice of $\gamma_j$ as above)
provide canonical positive weights on $\vec{\Gamma}_\momega$
\cite{FKL,Honda}. These weights can be used \cite{Gelbukh3} to describe
the set of Morse 1-forms $\momega$ which have the property that ${\bar
  \cF}_\momega=\bcF$ for a fixed singular foliation $\bcF$.

\paragraph{Expression for the weights in terms of $\mb$ and $\mf$.}

In our application, the vector field $n={\hat V}^\sharp\in
\Gamma(T\cU)$ is orthogonal to the leaves of $\cF$ and satisfies:
\ben
\label{nomega}
n\lrcorner \momega=4\kappa e^{3\Delta}||V||=n\lrcorner \mf -\partial_n \mb\geq 0~~
\een
as a consequence of \eqref{meq}. Equality with zero in the right hand
side occurs only at the points of $\cW=\Sing(\momega)$. It follows
that the orientation of the edges of the foliation graph is in the
direction of $n$ and that we can take $\gamma_j$ to be any integral
curve $\ell_j$ of the vector field $n|_{C_j^\M}$.  Relation
\eqref{nomega} gives:
\be
w_j=\mb_j(\gamma_j(1))-\mb_j(\gamma_j(0))+\int_{\gamma_j}\mf~~.
\ee
When $\cF_\momega$ is compactifiable, this relation implies that the
sum of weights along all edges of a cycle of the {\em oriented}
foliation graph $\vec{\Gamma}_\momega$ equals the period of $\f$ along
the corresponding homology 1-cycle $\alpha\in H_1(M)$ of $M$:
\be
\sum_{\vec{e}_j~\mathrm{in~a~cycle~of~}\vec{\Gamma}_\momega}w_j=\int_\alpha\f~~.
\ee

\subsection{The fundamental group of the leaf space}
Even though the quotient topology of the leaf space $M/{\bar
  \cF}_\momega$ can be very poor, one can use the classifying space
$\cG$ of the holonomy pseudogroup of the regular foliation
$\cF_\momega$ \cite{Hclass} to define the fundamental group of the
leaf space through \cite{Levitt3}:
\be
\pi_1(M/\bcF_\momega)\eqdef \pi_1(B\cG)~~.
\ee
Notice that $B\cG$ is an Eilenberg-MacLane space of type $K(\pi,1)$ \cite{Hclass},
(i.e. all its homotopy groups vanish except for the fundamental group)
 since $\cF_\momega$ is defined by a closed one-form and
hence the holonomy groups of its leaves are trivial. One finds
\cite{Levitt3}:
\be
\pi_1(M/\bcF_\momega)=\pi_1(M)/\cL_\momega~~,
\ee
where $\cL_\momega$ is the smallest normal subgroup of $\pi_1(M)$
which contains the fundamental group of each leaf of $\cF_\momega$.
Notice that $M\setminus \Sing(\momega)$ is connected (since $M$ is) and that the
inclusion induces an isomorphism $\pi_1(M\setminus
\Sing(\momega))\simeq \pi_1(M)$, since we assume $\dim M\geq 3$ and
hence $\Sing\momega$ has codimension at least 3 in $M$. In particular,
the period map of $\momega$ can be identified with that of
$\momega|_{M\setminus \Sing(\momega)}$. Since $\momega$ vanishes along
the leaves of $\cF_\momega$, this map factors through the projection
$\pi_1(M)\rightarrow \pi_1(M/\bcF_\momega)$, inducing a map
$\per_0(\momega):\pi_1(M/\bcF_\momega)\rightarrow \R$.

A minimal component $C^\m_a$ is called {\em weakly complete}
\cite{Levitt3} if any curve $\gamma\subset C^\m_a$ contained in
$C^\m_a$ and for which $\int_\gamma\momega$ vanishes has its two
endpoints on the same leaf of $\cF_\momega$; various equivalent
characterizations of weakly complete minimal components can be found
in loc. cit.  Let:
\begin{itemize}
\itemsep 0.0em
\item $N'_\m(\momega)$ denote the number of minimal components which are
  not weakly complete
\item $N''_\m(\momega)$ denote the number of minimal components which are
  weakly complete
\item $C^\m_{a_1},\ldots,C^\m_{a_k}$ (where $1\leq a_1<\ldots
  <a_{N''_\m(\momega)}\leq N_\m(\momega)$) denote those minimal components
  of the Novikov decomposition which are weakly complete
\item $\momega_j\eqdef \momega|_{C^m_{a_j}}$ denote the restriction of
  $\momega$ to the weakly complete minimal component $C^\m_{a_j}$
\item $\Pi_j(\momega) \eqdef \Pi(\momega_j)$ denote the period group of
  $\momega_j$. Then $\Pi_j(\momega)$ is a free Abelian group of rank 
  $\rk\Pi_j(\momega)=\rho(\omega_j)\geq 2$ \cite{Levitt3}.
\end{itemize}
With these notations, it was shown in \cite{Levitt3} that
$\pi_1(M/\bcF_\momega)$ is isomorphic with a free
product of free Abelian groups:
\be
\pi_1(M/\bcF_\momega)\simeq F_\momega \ast \Pi_1(\momega)\ast\ldots \ast \Pi_{N''_\m(\momega)}(\momega)~~,
\ee
where $\ast$ denotes the free product of groups. Furthermore
\cite{Levitt3, Gelbukh2}, the free group $F_\momega$ factors as:
\be
F_\momega\simeq \pi_1(\Gamma_\momega)\ast \Z^{\ast K(\momega)}~~,
\ee
where $\pi_1(\Gamma_\momega)\simeq \Z^{\ast c(\momega)}$ is the
fundamental group of the foliation graph and $K(\momega)$ is a
non-negative integer which satisfies $K(\momega)\geq N'_\m(\momega)$
and $K(\momega)+c(\momega)+N''_\m(\momega)\leq b'_1(M)$. Here,
$b'_1(M)$ denotes the first noncommutative Betti number of $M$
\cite{Levitt1}, whose definition is recalled in Appendix
\ref{app:fol} (which also summarizes some further information
on the topology of $\bcF$).

\subsection{On the relation to compactifications of M-theory on 7-manifolds}

One way in which one may attempt to think about our class of
compactifications is via a two-step reduction of eleven-dimensional
supergravity, as follows:

\begin{enumerate}

\item First, reduce eleven-dimensional supergravity along a leaf of
  the foliation down to a supergravity theory in four dimensions; this
  would of course be a {\em gauged} supergravity theory since the
  restrictions of $F$ and $f$ to a leaf are generally non-trivial.

\item Further reduce the resulting four-dimensional theory down to
  three dimensions, along the ``one-dimensional space'' orthogonal to
  the leaf.

\end{enumerate}

\noindent This way of thinking, which corresponds to an attempt at
generalizing the well-known, but much simpler case of ``generalized
Scherk-Schwarz compactifications with a twist'' (see, for example,
\cite{Vandoren}), turns out to be rather naive, for the following
reasons:
\begin{itemize}
\item In the general case when $\cW$ is nonempty and differs from $M$,
  there is no such thing as a ``typical leaf'' of the regular
  foliation $\cF$ of $\cU=M\setminus \cW$, in the sense that the
  leaves of this foliation are not all diffeomorphic with each
  other. As explained above, what happens instead is that the leaves
  of the restriction of $\cF$ to each of the maximal or minimal
  components of the Novikov decomposition of M are diffeomorphic with
  each other, which means that for each component of the Novikov
  decomposition one generally has a distinct diffeomorphism class of
  leaves. As such, it is unclear which of these seven-manifolds one is
  supposed to reduce on in step 1 above.  Furthermore, the extended
  foliation $\bcF$ also contains singular leaves, and it is not
  immediately clear (from a Physics perspective) how to correctly
  reduce eleven-dimensional supergravity, in the presence of fluxes,
  on such singular seven-manifolds. One should also note that the
  leaves of the restriction of $\cF$ to a minimal component of the
  Novikov decomposition are non-compact, so the reduction along such
  leaves cannot be understood as a Kaluza-Klein reduction in the
  ordinary sense.
\item In general, there is no nice ``one-dimensional space''
  transverse to the leaves. As explained above, the best candidate for
  such a space is a non-commutative space whose ``commutative parts''
  can be described by the foliation graph, but where some unknown
  non-commutative pieces have to be pasted in at the exceptional
  vertices. It is of course already unclear how to correctly reduce a
  four-dimensional supergravity theory on a graph, let alone on a
  non-commutative space.
\end{itemize}

As pointed out in \cite[Subsection 4.4.]{g2}, many of the issues
mentioned above already appear in the much simpler case when $\xi$ is everywhere
non-chiral. In that situation, the foliation graph is either a circle
(and the Novikov decomposition is reduced to a single maximal
component, all leaves being compact and mutually diffeomorphic, being the fibers of 
a fibration over the circle) or a
non-commutative torus of dimension given by the projective
irrationality rank of $\momega$ (in which case the Novikov
decomposition is reduced to a single minimal component, all leaves
being non-compact, mutually diffeomorphic and dense in $M$). Only the
first of these two cases has a chance at a meaningful interpretation
as a ``generalized Scherk-Schwarz compactification with a twist'',
where the twist is provided by the Ehresmann connection discussed in
\cite[Appendix E]{g2}, whose parallel transport generates the defining diffeomorphism $\phi_{a_\f}$
which presents $M$ as a mapping torus in that case (see \cite[Subsection 4.2]{g2}). A proper analysis of that case (which is the
simplest of this class of compactifications) is already considerably
more subtle than might seem at first sight, for the following
reason. As shown in \cite[Subsection 2.6]{g2}, the restriction of
$\xi$ to a leaf $L$ of $\cF$ induces the spinor $\eta_0$ of equation
\eqref{eta0} (see also \cite[eq. (2.21)]{g2}) which, as shown in
loc. cit., is the normalized Majorana spinor (in the seven-dimensional
sense) along the seven-manifold $L$ which induces its $G_2$ structure
and which should be used to perform the compactification of
eleven-dimensional supergravity on $L$ -- a reduction which would
constitute the first step outlined above. Notice,
however, that what one needs in our case is not the standard $\cN=1$
compactification of eleven-dimensional supergravity on a 7-manifold with $G_2$ structure
which is usually considered in the literature following
\cite{Minasian}, since the latter is a compactification down to
four-dimensional Minkowski space --- while what would be needed in our
case would be a compactification down to a space
which is related to $\AdS_3\times S^1$. Also recall from
\cite[Subsection 2.6]{g2} that $\eta_0$ is a Majorana (a.k.a. real)
spinor on $L$ (in the seven-dimensional sense) with respect to a real
structure which is dependent of the precise leaf $L$ under consideration
and not only of its diffeomorphism class. In particular, the $G_2$
structure depends on the leaf $L$ (it varies from leaf to leaf) in the
complicated manner described by Theorems 1 and 2 of \cite{g2} and
it is not invariant under the parallel transport of the
Ehresmann connection mentioned above, so proper analysis of the second
step of the reduction is considerably more involved than what one might expect based on analogy with
previous work on Scherk-Schwarz-like constructions.

A conceptually better (and more uniform) way to think of the
``relation to seven-dimensional compactifications'' (beyond the
results of \cite{g2} and of this paper, which can be viewed as already
providing such a relation since they express very explicitly the
geometry of $M$ in terms of the seven-dimensional geometry of the
leaves of the foliation) is to consider the ``partial
decompactification limit'' in which the leaf space is ``large''.  The
correct way to formulate this mathematically employs the theory of
adiabatic limits of foliations (see, for example, \cite{Kordyukov}),
which, in its most general form, concerns their behavior when the leaf
space (understood, in general, as a non-commutative space) is
``large'' in an appropriate spectral sense. This relates to extending
the ordinary adiabatic argument (which lies behind a proper
Kaluza-Klein formulation of the idea of ``two-step reduction'') to the
case of foliations. Though this subject is well-outside the scope of
the present paper, we mention that such a way of formulating the
problem leads to non-trivial mathematical questions given the fact
that the adiabatic limit of foliations is poorly understood for the
case of foliations which are not Riemannian, such as those which are
of interest in our case (see Remark 3 after Theorem 2 of reference
\cite{g2}). The adiabatic limit for the general situation when one has
to deal with a singular foliation $\bcF$ does not seem to have been
investigated in the Mathematics literature.

\subsection{A non-commutative description of the leaf space ?}

Recall from \cite{g2} that the leaf space of $\cF$ admits a very
explicit description as a non-commutative torus in the everywhere
non-chiral case (the case $\cU=M$, when the foliation graph is reduced
either to a circle or to a single exceptional vertex).  This leads to
the speculation \cite{Levitt3} that the topological information which
is lost when constructing the exceptional vertices of the foliation graph 
in the general case could be encoded
by some sort of non-commutative geometry, as expected from the fact
that such vertices are constructed by collapsing at least one minimal
component of the Novikov decomposition to a single point; since the
minimal components are foliated by dense leaves, the $C^\ast$-algebra
of their leaf space must be non-commutative. Unfortunately, it is
non-trivial to make this expectation precise, because one also has to
take into account the effect of the singular leaves of $\bcF$, so progress on this question would
first require giving a proper definition/construction of the
$C^\ast$-algebras of singular foliations in the sense of Haefliger, a
task which, to our knowledge, has not yet been carried out in the mathematics
literature. One may hope that some modification of the construction of
\cite{Androulidakis1, Androulidakis2} (which applies to singular
foliations in the sense of Stefan-Sussmann) would lead to a solution
of this problem for the case of Haefliger structures, a case which is
logically orthogonal to that considered in loc. cit.

\section{Conclusions and further directions}

We studied $\cN=1$ compactifications of eleven-dimensional
supergravity down to $\AdS_3$ in the case when the internal part $\xi$
of the supersymmetry generator is not required to be everywhere
non-chiral, but under the assumption that $\xi$ is not chiral
everywhere. We showed that, in such cases, the Einstein equations
require that the locus $\cW$ where $\xi$ becomes chiral must be a set
with empty interior and therefore of measure zero.  The regular
foliation of \cite{g2} is replaced in such cases by a singular
foliation ${\bar \cF}$ (equivalently, by a Haefliger structure on $M$)
which ``integrates'' a cosmooth singular distribution (generalized
bundle) $\cD$ on $M$.  The singular leaves of $\bcF$ are precisely
those leaves which meet the chiral locus $\cW$, thus acquiring
singularities on that locus.

We discussed the topology of such singular foliations in the generic
case when $\momega$ is a Morse one-form, showing that it is governed
by the foliation graph of \cite{MelnikovaThesis,MelnikovaGraph,
  FKL}. On the non-chiral locus, we compared the foliation approach of
\cite{g2} with the $\Spin(7)_\pm$ structure approach of
\cite{Tsimpis}, giving explicit formulas for translating between the
two methods and showing that they agree. It would be interesting to
study what supplementary constraints --- if any --- may be imposed on
the topology of $\bcF$ (and on its foliation graph) by the
supersymmetry conditions; this would require, in particular, a
generalization of the work of \cite{FKL, Honda}.

The singular foliation $\bcF$ is defined by a closed one-form
$\momega$ whose zero set coincides with the chiral locus.  Along the
leaves of $\bcF$ and outside the intersection of the latter with
$\cW$, the torsion classes are determined by the fluxes \cite{g2}. For
the singular leaves in the Morse case, this leads to a more
complicated version of the problems which were studied in
\cite{Kcones1, Kcones2} for metrics with $G_2$ holonomy (the case of
torsion-free $G_2$ structures).

The backgrounds discussed in this paper display a rich interplay
between spin geometry, the theory of G-structures, the theory of
foliations and the topology of closed one-forms \cite{Farber}. This
suggests numerous problems that could be approached using the methods
and results of reference \cite{g2} and of this paper --- not least of
which concerns the generalization to the case of singular foliations
of the non-commutative geometric description of the leaf space. In
this regard, we note that a complete solution of this problem requires
extending the construction of the $C^\ast$ algebra of regular
foliations to the case of singular foliations in the sense of
Haefliger --- a generalization which would be different from
(and, in fact, ``orthogonal'' to) that performed in
\cite{Androulidakis1, Androulidakis2} for the case of singular
foliations in the sense of Stefan-Sussmann. This problem is unsolved
already for the case of singular foliations defined by a Morse
one-form (the difficulty being in how to deal with the singular
leaves). It would be interesting to study quantum corrections to this
class of backgrounds, with a view towards clarifying their effect on
the geometry of $\bcF$. As mentioned in the introduction, the class of
backgrounds discussed here appears to be connected with the proposals
of \cite{Grana} and \cite{Bonetti}, connections which deserve to be
explored in detail.

One of the reasons why the class of backgrounds studied in this paper
may be of wider interest is because, as pointed out in \cite{Tsimpis},
the structure group of $M$ does {\em not} globally reduce to a a proper
subgroup of $\SO(8)$. This is the origin of the phenomena discussed in
this paper, which illustrate the limitations of the theory of
classical G-structures as well as of the theory of regular
foliations. In its classical form \cite{Chern}, the former does not
provide a sufficiently wide conceptual framework for a fully general
{\em global} description of all flux compactifications.

\acknowledgments{The work of C.I.L. is supported by the research grant
  IBS-R003-G1 while E.M.B. acknowledges support from the strategic
  grant POSDRU/159/1.5/S/133255, Project ID 133255 (2014), co-financed
  by the European Social Fund within the Sectorial Operational Program
  Human Resources Development 2007--2013.  This work was also financed
  by the CNCS-UEFISCDI grants PN-II-ID-PCE 121/2011 and 50/2011 and
  also by PN 09 37 01 02.  The authors thank M.~Grana, D.~Tsimpis and
  especially I.~Gelbukh for correspondence and suggestions.}

\appendix

\section{Proof of the topological no-go theorem}
\label{app:nogo}

\paragraph{Lemma.} If $\kappa=0$, then $F$ and $f$ must vanish and $\Delta$ must be constant on $M$. Furthermore, 
both $\xi^+$ and $\xi^-$ must be covariantly constant on $M$ (and
hence $\xi$ is also covariantly constant) and $b$ must be constant on
$M$.

\

\noindent{\bf Proof.} The scalar part of the Einstein equations takes
the form \cite{MartelliSparks}:
\be
e^{-9 \Delta}\Box e^{9\Delta}+72\kappa^2=\frac{3}{2}||F||^2+3||f||^2~~.
\ee
Integrating this by parts on $M$ when $\kappa=0$,
implies\footnote{This was first noticed in \cite{MartelliSparks}.}
that $F$ and $f$ must vanish while $\Delta$ must be constant on $M$. In
this case $Q=0$ and $\mathbb{D}=\nabla^S$ so the supersymmetry
conditions \eqref{par_eq} reduce to the condition that $\xi$ is
covariantly constant on $M$. Then \eqref{flatness} implies that each
of $\xi^+$ and $\xi^-$ are covariantly constant and hence $b$ is
constant on $M$ while $V,~Y$ and $Z$ are covariantly constant since
$\nabla^S$ is a Clifford connection in the sense of \cite{BGV}. Notice
that both $\xi^+$ and $\xi^-$ can still be non-vanishing so we can
still have $|b|<1$, in which case $V$ is also non-vanishing and we
still have a global reduction of structure group to $G_2$ on
$M$. $\blacksquare$

\

\noindent {\bf Proof of the Theorem.} The argument is based on the results of
\cite{Tsimpis}. Let us assume that $\Int \cW$ is non-empty. Then at
least one of the subsets $\cW^+$ and $\cW^-$ has non-empty interior
and we can suppose, without loss of generality, that $\Int \cW^+\neq
0$. Let $U$ be an open non-void subset of $\cW^+$. By the definition
of $\cW^+$, we must have $\xi=\xi^+$ and thus $b=+1$ and $V=0$ at any
point of $\cW^+$ and hence of $U$. Since the one-form $L$ of
\cite{Tsimpis} (which we denote by $L_+$) is given in terms of $V$
by expression \eqref{Ld}, it follows that $L_+$ vanishes at
every point of $U$. The second of equations (3.16) of \cite{Tsimpis}
(notice that we {\em can} use the differential equations of
\cite{Tsimpis} on the subset $U$ of $\cW^+$ since $U$ is open) shows
that the following relation holds on $U$:
\be
e^{-12\Delta}\ast \dd\ast \Big(e^{12\Delta}\frac{L_+}{1+L_+^2} \Big)-4\kappa\frac{1-L_+^2}{1+L_+^2}=0~~
\ee
and since $L_+|_U=0$ this gives $\kappa=0$. The Lemma now implies that
$b$ is constant on $M$ and since the set $\cW^+$ where $b$ equals $+1$
is non-void by assumption, it follows that $b=+1$ on $M$ i.e. that we
must have $\cW^+=M$, which is Case 1 in the Theorem. Had we assumed
that $\Int\cW^-$ were non-empty, we would have concluded in the same
way that $\cW^-=M$, which is Case 2 in the theorem.

The argument above shows that either Case 1 or Case 2 of the Theorem hold or
that both $\cW^+$ and $\cW^-$ must have empty interior. If at least
one of them is a non-empty set, then we are in Case 4 of the Theorem.
If both of them are empty sets, then $\cU$ coincides with $M$ by the
definition of $\cU,\cW^+$ and $\cW^-$ and we are in Case 3. In Case 4, 
the fact that $\cW^\pm$ have empty interiors and the fact that
they are both closed and disjoint implies immediately that they are
both contained in the closure of $U$ and hence so is their union
$\cW$. Since $M$ equals $\cU\cup \cW$, this implies that the closure
of $\cU$ equals $M$ i.e. that $\cU$ is dense in $M$.  By the
definition of $\cU$ and $\cW$ we have $\cW=M\setminus \cU$ and, since
$M$ is the closure of $\cU$, this means that $\cW$ is the 
frontier of $\cU$. $\blacksquare$

\section{The case $\kappa=0$}
\label{app:0}

For completeness, we briefly discuss the case $\kappa=0$ (which corresponds to compactifications down to Minkowski space $\R^{1,2}$).

\subsection{When $M$ is compact}

In this case, the lemma of Appendix \ref{app:nogo} implies that $F$
and $f$ vanish while $\xi$ (thus also $V$) are covariantly
constant on $M$ and hence $b$ (and thus the norm of $V$) are constant on
$M$; furthermore, $\Delta$ is constant on $M$. The operator $\mathbb{D}$ of \eqref{par_eq} reduces to $\nabla^S$ while
$Q$ reduces to zero.  Since $\nabla^S$ is a Clifford connection on $S$
while $\nabla \nu=0$, it follows that $\nabla^S$ commutes with
$\gamma(\nu)$. Therefore, the supersymmetry equation $\nabla^S\xi=0$
implies $\nabla^S\xi^\pm=0$. When $\xi$ is non-chiral, this shows 
that the space of solutions $\dim \cK(\nabla^S,0)$ must be at least
two-dimensional provided that it is non-trivial. In fact, existence of
a non-trivial and non-chiral solution of the supersymmetry equations
\eqref{par_eq} is equivalent with existence of two non-trivial chiral
solutions of opposite chirality. Due to this fact, the non-chiral case
corresponds to $\cN=2$ (rather than $\cN=1$) supersymmetry
of the effective 3-dimensional theory. Notice that this phenomenon is specific to
the Minkowski case $\kappa=0$, since the operator $\mathbb{D}$ does not commute with
$\gamma(\nu)$ when $\kappa$ is non-vanishing. We thus distinguish the cases: 

\begin{enumerate}
\item $\xi$ has definite chirality at some point (and hence at every
  point) of $M$, which amounts to $|b|=1$. Then the metric $g$ of $M$
  has holonomy contained either in $\Spin(7)_+$ (when $\xi=\xi_+$,
  i.e. $b=+1$ and $\cW=\cW^+=M$, $\cU=\cW^-=0$, which is Case 1 of the
  topological no-go theorem of Subsection 2.3) or in $\Spin(7)_-$
  (when $\xi=\xi^-$, i.e. $b=-1$ and $\cW=\cW^-=M$, $\cU=\cW^+=0$,
  which is Case 2 of the topological no-go theorem), while $V$
  vanishes identically on $M$. As a consequence, the distribution
  $\cD=\ker V$ has corank zero and coincides with $TM$; the foliation
  $\cF$ becomes a {\em codimension zero} foliation consisting of a
  single leaf equal to $M$.  These cases correspond to the classical
  limit of the well-known compactifications of \cite{Becker1}. The
  holonomy equals $\Spin(7)_\pm$ iff. $M$ is simply-connected (which
  is allowed in this case). Notice that this class of Minkowski
  compactifications {\em cannot} be viewed as the $\kappa\rightarrow
  0$ limit of the compactifications considered in \cite{g2} (which
  correspond to Case 3 of the topological no-go theorem) or in this
  paper (which correspond to Case 4 of the topological no-go theorem).
  In particular, one cannot take the limit $\kappa\rightarrow 0$ of
  the formulas given in Theorems 1, 2, 3 of \cite{g2} (which only
  apply to Case 3 or to the regular foliation $\cF$ of the non-chiral
  set $\cU$ of Case 4), since one encounters division by zero.

\paragraph{Remark.} One can also consider for example 
the case when $\cK(\nabla^S,0)\subset \Gamma(M,S^+)$ has dimension 2, which 
corresponds to the classical limit of the compactifications considered
in \cite{BeckerCY}. In this case, $M$ has holonomy contained in
$\SU(4)\subset \Spin(7)_+$, i.e. it is a Calabi-Yau fourfold. One can
turn on fluxes by considering the leading quantum correction to the
Bianchi identity for $\mathbf{G}$ in such a way as to preserve $\cN=2$
supersymmetry \cite{BeckerCY} (in which case $F$ must be a primitive
$(2,2)$ form) or $\cN=1$ supersymmetry (in which case $F$ satisfies a
weaker constraint, see \cite[Subsection 3.1]{TsimpisSU4}).

\item   $\xi$ is nowhere chiral on $M$, which amounts to $|b|<1$ and
  thus $||V||\neq 0$. Then $\cD=\ker V$ is the (corank one) kernel
  distribution of a non-trivial covariantly constant one-form. The
  metric $g$ of $M$ has holonomy contained in $G_2$, the $G_2$ group
  at every point $p\in M$ being contained in the subgroup of
  $\SO(T_pM,g_p)\simeq \SO(8)$ consisting of those proper rotations
  which preserve the one-form $V_p\in T_p^\ast M$ (rotations which
  form the group $\SO(\cD_p,g_p|_{\cD_p})\simeq \SO(7)$). The holonomy
  group at $p$ coincides with the intersection of the $\Spin(7)_+$ and
  $\Spin(7)_-$ subgroups given by the stabilizers of $\xi^\pm_p$. The
  regular foliation $\cF$ has codimension one, the restriction of the
  metric to each leaf having holonomy contained in $G_2$. As explained
  above, such compactifications lead to an effective action having
  $\cN=2$ supersymmetry in 3 dimensions. They can be viewed as the
  limit $\kappa\rightarrow 0$ of Case 3 of the topological no-go
  theorem (a case which was studied in \cite{g2}), the supersymmetry
  enhancement arising at the value $\kappa=0$. In fact, when
  $||V||\neq 0$, one can immediately take the limit $\kappa\rightarrow
  0$ in the formulas of Theorems 1,2 and 3 of loc. cit., since
  $\kappa$ appears at most linearly in those expressions (and hence the limit is manifest). 
  Using the fact that $b=0$ while $\Delta$ and
  $||V||$ are constant in the limit of interest here, one immediately
  checks, for example, that Theorem 1 of \cite{g2} reduces to a
  tautology (since $F=0$) while Theorem 2 gives $H=A=0$, which is
  equivalent with the fact that $V$ is covariantly constant ($\nabla
  V=0$) as well as $\vartheta=0$ (in the notations of loc. cit.),
  which is equivalent with $D_n\varphi=D_n\psi=0$ (which is a
  consequence of the fact that $\cD$ is the kernel distribution of a
  covariantly constant one-form) and
  $\boldsymbol{\tau}_0=\boldsymbol{\tau}_1=\boldsymbol{\tau}_2=\boldsymbol{\tau}_3=0$,
  which shows that the $G_2$ structure has trivial torsion classes,
  thus corresponding, as expected, to a metric whose restriction to
  the leaves of $\cF$ has $G_2$ holonomy. Since $H=A=0$, it follows by
  Reinhart's criterion that $\cF$ is a Riemannian foliation (the
  metric $g$ is bundle-like for $\cF$). Notice that such $\cN=2$
  Minkowski compactifications are different from the classical limit
  of the Calabi-Yau fourfold compactifications considered in
  \cite{BeckerCY}, which instead correspond to the case when the
  two-dimensional space of solutions $\cK(\nabla^S,0)$ consists of
  chiral spinors of the same chirality and hence arise as a particular case of 1. above.  
  In the case discussed here $\cK(\nabla^S,0)\subset \Gamma(M,S)$ is spanned by two chiral
  spinors of {\em opposite} chirality.  As in \cite{BeckerCY}, one
  could turn on fluxes in such Minkowski compactifications (while
  preserving $\cN=2$ supersymmetry) by considering the quantum
  correction to the Bianchi identity of $\mathbf{G}$ which is induced
  by 5-brane anomaly cancellation, leading to a class of ``first order
  corrected compactifications'' which, in our opinion, deserve further study.

\end{enumerate}
For reader's convenience, we reproduce below some results of
\cite[Theorem 3]{g2} which are relevant for this discussion, where we
use arrows to indicate the limit $\kappa=0$ with constant $\Delta$ and
constant $b$ (as required by the lemma of Appendix \ref{app:nogo}).
Solving the supersymmetry conditions, i.e. finding at least one
non-trivial solution $\xi$ for \eqref{par_eq} which is everywhere
non-chiral (and which can be taken to be everywhere of norm one) was
shown in \cite{g2} to give the following constraints:
\begin{itemize}
\item On the fluxes $f\in \Omega^1(M)$ and $F\in \Omega^4(M)$:
\beqa
f&=&4\kappa V+e^{-3\Delta}\dd(e^{3\Delta} b)\rightarrow 0,\\
F_\perp &=& \alpha_1\wedge\varphi-\hat  h_{ij}e^i\wedge\iota_{e^j}\psi ~~,\\
F_\top &=& -\iota_{\alpha_2}\psi +\chi_{ij}e^i\wedge\iota_{e^j}\varphi~~,
\eeqa
with:
\beqan
\label{G2param}
\alpha_1=\frac{1}{2||V||}(\dd b)_\perp\rightarrow 0~~&,&~~\alpha_2=-\frac{b}{2||V||}(\dd b)_\perp+\frac{3||V||}{2}(\dd\Delta)_\perp\rightarrow 0~~,\\
\tr_g(\hat\chi)=\kappa-\frac{1}{2||V||}(\dd b)_\top\rightarrow 0~~&,&~~\tr_g(\hat h)=2\kappa b-\frac{3||V||}{2}(\dd\Delta)_\top+\frac{b}{2||V||}(\dd b)_\top\rightarrow 0~~,~~\nn
\eeqan
where:
\be
\hat h_{ij}=h^{(0)}_{ij}+\frac{1}{7}\tr_g(\hat h)g_{ij}~~~,~~~ \chi_{ij}=\chi^{(0)}_{ij}+\frac{1}{7}\tr_g(\chi)g_{ij}~~,~~~\tr_g(\chi)=-\frac{4}{3}\tr_g(\hat \chi)~~.
\ee
In the above  limit, the fluxes vanish by the no-go theorem and one finds $\chi={\hat h}=0$. 

\item On the quantities $H$, $\tr A $ and $\vartheta$ of  the foliation $\cF$:
\beqan
\label{TgeomHB}
&&H_\sharp=3(\dd\Delta)_\perp-\frac{b}{||V||^2}(\dd b)_\perp \rightarrow 0~, \nn\\
&& \tr A =- \frac{8\kappa b}{||V||}+12(\dd\Delta)_\top-\frac{b(\dd b)_\top}{||V||^2} \rightarrow 0~,\nn\\
&& \vartheta=\frac{b}{2}(\dd\Delta)_\perp-\frac{1+b^2}{6||V||^2}(\dd b)_\perp \rightarrow 0~~,
\eeqan
\item On the  torsion classes  (notice that $\boldsymbol{\tau}_3$ is not constrained) of the leafwise $G_2$ structure:
\beqan
\label{TgeomTorsion}
&&\boldsymbol{\tau}_{0}=\frac{4}{7||V||}\Big[  2\kappa(3+ b^2)-\frac{3b}{2}||V||(\dd\Delta)_\top+\frac{1+b^2}{2||V||}(\dd b)_\top  \Big]\rightarrow 0~,\nn\\
&&\boldsymbol{\tau}_{1}= -\frac{3}{2}(\dd\Delta)_\perp\rightarrow 0~~,~~~\boldsymbol{\tau}_{2}=0~,~\nn\\
&&\boldsymbol{\tau}_{3}=\frac{1}{||V||}(\chi^{(0)}_{ij}-h^{(0)}_{ij})e^i\wedge\iota_{e^j}\varphi\rightarrow 0~.
\eeqan
\end{itemize}

\subsection{When $M$ is non-compact}

Even though our interest is specifically in the case when $M$ is
compact, it may be instructive to consider for the moment also the
non-compact case (this is the only place in this paper where we shall
do so).  Let us assume that $M$ is non-compact but connected and
paracompact.  In this case, the lemma of Appendix \ref{app:nogo} fails
to hold (and hence non-vanishing fluxes are allowed) but the first
part of the proof of the topological no-go theorem (which is
independent of the lemma) still applies, showing the the condition
$\kappa \neq 0$ still requires that the closed sets $\cW^+$ and
$\cW^-$ have empty interior.  When $\kappa=0$, however, any of these
sets may acquire interior points. In that case, one has a background
given by a warped product of $\R^{1,2}$ with the non-compact
Riemannian manifold $(M,g)$ and we have $\mf=\dd \mb$ on $M$. The
geometry can be described as follows upon using the chirality
decomposition $M=\cU\sqcup \cW^+\sqcup \cW^-$ (we must of course
assume that the warp factor $\Delta$ is smooth on $M$ also in the
Minkowski limit, in order to have a meaningful physical interpretation
in supergravity):
\begin{itemize}
\item The open submanifold $\cU$ of $M$ can support non-vanishing
  fluxes $F|_\cU$ and $f|_\cU$ and carries a regular codimension one
  foliation $\cF$ (the foliation which integrates the kernel
  distribution of the one-form $V|_\cU$) endowed with a longitudinal
  $G_2$ structure, whose geometry is determined by the case
  $\kappa=0$\footnote{On the locus $\cU$, one can set $\kappa=0$
    directly in all expressions given in \cite[Theorems 1, 2, 3]{g2}
    (in particular, in the expressions reproduced above), since
    $||V||$ does not vanish anywhere on $\cU$ and since those
    expressions depend at most linearly on $\kappa$.  Also notice that
    the one-form $e^{3\Delta}V$ is closed and that, when $\kappa$ is
    nonzero, it has the same kernel distribution as the one-form
    $\momega=4\kappa e^{3\Delta} V$. Since $\momega$ vanishes when
    $\kappa=0$, it must of course be replaced with $e^{3\Delta}V$ when
    considering the limit $\kappa\rightarrow 0$ of the distribution
    $\cD$. } of Theorems 2, 3 of reference \cite{g2}. This is the
  non-compact version of the solutions discussed at point 2 of the
  previous subsection. Notice that both $b$ and $\Delta$ may be
  non-constant in the non-compact case and hence the limit
  $\kappa\rightarrow 0$ (which is again trivial to take) is slightly
  different from that given in the previous subsection.
\item Up to a conformal transformation, the restriction $g|_{\Int
  \cW^+}$ has holonomy contained in $\Spin(7)_+$, the type of the
  solution along $\Int\cW^+$ being, locally, the classical limit (the
  limit when the effect of the tadpole term induced by 5-brane anomaly
  cancellation in M-theory is neglected) of the non-compact version of
  the solution considered in \cite{Becker1}; in particular, the
  restriction $F|_{\cW^+}$ is self-dual while the restriction of $f$
  to $\cW^+$ is completely determined by the warp factor, as in
  loc. cit.  These conclusions follow either from the computations of
  \cite{Becker1} (computations which are local in nature and hence
  apply on $\Int \cW^+$) or more directly by setting $\kappa=0, b=+1,
  V=L=0$ in the results of Appendix \ref{app:T}, which gives
  $F|_{\cW^+}=F^{[27]}|_{\cW^+}$ and $\theta_+=-6 \dd \Delta$, $T_+=\ast(\Phi^+\wedge \dd \Delta)$ on 
this locus (see the last remark in that appendix).

\item The restriction of $F|_{\cW^-}$ is anti-selfdual while $g|_{\Int
  \cW^-}$ is conformally of holonomy contained in $\Spin(7)_-$, the
  type of the solution along $\Int\cW^-$ being, locally, the classical
  limit of the non-compact version of the solutions considered in
  \cite{Becker1}, up to a change of orientation of $\Int\cW^-$.
\end{itemize}
Notice that the one-form $V$ vanishes along $\cW^+$ and $\cW^-$ but
that it does not vanish anywhere on $\cU$.  In general, the closures
$\cW^\pm$ of the open submanifolds $\Int\cW^\pm$ need not themselves
be manifolds, since the frontiers $\Fr(\cW^\pm)=\fr(\cW^\pm)$ could be
quite ``wild'', i.e. quite far from being immersed submanifolds of
$M$.  Globally, the geometry of $M$ can be described by saying that
$M$ admits\footnote{The concept of ``generalized G-structure''
  requires some abstract mathematical development, which will be taken
  up in a different publication.} a metric-compatible ``cosmooth
generalized G-structure of type $(G_2, \Spin(7)_+,\Spin(7)_-)$,
supported on $(\cU,\Int\cW^+,\Int\cW^-)$'', where the 
of the $\Spin(7)_\pm$ components are conformally parallel.  As in Subsection
\ref{subsec:Haefliger}, one can pack this information into a Haefliger
structure, which amounts, geometrically, to adding ``singular leaves''
to the foliation $\cF$, thus completing it to a singular foliation
$\bcF$. Namely, $\bcF$ will contain supplementary leaves of
codimension one which meet the frontier $\Fr(\cW)=\fr(\cW)$ (being
singular there) as well as supplementary leaves of codimension zero
(dimension eight) which are given by the connected components of the
open sets $\Int \cW^\pm$. The latter are open submanifolds of $M$ whose
induced metric has $\Spin(7)_\pm$ holonomy.  When the form $\momega$
is Morse, the sets $\cW^+$ and $\cW^-$ are finite (and hence have
empty interior) and the second kind of supplementary leaves do not
appear; in this case, one has a codimension one Morse form foliation
of the non-compact manifold $M$, which can again be described using a
foliation graph. The geometric description given above could be used,
in principle, to attempt a mathematical classification of all
non-compact backgrounds given by warped products $\AdS_3\times_\Delta M$, but
such a study lies well outside the scope of the present paper.

\paragraph{Remark.} 
Note that $F$ and $f$ need not be ``small'' on the locus $\cU$ in this class 
of non-compact Minkowski reductions.  The {\em small flux} approximation 
(with $M$ non-compact) along the locus $\Int\cW^+$
was studied in \cite{Tsimpis}.

\section{Comparison with the results of \cite{Tsimpis}}
\label{app:T}

Recall that the positive chirality component $\xi^+$ of $\xi$ is
non-vanishing along the locus $\cU^+$ and hence defines a $\Spin(7)_+$
structure on the open submanifold $\cU$ of $M$. The locus $\cU^+$ was
studied in \cite{Tsimpis} using this $\Spin(7)_+$ structure. In this
appendix, we show that the results of \cite{Tsimpis} agree with those
of \cite{g2} along the non-chiral locus $\cU$ when taking into account
the relation between $L$ and $V$ given in Subsection \ref{subsec:L}
and the relation between the $G_2$ and $\Spin(7)_+$ parameterizations
of the fluxes given in Subsection \ref{subsec:Spin7Param}.  Note that
reference \cite{Tsimpis} uses the notation $\Phi\eqdef \Phi^+$ and
$L\eqdef L^+$. Accordingly, in this appendix we work only with the
$\Spin(7)_+$ structure and we drop the ``+'' superscripts and
subscripts indicating this structure. Only the major steps of some
computations (many of which were performed using code based on the
package {\tt Ricci} \cite{Ricci} for {\tt
  Mathematica}$^{\textregistered}$) are given below.

\paragraph{Equations for $L$ ($V$).}

Using the relation $L=\frac{1}{1+b}V$, equations
\cite[(3.16)]{Tsimpis} take the following form when written in an
arbitrary local frame of $\cU$:
\ben
\dd\Big(e^{3\Delta} V\Big)=0~~,~~e^{-12\Delta}\ast \dd\ast (e^{12\Delta}V)-8\kappa b=0~~.
\een
These coincide with the equations discussed in the Remarks after
Theorem 3 of \cite{g2}.

\paragraph{Equations for fluxes in terms of $L$ ($V$).}
The first two and last of relations \eqref{TF1F7} take the following
coefficient form in the $\Spin(7)_+$ case, being equivalent with
equations \cite[(C.2)]{Tsimpis}:
\beqan
\label{TF1F7a}
&&F^{[\bf{1}]}_{a_1 a_2 a_3 a_4}=\frac{1}{42}\Phi_{a_1 a_2 a_3 a_4}\cF^{[\bf{1}]}~~,\nn\\
&&F^{[\bf{7}]}_{a_1 a_2 a_3 a_4}=\frac{1}{24}\Phi_{[a_1 a_2 a_3}{}^a\cF^{[\bf{7}]}_{a_4]a}~~,\\
&& F^{[\bf{35}]}_{a_1 a_2 a_3 a_4}=\frac{1}{6}\Phi_{[a_1 a_2 a_3}{}^a\cF^{[\bf{35}]}_{a_4]a}\nn~~.
\eeqan
Furthermore, the $\Spin(7)_+$ case of relation \eqref{cFF} has the following coefficient form,
which is equivalent with \cite[(C.1)]{Tsimpis}:
\ben
\label{cFFa}
F_{a_1 a_2 a_3 a_4}{\Phi^{a_1 a_2 a_3}}_{a_5}=g_{a_4 a_5}\cF^{[{\bf 1}]}+\cF^{[{\bf 7}]}_{a_4 a_5}+\cF^{[{\bf 35}]}_{a_4 a_5}~~.
\een
Reference \cite{Tsimpis} uses the notations:
\beqan
\label{Totimes}
&&(P^{\bf{7}})_{rs}^{pq}\eqdef \frac{1}{4}\left(\delta_{[r}^p\delta^q_{s]}-\frac{1}{2}\Phi_{rs}{}^{pq} \right)~~,\\
&&(L\otimes \cF^{[\bf{7}]})_{a_1 a_2 a_3}^{\bf{48}}=6\left(
L_{[a_1}\cF^{[\bf{7}]}_{a_2 a_3]}+\frac{1}{7}\Phi_{a_2 a_2 a_3}{}^b
L^a\cF^{[\bf{7}]}_{ab}\right)\Longleftrightarrow (L\otimes
\cF^{[\bf{7}]})^{\bf{48}}=2 L\wedge \cF^{[\bf{7}]}
-\frac{1}{7}\iota_{\iota_L \cF^{[\bf{7}]}}\Phi~,\nn\\
&&(L\otimes F^{[\bf{27}]})_{a_1 a_2 a_3 }^{\bf{48}}\eqdef L^aF^{[\bf{27}]}_{a a_1 a_2 a_3}~~~\mathrm{i.e.}~~~L\otimes F^{[\bf{27}]}\eqdef \iota_LF^{[\bf{27}]}~~.\nn
\eeqan
Using the relation $L=\frac{1}{1+b}V$ and the identity
$||V||^2=1-b^2$, one computes, for example:
\be
||L||^2\!\!=\frac{1- b}{1+b}~~,
~~1+||L||^2\!\!=\frac{2}{1+ b}~~,~~1-||L||^2\!\!= \frac{2b}{1+b}~~,~~\frac{1-||L||^2}{1+||L||^2}\!\!= b~~,~~\frac{L}{1+||L||^2}\!\!=\frac{1}{2}V~~.
\ee
Due to such identities, equations \cite[(3.17)]{Tsimpis} take the form:
\ben
\label{TF}
\begin{split}
&f=e^{-3\Delta}\dd (e^{3\Delta}b)+4\kappa V~~,\\
&\frac{1}{12}\cF^{[\bf{1}]}= \frac{||V||}{2(1+b)}e^{-3\Delta}[\dd (e^{3\Delta}(1+b))]_\top- \kappa (1+2b)~~,\\
&\frac{1}{96}\cF^{[\bf{7}]}_{pq}=-\frac{1}{2(1+b)}e^{-3\Delta}(P^{\bf{7}})^{rs}_{pq}V_r\partial_s(e^{3\Delta}(1+b))~~,\\
&\frac{1}{24}\cF^{[\bf{35}]}_{pq}=-\frac{||V||}{1+b}\nabla_{(p}{\hat V}_{q)}+\frac{1+b^2}{2(1+b)||V||}{\hat V}_{(p}\nabla_{q)}b+\frac{3(1-b)||V||}{2(1+b)}{\hat V}_{(p}\nabla_{q)}\Delta +\mathbb{T}_{pq}-\\
&-\frac{1}{14(1+b)}\left[ \frac{3(1-b)||V||}{1+b}(\dd b)_\top+9(1-b)||V||(\dd\Delta)_\top+8(1-b)(1+2b)\kappa\right]{\hat V}_p{\hat V}_q-\\
&-\frac{1}{14(1+b)}\left[\frac{(1-b)||V||}{2(1+b)}(\dd b)_\top+\frac{3}{2}(15-2b)||V||(\dd \Delta)_\top-(1+15b-2b^2)\kappa\right]g_{pq}~~,
\end{split}~~
\een
where the quantity $\mathbb{T}_{ab}$ (which appears in the last equation of
\cite[(3.17)]{Tsimpis}) can be expressed as:
\ben
\label{vanishing}
\mathbb{T}_{ab}\eqdef -\frac{1}{4}\Phi_{(a}{}^{cde}(L\otimes F^{[\bf{27}]})^{\bf{48}}_{b)cd}L_e =\frac{1}{4}\Phi_{(a}{}^{cde}L^fF^{[{\bf 27}]}_{b)fcd}L_e=
\frac{1-b}{2(1+b)}  (\iota_{e^{(a}}\Phi)\btu_3[(\iota_{e^{b)}}F^{[{\bf 27}]})_\parallel]~~.
\een
In an orthonormal local frame with $e_1=n$, we have:
\be
\mathbb{T}_{11}=\mathbb{T}_{1j}=\mathbb{T}_{j1}=0~~,~~\mathbb{T}_{ij}=\frac{1-b}{2(1+b)} (\iota_{e^{(i}}\varphi)\btu_2^{\perp}(\iota_{e^{j)}}F_\top^{[{\bf 27}]})=-\frac{1-b}{24(1+b)}\cF^{[{\bf 27}]}_{ij}~~.
\ee
The first equation in \eqref{TF} coincides with a relation given in
Theorem 3 of \cite{g2}.  The second equation in \eqref{TF} can be
written as:
\ben
\label{eq1}
\cF^{[{\bf 1}]}=12\left[\frac{3||V||}{2}(\dd\Delta)_\top+\frac{||V||}{2(1+b)}(\dd b)_\top-\kappa(1+2b)\right]~~,\\
\een
while the third relation in \eqref{TF} separates as follows into parts
parallel and perpendicular to $n$:
\beqan
\label{eq2}
&&\cF_\top^{[{\bf 7}]}=-6||V||\left[3(\dd\Delta)_\perp+\frac{(\dd b)_\perp}{(1+b)}\right]~~,\nn \\
&&\cF_\perp^{[{\bf 7}]}=6||V||\left[3\iota_{(\dd\Delta)_\perp}\varphi+\frac{1}{1+b}\iota_{(\dd b)_\perp}\varphi\right]~~.
\eeqan
In an orthonormal frame as above, we find that the last
equation in \eqref{TF} amounts to:
\beqan
\label{eq3}
&& \cF_{11}^{[\bf{35}]}=12\left[-\frac{3}{2}||V||(\dd \Delta)_\top-\kappa (1-2b)+\frac{1+b}{2||V||}(\dd b)_\top\right]~~,\nn\\
&& \cF_{1i}^{[\bf{35}]}e^i=12\left[\frac{1+b}{2||V||}(\dd b)_\perp-\frac{3}{2}||V||(\dd \Delta)_\perp\right]~~,\\
&& \frac{1}{2}\cF_{ij}^{[\bf{35}]}e^i\odot e^j=\frac{12}{7}\left[\frac{3}{2}||V||(\dd \Delta)_\top-\frac{1+b}{2||V||}(\dd b)_\top+\kappa(1-2b)\right]g-12(h^{(0)}-\chi^{(0)})~~.\nn
\eeqan
Substituting the expressions for $\alpha_1,\alpha_2$ and ${\hat h}, {\hat \chi}$
given in Theorem 3 of \cite{g2}, it is now easy to check that
relations \eqref{eq1}-\eqref{eq3} are equivalent with:
\beqan
&&\cF^{[{\bf 1}]}=-12\tr({\hat h}+{\hat \chi})~~,\nn\\
&&\cF_\top^{[{\bf 7}]}=-12(\alpha_1+\alpha_2)~~,\nn\\
&&\cF_\perp^{[{\bf 7}]}=~12\iota_{(\alpha_1+\alpha_2)}\varphi~~,\nn\\
&& \cF_{11}^{[\bf{35}]}=12\tr({\hat h}-{\hat \chi})~~,\\
&& \cF_{1i}^{[\bf{35}]}e^i=12(\alpha_1-\alpha_2)~~,\nn\\
&& \frac{1}{2}\cF_{ij}^{[\bf{35}]}e^i\odot e^j=-12({\hat h}-{\hat \chi})~~,\nn
\eeqan
which in turn are equivalent with \eqref{Spin7G2Param} when
$\cF^{[{\bf k}]}$ are expressed in the $\Spin(7)_+$ parameterization
using \eqref{Tparam} and \eqref{cF7perp}.  

\paragraph{Remark.} To arrive at equations \eqref{eq3}, one uses the relations:
\ben
{\hat V}_{(1;1)}=0~~,~~{\hat V}_{(1;j)}=\frac{1}{2}H_j~~,~~{\hat V}_{(i;j)}=-A_{ij}~~,
\een
which can be derived by using the local expressions given in Appendix
C of \cite{g2}. Notice that the tensor $\frac{1}{2}V_{(a,b)}e^a\odot
e^b=\frac{1}{2}V_{a;b}e^a\odot e^b={\hat V}_{(a;b)}e^a\otimes e^b$ is
the Hessian\footnote{We define the Hessian of an arbitrary one-form
  $\omega\in \Omega^1(M)$ to be the symmetric part of the tensor
  $H(\omega)\eqdef \nabla\omega\in\Gamma(M,T^\ast M\otimes T^\ast
  M)=\Omega^1(M)\otimes \Omega^1(M)$. Thus
  $H(\omega)(X,Y)=(\nabla_X\omega)(Y)=X(\omega(Y))-\omega(\nabla_XY)$
  and
  $H(\omega)_{ab}=\omega_{b;a}=e_a(\omega_b)-\Omega_{ab}^c\omega_c$ in
  any (generally non-holonomic) local frame $e_a$ of $M$, with the
  connection coefficients $\Omega_{ab}^c$ defined through
  $\nabla_{e_a}e_b=\Omega_{ab}^ce_c$.  We have
  $\Hess(\omega)_{ab}=\omega_{(a;b)}$. When $f\in \cinf$, the tensor
  $\Hess(\dd f)$ coincides with the usual Hessian of $f$.}
$\Hess({\hat V})$ of ${\hat V}$, where we remind the reader that we
use conventions \eqref{DetPerm}, which were also used in \cite{g2}.

\paragraph{Equations for the $\Spin(7)_+$ structure in terms of $V$ and of the fluxes.}

Reference \cite{Tsimpis} uses a one-form $\omega^1\in \Omega^1(M)$ and a three-form $\omega^2\in \Omega^3(M)$ which are given by \cite[eq. (3.18)]{Tsimpis}:
\ben
\label{Ttorsion}
\begin{split}
\!\!\!&\omega^1_m\!\!=\!\!\frac{\kappa}{2}L_m+\!\!\frac{3}{4}\partial_m\Delta +\!\!\frac{1}{168}(L_m\cF^{[\bf{1}]}-\!\!L^i\cF_{im}^{[\bf{7}]} )~~\Leftrightarrow~
\omega^1\!\!=\!\!\frac{\kappa}{2}L+ \!\!\frac{3}{4}\dd \Delta
+\!\!\frac{1}{168}(\cF^{[\bf{1}]}L-\!\!\iota_L\cF^{[\bf{7}]} )~~,\\
\!\!\!&\omega^2_{mpq}\!\!=\!\!\frac{1}{192}(L\otimes
\cF^{[\bf{7}]})^{\bf{48}}_{mpq}+\!\!\frac{1}{4}(L\otimes
F^{[\bf{27}]})^{\bf{48}}_{mpq}~\Leftrightarrow~ \omega^2\!\!=\!\!\frac{1}{192}(2 L\wedge
\cF^{[\bf{7}]} -\!\!\frac{6}{7}\iota_{\iota_L
  \cF^{[\bf{7}]}}\Phi)+\!\!\frac{1}{4}\iota_LF^{[\bf{27}]}~.
\end{split}
\een
These forms satisfy the equation (cf. \cite[eq. (3.15)]{Tsimpis}):
\ben
\label{Tclasses}
\partial_{[m}\Phi_{pqrs]}=-8\Phi_{[mpqr}\omega^1_{s]}-\frac{4}{15}\varepsilon_{mpqrs}{}^{ijk}\omega^2_{ijk}~~\Longleftrightarrow~~ \dd \Phi=-8\Phi\wedge \omega^1+8 \ast\omega^2 ~~,
\een
where to arrive at the coordinate-free relation we used the expression: 
\be
(\ast \omega^2)_{mpqrs}=-\frac{1}{5!}\epsilon_{mpqrs abc}(\omega^2)^{abc}~~.
\ee
Defining $\theta'\in \Omega^1(M)$ and $T'\in \Omega^3(M)$ through:
\ben
\label{omega12}
\omega^1\eqdef -\frac{7}{48}\theta'~~,~~\omega^2\eqdef -\frac{1}{8} T'~~,
\een
equations \eqref{Tclasses} take the form: 
\ben
\label{Spin7eq}
\dd \Phi=\frac{7}{6}\theta'\wedge \Phi-\ast T'~~.
\een
Relation \eqref{Spin7TorsionEq} tells us that the Lee-form $\theta$ and the characteristic torsion form $T$ of the 
$\Spin(7)_+$ structure form the particular solution of this inhomogeneous equation which also satisfies condition 
\eqref{thetaT}. It follows that $(\theta',T')$ must differ from $(\theta,T)$ through a solution $(\theta^0,T^0)$ of the homogeneous 
equation associated with \eqref{Spin7eq}, i.e. we must have:
\ben
\label{thetasol}
\theta'=\theta+\theta^0~~,~~T'=T+T^0~~\mathrm{with}~~T^0=-\frac{7}{6}\iota_{\theta^0}\Phi~~,
\een
where $\theta^0\in \Omega^1(M)$. Using \eqref{PhiT}, we find:
\ben
T^0_\perp=-\frac{7}{6}(\theta_\top^0\varphi+\iota_{\theta^0_\perp}\psi)~~,~~T^0_\top=\frac{7}{6}\iota_{\theta^0_\perp}\varphi~~
\een
and hence: 
\ben
\label{omega}
\boxed{
\begin{split}
&\omega^1_\top=-\frac{7}{48}(\theta_\top+\theta^0_\top)~~,~~\omega^2_\top=-\frac{1}{8}\left(T_\perp+\frac{7}{6}\iota_{\theta^0_\perp}\varphi\right)~~\\
&\omega^1_\perp=-\frac{7}{48}(\theta_\perp+\theta^0_\perp)~~,~~\omega^2_\perp=-\frac{1}{8}\left[T_\top-\frac{7}{6}(\iota_{\theta^0_\perp}\psi+\theta^0_\top\varphi)\right]
\end{split}}~~.
\een
Using the refined $\Spin(7)_+$ parameterization given in Table
\ref{table:Spin7param} and relations \eqref{Spin7G2Param}, equations
\eqref{Ttorsion} can be seen to be equivalent with:
\beqan
\label{int}
&&\omega^1_\top\!\!=\!\!\frac{3}{4}(\dd\Delta)_\top\!+\!\frac{\kappa||V||}{2(1+b)}\!-\!\frac{||V||}{14(1+b)}\tr_g({\hat h}+{\hat \chi})~~,~~\omega^2_\top\!\!=\!\!\frac{||V||}{14(1+b)}\iota_{(\alpha_1+\alpha_2)}\varphi~~,\\
&&\omega^1_\perp\!\!=\!\!\frac{3}{4}(\dd\Delta)_\perp\!+\!\frac{||V||}{14(1+b)}(\alpha_1+\alpha_2)~~,~~\omega^2_\perp\!\!=\!\!\frac{3||V||}{56(1+b)}\iota_{(\alpha_1+\alpha_2)}\psi\!+\!\frac{||V||}{8(1+b)}(h^{(0)}_{ij}\!
+\!\chi^{(0)}_{ij})e^i\wedge \iota_{e^j}\varphi~.\nn
\eeqan
Combining \eqref{omega} and \eqref{Ttheta}, we find that equations \eqref{int} agree with the relations given for the torsion classes of the $G_2$ structure in Theorems 2 and 3 of \cite{g2} provided that:
\ben
\label{theta0}
\boxed{\theta^0=-\frac{1}{7}\theta}~~.
\een

\paragraph{Conclusion.}
Combining the results of the paragraph above, we conclude that
equations \cite[(3.16), (3.17), (3.18)]{Tsimpis} are {\em equivalent}
on the non-chiral locus with the results of Theorems 2 and 3 of
\cite{g2}. Furthermore, the results of Section \ref{sec:G2Spin7} and
of this appendix provide a complete dictionary which allows one to
translate between the language of \cite{g2} and that of \cite{Tsimpis}
along the non-chiral locus.

\paragraph{Remark.} 
When $M$ is non-compact and $\kappa=0$, setting $V=L=0$ and $b=+1$ in
the relations above allows us to determine the nature of the solution
along the locus $\Int\cW^+$ \footnote{Some of the relations obtained 
  extend to $\cW^+$ by continuity.}. Doing so in \eqref{TF} and
\eqref{Ttorsion} and using \eqref{TF1F7a} gives
$f=3\dd \Delta$ and 
$F^{[1]}=F^{[7]}=F^{[35]}=0$ (thus 
$F|_{\cW^+}=F^{[27]}|_{\cW^+}$ is self-dual) and
$\omega^1=\frac{3}{4}\dd \Delta$, $\omega^2=0$. Relations
\eqref{thetasol} and \eqref{theta0} give $\theta'=\frac{6}{7}\theta$,
so \eqref{omega12} implies $\theta=-6\dd \Delta$. Relation
\eqref{Tclasses} gives $\dd \Phi=-6 \Phi\wedge \dd \Delta$, hence
\eqref{Spin7Torsion} implies $T=\ast (\Phi \wedge \dd \Delta)$ on
$\Int \cW^+$.  It follows that the conformally transformed metric
$e^{3\Delta} g|_{\Int \cW^+}$ has holonomy contained in $\Spin(7)_+$
(the transformation rules of $T$ and $\theta$ under a conformal
transformation can be found, for example, in \cite[Proposition
  4.1]{Ivanov}).

\section{Generalized bundles and generalized distributions}
\label{app:gendist}
Let $M$ be a connected and paracompact Hausdorff manifold. 
Recall that a {\em generalized subbundle} $F$ of a vector bundle $E$
on $M$ is simply a choice of a subspace of each fiber of that
bundle. A {\em (local) section} of $F$ is a (local) section $s$ of $E$
such that $s(p)\in F_p$ for any point $p$ lying in the domain of
definition of $s$; such a section is called {\em smooth} when it is
smooth as a section of the bundle $E$. The set of smooth sections of
$E$ over any open subset $U$ of $M$ forms a module over
$\cC^\infty(U,\R)$ which we denote by $\cC^\infty(U,F)$.  The modules
$\cC^\infty(U,F)$ need not be finitely generated; furthermore the
module $\cC^\infty(M,F)$ of global smooth sections of $F$ need not be
projective or finitely generated
\footnote{When $F$ is an ordinary subbundle of $E$, the module of
  global sections is finitely generated and projective since we assume
  $M$ to be connected, Hausdorff and paracompact.}. We say that $F$ is
         {\em algebraically locally finitely generated} if every point
         of $M$ has an open neighborhood $U$ such that
         $\cC^\infty(U,F)$ is finitely generated as a
         $\cC^\infty(U,\R)$-module. A generalized subbundle of $E$ is
         called {\em regular} if it is an ordinary smooth subbundle of
         $E$. Some references for the theory of generalized subbundles
         are \cite{BulloLewis, Drager}.

The {\em rank} of a generalized sub-bundle $F$ is the map $\rk F:
M\rightarrow \N$ which associates to each point of $M$ the dimension
of the fiber of $F$ at that point. The {\rm corank} of $F$ is the
function $\corank F\eqdef \dim M-\rk F:M\rightarrow \N$. A point $p\in
M$ is called a {\em regular point} for $F$ if the rank function is
locally constant at $p$. The {\em regular set} of $F$ is the open
subset of $M$ consisting of all regular points, while its closed
complement is the {\em singular set} of $F$; this is the set of points
where the rank of the fiber of $F$ `jumps'.  Notice that $F$ is
regular iff all points of $M$ are regular for $F$, i.e. (since $M$ is
connected) iff the rank function of $F$ is constant on $M$.

$F$ is called {\em smooth} if its fiber at any point $p$ of $M$ is
generated as a vector space by the values at $p$ of some finite
collection of smooth local sections of $E$ (equivalently, if any point
of $F_p$ is the value at $p$ of a smooth local section of $E$). It is
called {\em cosmooth} if, for all $p\in M$, the fiber $F_p$ can be
presented as the intersection of the kernels of the values at $p$ of
the elements of a finite collection of smooth local sections of the
bundle $E^\ast$ dual to $E$; this amounts to the condition that $F$ is
the polar of a smooth generalized subbundle of $G$ of $E^\ast$,
i.e. that each of its fibers $F_p$ coincides with the subspace of
$E_p$ where all linear functionals from $G_p\subset
E_p^\ast=\Hom_\R(E_p,\R)$ vanish. It is easy to see that the rank of a
smooth generalized bundle is a lower semicontinuous function, while
the rank of a cosmooth generalized subbundle is upper semicontinuous.
As a consequence, the set of regular points of a generalized subbundle
$F$ is open and dense in $M$ (hence the singular set is nowhere dense)
when $F$ is either smooth or cosmooth. Also notice that $F$ is both
smooth and cosmooth iff its rank function is constant on $M$
i.e. iff $F$ is regular.

It was shown in \cite{Drager} that a generalized subbundle $F$ of $E$
is smooth iff there exists a finite collection $s_1\ldots s_N$ of
smooth {\em global} sections of $E$ such that $F_p$ is the linear span of
$s_1(p),\ldots, s_N(p)$ for all $p\in M$; furthermore, the number $N$
of sections needed to generate all fibers of $F$ is bounded from above
by $(1+\dim M)\rk E $. Hence any generalized subbundle of $E$ is {\em
  pointwise} globally finitely-generated in this manner\footnote{This,
  of course, does {\em not} imply that it is globally or locally {\em
    algebraically} finitely generated. See \cite{Drager} for a
  counter-example.}.

A generalized subbundle of $T M$ is called a {\em singular (or
  generalized) distribution} on $M$ while a generalized subbundle of
$T^\ast M$ is called a {\em singular (or generalized) codistribution}
on $M$. Notice that a regular generalized (co)distribution is the same as
a Frobenius (co)distribution (a subbundle of the (co)tangent bundle).

\paragraph{Remark.} Given a smooth generalized codistribution which is algebraically 
locally finitely generated, its polar need not be algebraically
locally finitely generated.  To see this, consider the following:

\paragraph{Example.} 
Let $M=\R$ and take the smooth generalized codistribution generated by
the one-form $V=f(x)\dd x$, where $f\in \cC^\infty(\R,\R)$ is a smooth
function which is everywhere non-vanishing outside the interval
$[0,1]$ and vanishing on $[0,1]$. The dual $\cD$ of this
codistribution has rank one on the interval $[0,1]$ and rank zero on
its complement. For $p=0\in [0,1]$ and $I$ any open interval
containing $p$, the space $\cC^\infty(I,\cD)\subset \cC^\infty(I,\R)$
consists of all functions $h\in \cC^\infty(I,\R)$ whose open support
$\supp(h)\eqdef \{x\in \R|h(p)\neq 0\}$ is contained in the open
interval $I_+\eqdef I\cap (0,+\infty)$. Such functions form an ideal
of $\cC^\infty(I,\R)$ which is not finitely generated.

\

\noindent A generalized distribution $\cD\subset TM$ with polar generalized 
codistribution $\cD^{\mathrm{o}}\subset T^\ast M$ is called:
\begin{itemize}
\item Cartan integrable at a point $p\in M$ if there exists an
  immersed submanifold $N$ of $M$, passing through $p$, such that $T_p
  N=\cD_p$
\item Cartan integrable, if it is Cartan integrable at every point of
  $M$
\item Pfaff integrable, if the $\cinf$-module of global smooth
  sections $\cC^\infty(M,\cD^{\mathrm{o}})\subset \Omega^1(M)$ is globally
  generated by a finite number of exact forms (in particular, this
  requires that $\cD^{\mathrm{o}}$ is globally algebraically finitely generated). It is is
  easy to see that Pfaff integrability implies that
  $\cC^\infty(M,\cD^{\mathrm{o}})$ is a differential ideal of the
  (graded-commutative) differential graded ring $(\Omega(M),d,
  \wedge)$. This in turn implies (but generally is not equivalent
  with) Pfaff's condition, which states that any finite set $\momega_1,\ldots,
  \momega_N$ of generators of $\cC^\infty(M,\cD^{\mathrm{o}})$ over $\cinf$ has the
  property that $\dd \momega\wedge \momega_1\wedge \ldots \wedge
  \momega_N=0$ for all $\momega\in \cC^\infty(M,\cD^{\mathrm{o}})$.
\end{itemize}
Cartan integrability and Pfaff integrability are logically independent
conditions when $\cD$ is not regular, i.e. there exist Pfaff
integrable generalized distributions which are not Cartan integrable
and Cartan integrable generalized distributions which are not Pfaff
integrable. Furthermore, Pfaff's condition is no longer equivalent
with Pfaff integrability, unlike the case when $\cD$ is regular.
Conditions for Cartan integrability of cosmooth generalized
distributions were given in \cite{Freeman}.

Almost all cosmooth generalized distributions arising in practice fail to be
globally Cartan integrable. Due to this fact, one usually adopts the
following definition. A {\em leaf} of a cosmooth distribution $\cD$
is a maximal connected subset $\cL$ of $M$ with the property that any
two points $p,q$ of $\cL$ can be connected by a smooth curve
$\gamma:[0,1]\rightarrow M$ ($\gamma(0)=p, \gamma(1)=q$) such that the
tangent vector of $\gamma$ at each $t\in (0,1)$ lies inside the
subspace $\cD_{\gamma(t)}$. With this definition, the leaves can
be singular (i.e. they need not be immersed submanifolds of $M$) and
Cartan integrability at a point insures existence of a leaf through
that point which is locally an immersed submanifold of dimension equal
to $\dim \cD_p$. When $\cD$ fails to be Cartan integrable at $p$, the
leaf through $p$ is singular at $p$.

\paragraph{Remark.} 
Our terminology agrees with that of \cite{BulloLewis} but differs from
the notion used by other authors. For example:
\begin{itemize}
\item A {\em Stefan-Sussmann distribution} (i.e. a singular
  distribution in the sense of \cite{Stefan} and \cite{Sussmann}) is
  what we call a {\em smooth} singular distribution. For such singular
  distributions Stefan and Sussmann proved a generalization of the
  Frobenius integrability theorem (see \cite{Michor} and \cite{Ratiu}
  for textbook treatments).
\item What the authors of \cite{Androulidakis1, Androulidakis2} call
  singular distribution is what we call an algebraically locally
  finitely generated smooth distribution.  For such singular
  distributions, the Stefan-Sussmann integrability theorem states
  (similarly to the Frobenius theorem) that $\cD$ is integrable
  iff it is locally involutive with respect to the Poisson bracket
\footnote{For singular smooth distributions which are not
  algebraically finitely generated the integrability condition is more
  complicated --- see \cite{Stefan,Sussmann,Michor,Ratiu}.}.
\item The integrability conditions for a non-regular cosmooth
  distribution (equivalently, for a non-regular smooth codistribution)
  are much more complicated \cite{Freeman} than those given by Stefan
  and Sussmann for smooth distributions.
\end{itemize}

\paragraph{The cosmooth singular distribution defined by $V$.} 

Consider the codistribution $\cV\subset T^\ast M$ on $M$ which is
generated at every point by $V$, i.e. $\cV_p=\R V_p\subset T_p
M$. This distribution is smooth (since $V$ is) as well as globally
algebraically finitely generated by the single smooth section $V$ of
$T^\ast M$.  Let $\cD\subset TM$ be the polar of this
codistribution. Thus $\cD$ is the generalized subbundle of $TM$
defined by associating to a point $p$ of $M$ the kernel of the
one-form $V_p$ (which coincides with the orthogonal complement in
$T_pM$ of the dual vector $n_p=V^{\sharp}_p$ at that point). It
follows that $\cD$ is {\em cosmooth} (as the polar of a smooth
codistribution) but that it need not be algebraically locally finitely
generated (see the example above). Notice that $\cD$ is smooth iff it
is a regular Frobenius distribution, which happens only when $V$ is
everywhere non-vanishing, i.e. when the Majorana spinor $\xi$ is
everywhere non-chiral.  The fiber $\cD_p=\ker V_p \subset T_p M$ of
$\cD$ at a point $p\in M$ has rank seven when $V_p\neq 0$ and rank
eight when $V_p=0$. Since $\cD$ is cosmooth, its rank function $\rk
\cD=8-\rk\cV:M\rightarrow \N$ is upper semicontinuous; its value at
$p$ equals $7$ when $V_p \neq 0$ and equals 8 otherwise.  Assuming
that we are in Case 4 of the topological no-go theorem of Subsection
\ref{subsec:nogo}, it follows that $\corank \cD$ equals $1$ on the
non-chiral locus $\cU$ and vanishes on the chiral locus $\cW$. The set of
regular points of $\cD$ coincides with $\cU$.

\section{Some topological properties of singular foliations defined by a Morse one-form}
\label{app:fol}

\subsection{Some topological invariants of $M$}

Let $b_1'(M)$ denote the {\em first noncommutative Betti number}
\cite{Levitt1} of $M$, i.e. the maximum rank of a quotient group of
$\pi_1(M)$ which is a free group\footnote{Such quotient groups are
  allowed to be non-Abelian.}.  Let $\H(M)$ denote the largest rank
of a subgroup of $H^1(M,\Z)$ on which the cup product vanishes
identically. It was shown in \cite{Gelbukh2} that $b_1'(M)\leq
\H(M)$. Moreover, $\H(M)$ has the following properties which are
useful in computations \cite{Melnikova1, Melnikova2}:
\begin{enumerate}
\itemsep0em
\item $\H(M_1 \times M_2 ) = \max(\H(M_1 ), \H(M_2))$.
\item $\H(M_1 \# M_2 ) = \H(M_1 ) + \H(M_2)$ for $\dim M_i \geq
  2$, where $\#$ denotes the connected sum.
\item Let $r = \rk(\ker \cup)$, where $\cup$ is the cup product on
  $H^1 (M, \Z)$. Then:
\be
\frac{b_1(M) + b_2(M) r}{b_2(M) + 1}\leq \H(M) \leq \frac{b_1(M)
  b_2(M) + r}{b_2(M) + 1}~~. 
\ee
Since $r\leq b_1(M)$, this gives $\H(M)\leq b_1(M)$.
\item One has $\H(T^n)=1$ and $\H(M^2_g)=g$ where $T^n$ is the
  $n$-torus and $M^2_g$ an orientable closed surface of genus $g$.
\end{enumerate}
Combining the inequalities above gives:
\be
\boxed{b_1'(M)\leq \H(M)\leq b_1(M)}~~.
\ee
Notice that $H_{n-1}(M,\Z)$ is torsion free since it is isomorphic to
$H^1(M,\Z)\simeq \Hom(\pi_1(M,\Z),\Z)$ by Poincar\'{e} duality --- since
$M$ is a manifold, both groups are finitely generated and thus free
Abelian.  If $A\subset H_{n-1}(M,\Z)$ is any subgroup, we let
$A^\perp\subset H_1^{\rm tf}(M,\Z)$ denote the polar of $A$ with
respect to the intersection pairing $(~,~):H^{\rm tf}_1(M,\Z)\times
H_{n-1}(M,\Z)\rightarrow \Z$ (which is a perfect pairing).

\subsection{Estimate for the number of splitting saddle points}

Define:
\be
\boxed{D(\momega)\eqdef 1+\frac{|\Sigma_1^\sp(\momega)|-|\Sigma_0(\momega)|}{2}\in \frac{1}{2}\Z}~~,
\ee
where the numbers appearing in the right hand side where defined in
Section \ref{sec:Morse}.  It was shown in \cite{Gelbukh7} that
$D(\momega)\geq 0$, equality being attained iff $\momega$ is
exact. When $\momega$ is not exact, one further has $D(\momega)\geq
1$, i.e. $D(\momega)$ can never take the value $\frac{1}{2}$. All
greater integer and half-integer values can be realized for some Morse
form $\momega$ belonging to any given nontrivial cohomology class
$\f\in H^1(M,\R)\setminus\{0\}$.

\subsection{Estimates for $c$ and $N_\m$}

It was shown in \cite{Gelbukh2} that: 
\ben
\label{E1}
\boxed{c(\momega)+N_\m(\momega)\leq b_1'(M)}
\een 
and that every value of $c(\momega)$ between zero and $b_1'(M)$ is
attained by some $\momega$ which is generic and which has compactifiable
foliation $\cF_\momega$ (i.e. which has $N_\m(\momega)=0$).  This inequality
implies the non-exact estimate $c(\momega)+N_\m(\momega)\leq \H(M)$ of
\cite{Gelbukh8}. The latter reference also gives the following
estimate which is independent from \eqref{E1}:
\ben
\label{E2}
\boxed{c(\momega)+2N_\m(\momega)\leq b_1(M)}~~.
\een 
Finally, the following inequality holds \cite{Gelbukh7}: 
\ben
\label{E3}
\boxed{c(\momega)+N_\m(\momega)\leq D(\momega)}~~. 
\een
This implies an older estimate of \cite{Melnikova1}. Notice that
$D(\momega)$ can be smaller, equal to or larger than $b_1'(\momega)$ so
\eqref{E3} is independent of \eqref{E1} unless one has more
information about the form $\momega$.

\subsection{Criteria for existence and number of homologically independent compact leaves}

\paragraph{Theorem \cite{Gelbukh5}.} The following statements are equivalent: 
\begin{enumerate}[(a)]
\itemsep0em
\item $\cF_\momega$ has at least one compact leaf $L$
\item There exists a smooth non-constant function $ h\in \cinf$ (which
  need not be Morse !) such that $\momega\sim \dd h$
\item There exists a closed one-form $\alpha$ (which need not be Morse
  !) such that $\alpha \wedge \momega=0$, $\alpha$ has integer periods
  (i.e. $[\alpha]\in H^1(M,\Z)$) and $\alpha$ is not identically
  zero. Moreover, $L$ can be chosen with $[L]\neq 0$ in
  $H_{n-1}(M,\Z)$ iff $\alpha$ can be chosen with $[\alpha]\neq 0$ in
  $H^1(M,\R)$.
\end{enumerate}

\paragraph{Theorem \cite{Gelbukh5}.} The following statements are equivalent: 
\begin{enumerate}[(a)]
\itemsep0em
\item $\cF_\momega$ has $c$ homologically independent compact leaves
\item There exist $c$ cohomologically independent (over $\R$) closed
  one-forms $\alpha_i$ with {\em integer} periods, each of which
  satisfies $\alpha_i\wedge \momega=0$.
\end{enumerate}

\subsection{Generic forms} 

Recall that the Morse form $\momega$ is called generic if each
singular leaf of $\cF_\momega$ contains exactly one singular
point. Some special properties of such Morse forms are summarized in
the following:

\paragraph{Proposition \cite{Gelbukh7}.} Let $\momega$ be a generic Morse one-form. Then: 
\begin{enumerate}
\itemsep0em 
\item $D(\momega)$ is an integer and satisfies $D(\momega)\leq b_1'(M)$. Furthermore, 
  any value between $0$ and $b_1'(M)$ can be realized on $M$ by some
  generic Morse 1-form $\momega$.
\item All regular (a.k.a. type I) vertices of $\Gamma_\momega$ have degree at most 3 while each
  exceptional (a.k.a. type II) vertex contains exactly one minimal component.
\item If each of the minimal components of $\momega$ is weakly
  complete, then equality holds in \eqref{E3}.
\end{enumerate}

\subsection{Exact forms} 

Let the Morse one-form $\momega$ be exact, thus $\rho(\momega)=0$. In
this particular case, we have $\momega=\dd h$ for some globally-defined Morse function
$h\in \cinf$.  Since $M$ is compact and connected, $h$ attains its
maximum and minimum on $M$ and the image $h(M)\subset \R$ is a closed
interval $[a_1,a_N]$, where $a_1<\ldots < a_N$ are the critical values
of $h$.  We have $\Sing(\momega)=\cup_{j=1}^N S_j$, where $S_j\eqdef
\Sing(\momega)\cap h^{-1}(a_j)$ is the set of those critical points of
$h$ having critical value $a_j$. The leaves of the singular foliation
$\bcF_\momega$ are the connected components of the level sets
$h^{-1}(\{x\})$, where $x\in [a_1,a_N]$. The singular leaves are those
connected components of $h^{-1}(a_j)$ which contain at least one point
of $S_j$.  Hence the foliation $\cF_\momega$ is compactifiable and its
foliation graph projects onto the chain graph which has $a_j$ as its
vertices. The singular points belonging to $S_1$ and $S_N$ are
centers, while the remaining critical points are saddle points. The geometry
of such foliations is a classical subject in Morse theory
\cite{Milnor, Morse1, Morse2}. In this case, the form $\momega$ is generic 
iff $h$ is generic in the sense of Morse theory, i.e. iff $|S_j|=1$ for all
$j=1,\ldots, N$. In this case, $M$ can be constructed by successively
attaching handles starting from the ball $h^{-1}([0,a_1))$.

\subsection{Behavior under exact perturbations}

Fix $\f\in H^1(M)$ and let $\Omega(\f)\eqdef \{\momega\in \Omega(M)|\dd
\momega=0~~\mathrm{and}~~\momega\in \f\}$ be endowed with the
$\cC^\infty$ topology.  Define:
\begin{itemize}
\itemsep0em 
\item $\Omega_\cM(\f)\eqdef \{\momega\in \Omega(\f)|\momega~\mathrm{is~Morse}\}$
\item $\Omega_\cK(\f)\eqdef \{\momega\in \Omega_\cM(\f)|\cF_\momega~\mathrm{has~at~least~one~compact~leaf}\}$
\item $\Omega_\cy(\f)\eqdef \{\momega\in \Omega_\cM(\f)|\cF_\momega~\mathrm{is~compactifiable}\}$
\item $\Omega_\gen(\f)\eqdef \{\momega\in \Omega_\cM(\f)|\cF_\momega~\mathrm{is~generic}\}$
\end{itemize}

\paragraph{Theorem \cite{Gelbukh6}.} We have:
\begin{enumerate}
\itemsep0em 
\item $\Omega_\cM(\f)$ is open and dense in $\Omega(\f)$ while
  $\Omega_\gen(\f)$ is dense (but not necessarily open) in
  $\Omega(\f)$ (and hence also in $\Omega_\cM(\f)$).
\item $\Omega_\cK(\f)$ and $\Omega_\cy(\f)$ are open in $\Omega(\f)$
\item $\Omega_\cy(\f)\cap \Omega_\gen(\f)$ is open in $\Omega(\f)$
\item The restriction of the function $c$ (which counts the number of
  homologically independent compact leaves) to $\Omega_\cK(\f)$ is
  lower semicontinuous.
\end{enumerate}


\end{document}